\documentclass[twocolumn,amsmath,amssymb,prd,10pt,nofootinbib,superscriptaddress]{revtex4-2}

\usepackage{comment}
\usepackage{blindtext}
\usepackage{subcaption}
\usepackage{graphicx}
\usepackage{dcolumn}
\usepackage{bm}
\usepackage{romannum}

\usepackage{amsmath}
\usepackage{amssymb}
\usepackage{float}
\usepackage[export]{adjustbox}
\usepackage[labelsep=period]{caption}
\usepackage[normalem]{ulem}
\usepackage{mathtools}
\usepackage{siunitx}
\usepackage{soul}
\usepackage{physics}
\usepackage{minitoc}

\usepackage{bm}
\usepackage{xcolor}

\makeatletter
\newcommand*{\rom}[1]{\expandafter\@slowromancap\romannumeral #1@}
\makeatother

\newcommand{\EA}[1]{{\color{blue}\ EA: #1} }

\newcommand{\rvline}{\hspace*{-\arraycolsep}\vline\hspace*{-\arraycolsep}}

\begin{document}

\preprint{APS/123-QED}

\title{Gravitational waves from high-power twisted light}

\author{Eduard Atonga}%
\email{eduard.atonga@univ.ox.ac.uk}
\affiliation{%
Department of Physics, Atomic and Laser Physics sub-Department, Clarendon Laboratory, University of Oxford,\\
Parks Road, Oxford OX1 3PU, United Kingdom
}%

\author{Killian Martineau}%
\email{martineau@lpsc.in2p3.fr}
\affiliation{%
Laboratoire de Physique Subatomique et de Cosmologie, Universit\'e Grenoble-Alpes, CNRS-IN2P3\\
53, avenue des Martyrs, 38026 Grenoble cedex, France
}%

\author{Ramy Aboushelbaya}%
\email{ramy.aboushelbaya@physics.ox.ac.uk}
\affiliation{%
Department of Physics, Atomic and Laser Physics sub-Department, Clarendon Laboratory, University of Oxford,\\
Parks Road, Oxford OX1 3PU, United Kingdom
}%

\author{Aur\'elien Barrau}%
\affiliation{%
Laboratoire de Physique Subatomique et de Cosmologie, Universit\'e Grenoble-Alpes, CNRS-IN2P3\\
53, avenue des Martyrs, 38026 Grenoble cedex, France
}%

\author{Marko von der Leyen}%
\affiliation{%
Department of Physics, Atomic and Laser Physics sub-Department, Clarendon Laboratory, University of Oxford,\\
Parks Road, Oxford OX1 3PU, United Kingdom
}%
\affiliation{%
John Adams Institute for Accelerator Science, Denys Wilkinson Building, University of Oxford\\
Keble Road, Oxford OX1 3RH, United Kingdom
}%

\author{Sunny Howard}%
\affiliation{%
Department of Physics, Atomic and Laser Physics sub-Department, Clarendon Laboratory, University of Oxford,\\
Parks Road, Oxford OX1 3PU, United Kingdom
}%

\author{Abigail James}%
\affiliation{%
Department of Physics, Atomic and Laser Physics sub-Department, Clarendon Laboratory, University of Oxford,\\
Parks Road, Oxford OX1 3PU, United Kingdom
}%
\affiliation{%
John Adams Institute for Accelerator Science, Denys Wilkinson Building, University of Oxford\\
Keble Road, Oxford OX1 3RH, United Kingdom
}%
\author{Jordan Lee}%
\affiliation{%
Department of Physics, Atomic and Laser Physics sub-Department, Clarendon Laboratory, University of Oxford,\\
Parks Road, Oxford OX1 3PU, United Kingdom
}%

\author{Chunshan Lin}%
\affiliation{%
Faculty of Physics, Astronomy and Applied Computer Science, Jagiellonian University,\\ 30-348 Krakow, Poland
}%

\author{Heath Martin}%
\affiliation{%
Department of Physics, Atomic and Laser Physics sub-Department, Clarendon Laboratory, University of Oxford,\\
Parks Road, Oxford OX1 3PU, United Kingdom
}%
\author{Iustin Ouatu}%
\affiliation{%
Department of Physics, Atomic and Laser Physics sub-Department, Clarendon Laboratory, University of Oxford,\\
Parks Road, Oxford OX1 3PU, United Kingdom
}%

\author{Robert Paddock}%
\affiliation{%
Department of Physics, Atomic and Laser Physics sub-Department, Clarendon Laboratory, University of Oxford,\\
Parks Road, Oxford OX1 3PU, United Kingdom
}%

\author{Rusko Ruskov}%
\affiliation{%
Department of Physics, Atomic and Laser Physics sub-Department, Clarendon Laboratory, University of Oxford,\\
Parks Road, Oxford OX1 3PU, United Kingdom
}%

\author{Robin Timmis}%
\affiliation{%
Department of Physics, Atomic and Laser Physics sub-Department, Clarendon Laboratory, University of Oxford,\\
Parks Road, Oxford OX1 3PU, United Kingdom
}%
\affiliation{%
John Adams Institute for Accelerator Science, Denys Wilkinson Building, University of Oxford\\
Keble Road, Oxford OX1 3RH, United Kingdom
}%

\author{Peter Norreys}%
\email{peter.norreys@physics.ox.ac.uk}
\affiliation{%
Department of Physics, Atomic and Laser Physics sub-Department, Clarendon Laboratory, University of Oxford,\\
Parks Road, Oxford OX1 3PU, United Kingdom
}%
\affiliation{%
John Adams Institute for Accelerator Science, Denys Wilkinson Building, University of Oxford\\
Keble Road, Oxford OX1 3RH, United Kingdom
}%



\begin{abstract}
Recent advances in high-energy and high-peak-power laser systems have opened up new possibilities for fundamental physics research. In this work, the potential of twisted light for the generation of gravitational waves in the high frequency regime is explored for the first time. Focusing on Bessel beams, novel analytic expressions and numerical computations for the generated metric perturbations and associated powers are presented. The gravitational peak intensity is shown to reach $1.44 \times 10^{-5}~\text{W.m}^{-2}$ close to the source, and $1.01 \times 10^{-19}~\text{W.m}^{-2}$ ten meters away. Compelling evidence is provided that the properties of the generated gravitational waves, such as frequency, polarisation states and direction of emission, are controllable by the laser pulse parameters and optical arrangements.
\end{abstract}

\maketitle
\pagenumbering{arabic}
\section{\label{sec:Introduction}Introduction}

The importance of laboratory control of gravitational fields was first outlined by Weber \cite{PhysRev.117.306} in 1960. Indeed, the successful generation and detection of terrestrially controlled gravitational waves (\textit{e.g.} via mechanical or electromagnetic experimental schemes \cite{PhysRev.117.306, HalpernLaurent, 1974GrishchukSazhin, 1975GrischukSazhin, LeonidPGrishchuk_1977, Grishchuk:2003un, Kolosnitsyn:2015zua, Hertz, 2021Morozov}) would undoubtedly usher in major advances in scientific research, in the same manner as the control of electromagnetic radiation. In addition, it is intriguing to note that, even though the action of gravity on light is well known and has been extensively studied over the past century, the converse - i.e. the way light acts as a source of gravity - remains, to a large extent, unexplored. 

Of course, the idea of studying gravitational properties of electromagnetic fields is not an entirely new subject of investigation. Already in the early 1930's, Tolman \textit{et al.} \cite{Tolman:1931zza} showed that parallel beams of light do not interact gravitationally, whereas anti-parallel beams do interact with twice the naive Newtonian force. Following that pioneering study, it has become clear that subtle effects are to be expected.  In 1962, as another example, Gertsenshtein posited that it was possible to convert electromagnetic and gravitational waves into each other via the wave resonance mechanism in a static magnetic field \cite{Gert}. Then, just over a decade later, in 1975, Grishchuk and Sazhin proposed an elegant method to generate standing gravitational waves using excitations of an alternating electromagnetic field located inside a toroidal electromagnetic resonator \cite{1975GrischukSazhin}. It has been demonstrated in recent literature that confining electromagnetic fields within electromagnetic cavities and waves guides are able to excite transverse gravitational waves. Gravitational strains ranging between $10^{-44}$ and $10^{-42}$ have been estimated to be generated by Fabry-Perot cavities and toroidal transverse magnetic or transverse electric cavities respectively, for an electric field strength of $10^6$ V/m \cite{2021Morozov, füzfa2018electromagnetic}. However, there is limited potential for strain improvement due to optical damage thresholds and dielectric strength of materials, which constrain the electric field strength to approximately $10$ MV/m within electromagnetic waveguides and cavities.  This limitation does not exist for electric fields in vacuum.

 Since the turn of the new millennium, there has been a large number of both high-energy and high-peak-power lasers being commissioned around the world \cite{Dawson}. Access to these facilities, along with advances in smaller-scale, ultra-short pulse laser systems, has enabled expertise in both the classical and quantum properties of light to increase dramatically. These extraordinary advances in both the  quantum optics and high-energy density regimes make the current situation a conducive one in which to re-examine outstanding questions linking  electromagnetic and gravitational concepts.
High-energy lasers, in particular, provide an attractive platform to study gravitational aspects of light in laboratory settings, as the properties of the gravitational waves they generate are potentially controllable by specific optical arrangements and pulse properties. 

The goal of this paper is to provide a new, first-principles calculation of gravitational waves associated with laser pulses that are generated by current and next generation high-energy laser facilities. 

Previous investigations have predominantly examined the gravitational effects of light modes characterized by radial extents negligible compared to their longitudinal dimensions, treating the electromagnetic wave as a spatially one dimensional object \cite{Rätzel_2016, PhysRevD.19.3582, PhysRevD.105.104052, Tolman:1931zza}. However, contemporary investigations has ventured into the realm of light modes with non-trivial transverse profiles. Particularly, research has been directed towards the gravitational properties of Gaussian and Bessel modes \cite{10.1063/1.4803644,Schneiter_2018}. However, these studies have remained focused on relatively-simple zeroth-order transverse modes. Thus, the examination of the full effects of transverse field structures, particularly phase structures, remains incomplete. As such, we have focused our attention on the higher-order Bessel modes endowed with orbital angular momentum (OAM) \cite{
doi:10.1080/0010751042000275259,VolkeSepulveda_2002} within the framework of gravitational wave emission. The effects of different optical parameters on the gravitational wave amplitudes as well as on the associated radiated power are presented. The angular distributions of these quantities around the source are also studied in detail.

Before delving more deeply into the calculations, it is emphasized here that the very concept of gravitational wave generation exhibits many non-intuitive effects as soon as one steps aside from usually considered astrophysical sources. Many subtle -- and often implicit -- hypotheses have to be made to unambiguously extract the wave part from the static part, keeping in mind that some steps remain somehow arbitrary \cite{Gomes:2023xda}. Those conceptual issues will not be further discussed in this work but it is worth underlying that the whole reasoning raises complicated questions, in particular about energy which, in gravity, can only be defined in some specific regimes.

The article is organised as follows. In Section II, an introduction to the mathematical formulation for gravitational wave generation is provided. A discussion then follows, in Section III, on the properties of Bessel beams as a source of gravitational waves, broken down into different sub-sections highlighting: the proposed experimental set-up (A); the electromagnetic mode properties of Bessel beams (B); the spacetime deformations induced by Bessel pulses (C). There follows in Section IV further experimental considerations. In Section V, the results of numerical calculations are presented for current and future inertial confinement fusion facilities, including the limiting case of Bessel beams with no orbital angular momentum and finally specific cases where the Bessel beam does have orbital angular momentum. The implications of those results are then further discussed in Section VI. Section VII then summarises and concludes this paper.

\section{Formalism}

\textit{The purpose of this section is to introduce readers unfamiliar with gravitational physics to the main concepts and equations that will be used to compute gravitational waves emitted by stress-energy distributions composed entirely of electromagnetic radiation.}\\

The linearised Einstein field equations allow the description of weak gravitational fields in a perturbative approach and play a critical role in the study of gravitational waves \cite{buonanno_sathyaprakash_2015, Aggarwal:2020olq, Maggiore:2007ulw}. \\

The Einstein field equations without cosmological constant are given by:

\begin{equation}
    R_{\mu \nu} -\frac{1}{2}g_{\mu \nu}R = -\kappa T_{\mu \nu} ~,
    \label{Full EFEs}
\end{equation}
$g_{\mu \nu}$ being the metric tensor, $T_{\mu \nu}$ the stress-energy tensor and $R_{\mu \nu}$ and $R$ being respectively the Ricci tensor and scalar, that can be written in terms of second derivatives of $g_{\mu \nu}$. As usual $\left\lbrace \mu , \nu \right\rbrace$ indices range from $0$ to $3$. The smallness (compared to unity) of the prefactor $\kappa = 8 \pi G c^{-4} \sim 10^{-43} \; \text{N}^{-1}$ underlines how difficult it is to significantly bend spacetime. 

Because of the magnitude of the prefactor $\kappa$, any terrestrial stress-energy distribution necessitates only minor perturbations around flat spacetime. Thus we may write:
\begin{equation}
        g_{\mu \nu} = \eta_{\mu \nu} +  h_{\mu \nu} ~,~~  |h_{\mu \nu}| \ll 1 ~,
\end{equation}

with $\eta_{\mu \nu} = \text{diag}(1,-1,-1,-1)$.

The linearised Ricci tensor is thus given by:

\begin{equation}
    R_{\mu \nu} = \frac{1}{2}\left( \partial_{\mu}\partial_{\nu}h + \Box h_{\mu \nu} - \partial_{\mu}\partial_{\rho}h^{\rho}_{\nu} - \partial_{\nu}\partial_{\rho}h^{\rho}_{\mu} \right) + \mathcal{O} \left(h^2\right),
    \label{Riccitensor}
\end{equation}

Where $\Box \equiv \partial_{\mu}\partial^{\mu} = c^{-2} \partial^2/\partial t^2 - \nabla^2$ denotes the d’Alembertian operator. 
The linearised Ricci scalar is further obtained by contraction of the previous tensor:

\begin{equation}
    R = \Box h -\partial_{\rho}\partial_{\mu}h^{\mu \rho} + \mathcal{O} \left(h^2\right)~.
    \label{Ricciscalar}
\end{equation}

One useful quantity for simplifying the writing of the linearised Einstein field equations is the trace-reversed metric perturbation $\Bar{h}_{\mu \nu}$:

\begin{equation}
    \Bar{h}_{\mu \nu} \equiv h_{\mu \nu}-\frac{1}{2}\eta_{\mu \nu} h^{\alpha}_{\alpha} ~.
    \label{Trace-reverse}
\end{equation}

Substituting equations (\ref{Riccitensor}), (\ref{Ricciscalar}) and (\ref{Trace-reverse}) into equation (\ref{Full EFEs}) leads to the following linearisation of Einstein's set of equations:

 \begin{equation}
     \Box \Bar{h}_{\mu \nu} + \eta_{\mu \nu}\partial_{\sigma}\partial_{\rho}\Bar{h}^{\sigma \rho}-\partial_{\nu}\partial_{\rho}\Bar{h}^{\rho}_{\mu}-\partial_{\mu}\partial_{\rho}\Bar{h}^{\rho}_{\nu} = -2\kappa T_{\mu \nu} ~.
 \end{equation}

Further simplifications can be performed by noticing that, under infinitesimal general coordinate transformations $x'^\mu = x^\mu + \zeta^\mu(x)$, where $\zeta^\mu(x)$ are four arbitrary functions of the position satisfying $\vert \partial_\mu \zeta_\nu \vert \ll 1$, the quantity $\partial_\rho h^{\mu \rho}$ transforms as $\partial_\rho h'^{\mu \rho} = \partial_\rho h^{\mu \rho} - \Box \zeta^\mu$. It is therefore possible to choose the functions $\zeta^\mu(x)$ such that $\partial_\rho h^{\mu \rho} = \Box \zeta^\mu$, leading to the Lorenz (or harmonic) gauge condition:

\begin{equation}
\partial_{\mu}\Bar{h}^{\mu \nu}=0 ~,
\end{equation} 

where the prime has been removed for clarity and under which previous equations simplify considerably. They reduce to: 

\begin{equation}
    \Box \Bar{h}_{\mu \nu} = -2\kappa T_{\mu \nu} ~.
    \label{GWscat}
\end{equation}
In block matrix form, the stress-energy tensor for a generic electromagnetic source is given by:

\begin{equation}
   T^{\mu \nu} = \begin{pmatrix}
 u
  & \rvline & \vec{N}/c \\
\hline
  \vec{N}/c \ & \rvline &
  -\sigma_{i,j}
\end{pmatrix} ~,
\label{Eq.TmunuEM}
\end{equation}

With $u$, $\Vec{N}$ and $\sigma_{ij}$ being respectively the electromagnetic field energy density, the associated Poynting vector and the rank 2 Maxwell tensor, whose expressions are given by:

\begin{eqnarray}
    u &=& \frac{\epsilon_0 c}{2}(E^2 + c^2 B^2) ~, \\
    \vec{N} &=& \frac{\Vec{E}\times \vec{B}}{\mu_0} ~, \\
        \sigma_{ij} &=& \epsilon_0 c(E_iE_j+c^2B_iB_j)-u\delta_{i,j} ~.
\end{eqnarray}

Taking care to consider sources of finite dimensions, equations.(\ref{GWscat}) can be solved using the standard Green’s function method \cite{Hobson, Jackson, Maggiore:2007ulw} and thus the metric perturbation $\bar{h}_{\mu \nu}$ outside the source takes the form:

\begin{equation}
    \bar{h}^{\mu \nu} \left( t, \vec{x} \right) = \frac{4G}{c^4} \int_\mathcal{V} \frac{T^{\mu \nu}\left( t-\frac{|\Vec{x}-\Vec{x'}|}{c}, \Vec{x'} \right)}{|\Vec{x}-\Vec{x'}|} d^3x' ~,
    \label{sol}
\end{equation}

Where $\mathcal{V}$ corresponds to the volume of the source, $\vec{x'}$ denotes the spatial coordinates of any source point and $|\vec{x}-\vec{x'}|$ therefore corresponds to the distance between a certain point of the source and the location at which the field $\bar{h}_{\mu \nu}$ is evaluated. 

One assumes the compact source approximation, \textit{i.e.} at a distance $r$ from a source of characteristic length $L$ such that $r \gg L$. Introducing $\vec{x} \equiv r \hat{n}$ one can therefore approximate  $|\Vec{x}-\Vec{x'}| = r - \Vec{x'} \cdot \hat{n}  + \mathcal{O}((\frac{L}{r})^2)$, leading to:

\begin{equation}
    \bar{h}^{\mu \nu} \left( t, \vec{x} \right) = \frac{1}{r} \frac{4G}{c^4} \int_\mathcal{V} T^{\mu \nu}\left( t-\frac{r}{c} + \frac{\Vec{x'} \cdot \hat{n}}{c},\vec{x'} \right) d^3x' ~,
    \label{Compactsource}
\end{equation}

At first order in $L/r$. In equation (\ref{Compactsource}), $\hat{n}$ is a unit vector connecting the center of the source to any coordinate point in the far-field.\\

The reader's attention is drawn to a very well-known problem: the symmetric metric perturbation $h_{\mu \nu}$ contains 10 degrees of freedom which, in the treatment provided here, have been reduced to 6 using the Lorenz gauge condition $\partial_\mu \bar{h}^{\mu \nu} =0$. The residual gauge freedoms have caused a lot of confusion in the early days of the field. In a 1922 paper \cite{doi:10.1098/rspa.1922.0085}, A.S. Eddington stated that “Weyl has classified plane gravitational waves into three types: (1) longitudinal-longitudinal modes, (2) longitudinal-transverse modes, and (3) transverse-transverse modes.” Those modes correspond to the six remaining degrees of freedom of the metric perturbation in the Lorenz gauge. It was shown that only the transverse-transverse modes are radiative\footnote{This is in line with Eddington’s result, where he demonstrated that waves of the first and second types do not possess fixed velocities at which they propagate through space \cite{doi:10.1098/rspa.1922.0085}.} \cite{doi:10.1098/rspa.1922.0085, Flanagan_2005}. 
In general the pure gauge modes are not associated with a spacetime curvature and thus cannot be measured by any apparatus. A general treatment of the classification of metric perturbations was later conducted by Liftshitz in 1946 \cite{lifshitz_2017}, while more modern treatments of this topic can be found in \cite{Flanagan_2005,bertschinger2000cosmological}. The physical components of a metric perturbation are extracted by transforming the perturbation into the so-called transverse-traceless (TT) gauge \cite{Hobson}. To do so, one can use the lambda projector tensor \cite{Hobson, Maggiore:2007ulw}: 

\begin{eqnarray}
    \Lambda (\hat{n})_{ij,kl} &=& P_{ik}P_{jl}-P_{ij}P_{kl} ~, \\
    P_{ij} &=& \delta_{ij}-n_in_j ~,
\end{eqnarray}

With indices $\left\lbrace i,j \right\rbrace$ running from 1 to 3, such that:

\begin{equation}
    \bar{h}_{i j}^{TT} =  \Lambda (\hat{n})_{ij,kl} \bar{h}^{k l} ~.
    \label{TT}
\end{equation}

It should be emphasised here that solutions to the linearised Einstein field equations presented in (\ref{sol}) or (\ref{Compactsource}), give the form of the trace-reversed metric $\bar{h}_{\mu \nu}$ which differs \textit{a priori} from the spacetime deformation $h_{\mu \nu}$ that one was initially interested in. However if $h^{\mu}_{\mu}=0$, as in the TT gauge, then $\bar{h}^{\mu \nu} = h^{\mu \nu}$. The quantities $\bar{h}_{\mu \nu}^{TT}$ and $h_{\mu \nu}^{TT}$ are therefore equal. We note that a similar argument utilizing the vanishing trace property of the stress-energy tensor $T^{\mu}_{\mu} = 0 \Rightarrow h^{\mu}_{\mu} = 0$ yields the same result, $\bar{h}^{\mu \nu} = h^{\mu \nu}$. \\ 

The writing of perturbations in the $TT$ gauge is highly advantageous as it allows for the retention of solely the physical degrees of freedom, \textit{i.e.} the two allowable polarisation states. In this gauge the linearised Riemann curvature tensor $R_{\mu \nu \alpha \beta}$ is linked to $h_{\mu \nu}^{TT}$ such that

\begin{equation}
    R_{i0j0} = -\frac{1}{2} \frac{\partial^2 {h}_{ij}^{TT}}{\partial t^2} ~.
\end{equation}

In vacuum, all other non-zero components of the Riemann tensor can be deduced from $R_{i0j0}$ by the mean of its symmetries and Bianchi identities. It should be noted that by vacuum, we are referring to a region of space where there is only electromagnetic radiation. Moreover, in the linearised theory, the linearised Riemann tensor is invariant under residual gauge (\textit{i.e.} coordinate) transformations.\\ 

As presented in \cite{Hobson, Maggiore:2007ulw}, a gravitational wave stress-energy tensor is defined from the Riemann tensor at quadratic order in $h$, leading to the following expression for the power per unit solid angle:

\begin{equation}
    \frac{dP}{d\Omega} = \frac{r^2 c^3}{32 \pi G}\left< \frac{\partial h_{ij}^{TT}}{\partial t} \frac{\partial h^{ij,TT}}{\partial t} \right>,
    \label{Eq:PPUSA}
\end{equation}

Where the average is performed over many reduced wavelengths $\lambdabar \sim h / \partial h$ and $r$ denotes the radial  distance to the source. This quantity corresponds to the radiation flux measurable by an observer located along any radial direction with respect to the center of the source. Interestingly, this quantity scales with the square of the gravitational wave frequency,  which is of importance when studying optical frequency gravitational waves from laser fields. \\

\section{Bessel pulses as sources of gravitational waves}

Now that the general formalism has been established, it is now possible to consider the generation of gravitational waves from so-called twisted light. This is a type of laser mode which has attracted considerable interest over the past couple of decades due to its ability to carry orbital angular momentum (OAM) in addition to the well known spin angular momentum caused by their polarisation \cite{Torres}. The two most well studied twisted light modes are the Laguerre-Gaussian \cite{PhysRevA.45.8185, BARNETT1994670} and Bessel modes \cite{Durnin1, Durnin2, Quin}. They are characterised by an azimuthal phase-dependence in the transverse profile of their field, $E(x^\mu) \propto e^{il\phi}$, where $l \in \mathbb{Z}$ is the topological index which quantises the OAM state. This can be seen by calculating the total angular momentum of the electromagnetic fields of these modes, defined as:
\begin{eqnarray}
    \Vec{J} &=& \iint \Vec{\mathcal{J}} dS~, \\
    \Vec{\mathcal{J}} &=& \Vec{r}\cross(\Vec{E}\cross\Vec{B})~; \label{Angular momentum density}
\end{eqnarray}
where $\Vec{J}$ is the total angular momentum of the electromagnetic field and $\vec{\mathcal{J}}$ is its angular momentum density \cite{Jackson}. The latter can be rigorously derived by the standard method of applying Noether's theorem to the electromagnetic Lagrangian density and considering the conserved quantities due to its rotational invariance. It can then be readily shown that, for example, the time-averaged total angular momentum of a Laguerre-Gaussian mode is \cite{allen_1992}:
\begin{equation}
    \langle\Vec{J} \rangle \propto (l + \sigma)\hat{z} ~;
\end{equation}
where $\hat{z}$ is the unit vector of the direction of propagation of the mode, $l$ is the aforementioned azimuthal topological index, quantifying the contribution to the angular momentum from its orbital component, and $\sigma= \left\lbrace 0,\pm 1 \right\rbrace$ characterises the spin component and depends on the polarisation of the mode, be it linear, left- or right-circular. It should be noted that, unlike the convention in atomic physics, $l$ does not denote the total OAM but represents its projection along the propagation direction of the laser. 

Since they were first characterised, twisted light modes have found applications in various fields spanning from non-linear vacuum  and high-power laser-plasma interactions \cite{PhysRevLett.123.113604,PhysRevLett.118.033902,shi_blackman_zhu_arefiev_2022,zhang_ji_shen_2022,PhysRevLett.121.145002} to quantum and classical communication technologies \cite{Mirhosseini,Krenn_2014,7835197}. Consequently, scientists have devoted efforts to push the development of techniques enabling their generation and control \cite{s21196690,8894467, Wei:15}. Our motivation for analysing the spacetime perturbations caused by these modes is that this additional OAM  may be used to generate gravitational waves and control their characteristics. 

Although most investigations of the effects of orbital angular momentum consider Laguerre-Gaussian modes, within the paraxial approximation these modes fail to induce transverse gravitational waves along the optical axis in vacuum due to the nearly parallel propagation of the constituent rays of the Laguerre-Gaussian mode \cite{Brooker}.  As such, our focus will be directed towards exploring Bessel modes as a source of gravitational waves.\\

\subsection{Bessel modes}

The interest in Bessel modes as solutions to the electromagnetic wave equation started mainly due to one specific feature that attracted a lot of attention, their "non-diffractive" propagation \cite{Durnin1}. In essence, unlike Laguerre-Gaussian modes whose transverse spatial profile changes as they propagate, Bessel modes maintain a constant spatial profile, thereby maintaining a constant peak intensity. Additionally, they are "self-healing", their transverse spatial profile will re-establish itself after any partial obstruction. These incredible properties have made them very attractive for a variety of use cases in optics and laser-matter interactions.     

The original formulation of Bessel modes relied on solving the scalar version of the electromagnetic wave equation in cylindrical coordinates \cite{Durnin1}. However, one can achieve exact vector solutions that have the same properties by considering a particular linear superposition of plane wave solutions. Specifically, the infinite superposition of plane waves whose wave-vectors lie tangent to surface of a cone of half-angle (see figure \ref{fig:BBsetup} and \ref{fig:Region}), and are arranged around the circular base of that cone: 

\begin{equation}
    \Vec{\mathcal{E}}{(x^\mu,\psi)} = i^l E_0 e^{i(k_{\mu}(\psi)x^{\mu}+ l \psi)}\hat{n}(\psi) ~;
\end{equation}

where $l$ is the azimuthal mode number whose significance will become apparent later on, $k^{\mu}(\psi) = (\omega/c, \Vec{k}(\psi))$ and $ x^{\mu} = (ct,\vec{x}) $ denote the wave and position four-vectors respectively, and $\psi$ is a dummy variable, that  parameterises the azimuthal position of the wave-vector on the conical surface. The the wave four-vector and polarisation unit vector are given by:

\begin{eqnarray}
      \Vec{k}(\psi) &=& -k_0 \sin(\alpha) \hat{m}(\psi) + k_0 \cos(\alpha)\hat{z}~,  \\ 
     \hat{n}(\phi) &=& -\hat{\Psi} ~;
\end{eqnarray}

where $\hat{m}(\psi)=\cos(\psi)\hat{x}+\sin(\psi)\hat{y}$, $\hat{\Psi}=-\sin(\psi)\hat{x}+\cos(\psi)\hat{y}$ and $\hat{x}$, $\hat{y}$ and $\hat{z}$ are the standard Cartesian unit vectors. It can be easily checked that the spatial components of $k^{\mu}(\psi)$ always satisfy the transverse propagation condition $\hat{n}(\psi)\cdot\Vec{k}(\psi) = 0$. The electric field of a vector Bessel beam can thus be written as \cite{Nowack,mcdonald2000bessel,PhysRevLett.123.113604}: 

\begin{equation}
    \vec{E}(\Vec{x},t)=\frac{1}{2\pi} \int_0^{2\pi} \Vec{\mathcal{E}}{(x^\mu, \psi)} d\psi ~.
\end{equation}

The corresponding magnetic field is computed from the Maxwell-Faraday equation ($\nabla \times \Vec{E} = -\partial \vec{B} / \partial t $) \cite{Jackson}. The components of both fields are:

\begin{widetext}
    \begin{eqnarray}
    E_x &=& \frac{E_0}{2}\left[ J_{l+1}(\beta \rho) \sin(\omega t -k_zz +(l+1)\phi)-J_{l-1}(\beta \rho) \sin \left( \omega t -k_zz +(l-1)\phi \right)    \right]~, \label{Eq. Ex BP}\\
    E_y &=& -\frac{E_0}{2}\left[ J_{l+1}(\beta \rho) \cos(\omega t -k_zz +(l+1)\phi)+J_{l-1}(\beta \rho) \cos \left( \omega t -k_zz +(l-1)\phi \right)    \right]~, \\
    E_z &=& 0~, \\ 
    B_x &=&   \cos(\alpha) \frac{E_0}{2c}\left[ J_{l+1}(\beta \rho) \cos(\omega t -k_zz +(l+1)\phi)+J_{l-1}(\beta \rho) \cos(\omega t -k_zz +(l-1)\phi)    \right]~, \\
    B_y &=& \cos(\alpha)\frac{E_0}{2c}\left[ J_{l+1}(\beta \rho) \sin(\omega t -k_zz +(l+1)\phi)-J_{l-1}(\beta \rho) \sin(\omega t -k_zz +(l-1)\phi)    \right] ~,\\
    B_z &=& \sin(\alpha) \frac{E_0}{c} J_{l}(\beta \rho)  \cos(\omega t -k_zz +l\phi) ~, \label{Eq. Bz BP}
\end{eqnarray}
\end{widetext}

where $\rho = \sqrt{x^2 + y^2}$  denote the radial distance in cylindrical coordinates, $\phi=\arctan(y/x)$ is the azimuthal position in cylindrical coordinates, $k_z = k_0 \cos(\alpha)$ is the propagation constant, and $\beta = k_0 \sin(\alpha)$ is the transverse wave-vector which characterises the beam's "waist". The latter two satisfy the dispersion relation $\omega^2 = c^2(\beta^2 + k_z^2 )$. The functions of the form $J_{n}(\tau), n \in \mathbb{Z}$ are the n-th order Bessel functions of the first kind. From the mathematical expression of the spatial distribution of the fields, it is clear that the transverse profile of the Bessel modes is invariant along their propagation direction since the transverse wave vector $\beta$ is completely independent from the longitudinal $z$-coordinate. 

One calculates the angular momentum density of these fields using equation (\ref{Angular momentum density}), for which the explicit expression of the angular momentum density is given by:

\begin{eqnarray}
        <\vec{\mathcal{J}}> = \frac{\epsilon_0 E_0}{c} \left[ \frac{l}{k_0} J_{l}^2(\beta \rho) \hat{z} - \frac{lz \sin(\alpha)}{\beta \rho}J_{l}^2(\beta \rho) \hat{\rho}\right. \nonumber \\
    - \left. \rho \cos(\alpha)\frac{J_{l+1}^2(\beta \rho)-J_{l-1}^2(\beta \rho)}{2} \hat{\phi} \right].
\label{bessel_oam}
\end{eqnarray}

By integrating formula (\ref{bessel_oam}) over the transverse profile of the mode, the radial and azimuthal components integrate to 0 and the total angular momentum is thus quantised by the azimuthal mode number $l$ and lies along the propagation direction $\hat{z}$. As in the case of the Laguerre-Gaussian modes, this observation shows that the electromagnetic field distribution's orbital angular momentum (OAM) is indeed described by the topological index $l$.\\

\subsection{Proposed experimental setup}

\begin{figure}[h]
\centering
\hspace{-0.3cm}
\includegraphics[scale=0.15]{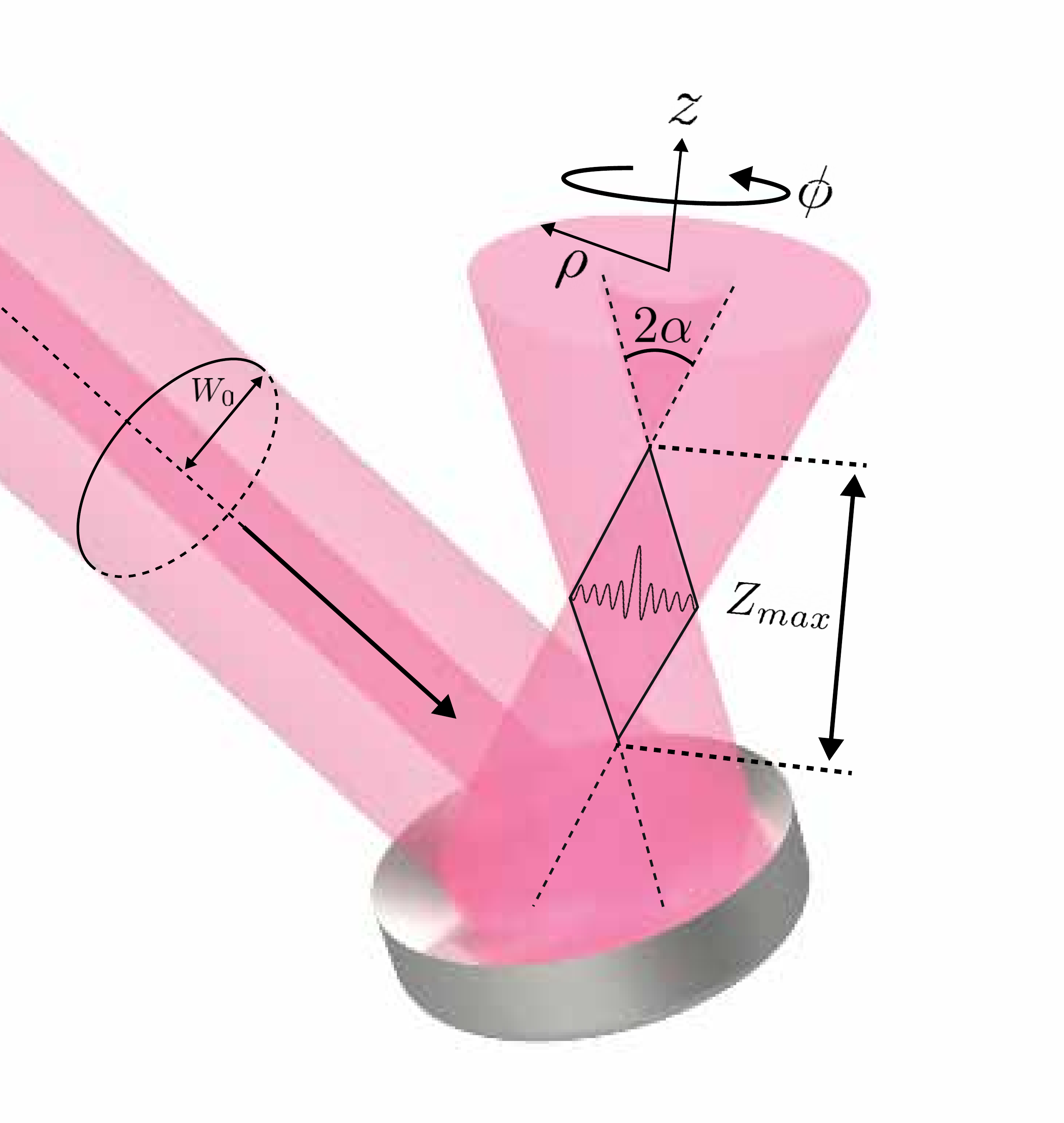}
\caption{Schematic example of Bessel pulse generation.   $W_0$ characterises the incident beam radius, $Z_{max}$ denotes the depth of focus and $\alpha$ the half-cone angle.}
\label{fig:BBsetup}
\end{figure}

Ideal Bessel beams with infinite extent are obviously non-physical as they carry infinite energy.  Realistic Bessel beams can however be generated over a finite region of space by a variety of methods, see \textit{e.g.} \cite{s21196690, 8894467}. In practice in the context of high-power laser systems, it is important to consider the laser-induced damage threshold of transmission optics. 
 Transmissive optics under the passage of a high-power laser pulse are prone to optical damage. In order to overcome the limitations associated with optical damage in transmission optics, the generation of Bessel beams can be effectively achieved by employing reflective axicon optics \cite{s21196690,Boucher:18,McLeod:54}, for which optical damage can be mitigated with the use of suitably sized laser beams.  Reflective axicons, by crossing the wavefront of the incident Laguerre-Gaussian laser pulse over a conical volume, produce an interference pattern that results in a Bessel mode over a conical region of space. Such a setup is depicted in figure \ref{fig:BBsetup}, for which the curve inside the conical region of created by the crossed wave fronts is a graphical illustration of the emergent Bessel mode. In this figure $z$, $\rho$ and $\phi$  are the usual cylindrical co-ordinates.

 The depth of focus (DOF) $Z_{max}$ is defined as the region of space over which the beam waist is assumed to remain constant. It can be expressed as:

\begin{equation}
    Z_{max} = \frac{W_0}{2 \sin(\alpha)} ~;
\end{equation}

where $W_0$ is the incident beam radius (that can reach 1 meter \cite{Dawson}) and $\alpha$ is the half cone angle appearing in figure \ref{fig:BBsetup}. Unlike their Laguerre-Gaussian counterparts whose energy distributions are localised to the central lobe, Bessel modes are able to propagate without diffraction due to their delocalised energy distribution \cite{Durnin:88}.  Bessel modes depth of field can therefore exceed the Rayleigh range of a Laguerre-Gaussian mode \cite{Durnin1, Durnin2,Brooker,PhysRevA.45.8185,BARNETT1994670}. 

\begin{figure}[h]
\centering
\includegraphics[scale=0.13]{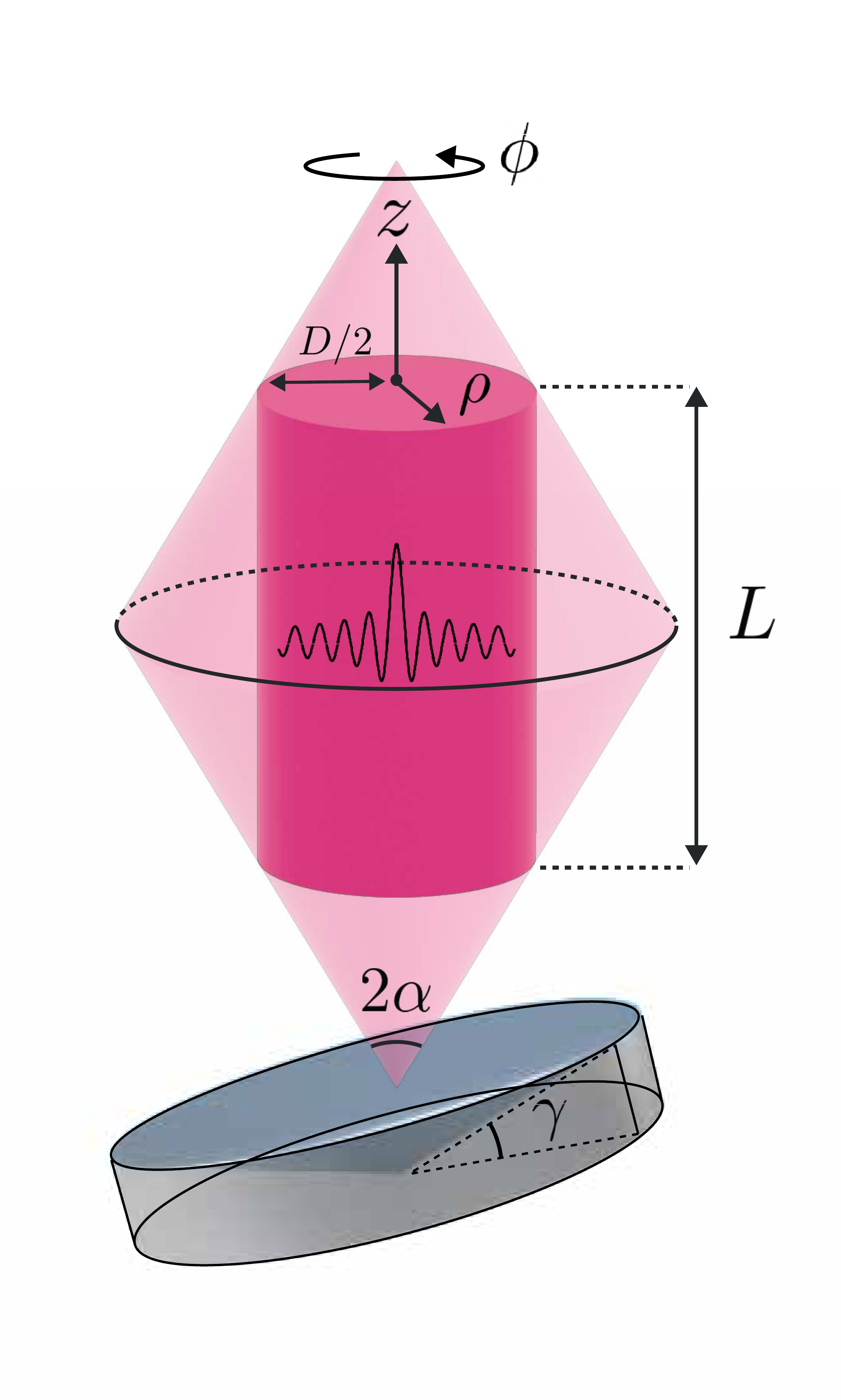}
\caption{Cylindrical integration region.}
\label{fig:Region}
\end{figure}

The radial extent of this cylinder is normally not constant along the entire propagation path. However, to facilitate the computation of the characteristics of the emitted gravitational wave radiation, it is assumed that the Bessel mode  is confined within a  cylindrical subset of the conical volume over which the Bessel mode is normally defined. An explicit depiction of the cylindrical approximation of the source volume $\mathcal{V}$ generated by a Laguerre-Gaussian laser pulse on a reflective axicon whose inner surface is inclined at an angle $\gamma$
is given in figure \ref{fig:Region}.

\begin{equation}
    \mathcal{V} \coloneqq (\rho,\phi,z) \in [0,D/2]\times [0,2\pi]\times [-L/2,L/2] \nonumber 
\end{equation} 

The axicon base angle $\gamma$ has been incorporated into figure \ref{fig:Region}, as it is  occasionally mentioned in the literature \cite{s21196690,10.1117/1.3530080}. It is related to the half-cone angle by $\alpha = 2\gamma$, however in the context of this article it will not be used. The interaction length $L$ is bounded such that $L \leq Z_{\text{max}} $. As an example, for $W_0 = 1~\text{m}$ and $\alpha = 0.6$, the corresponding $Z_{\text{max}}$ is around $1.77$ m,  for which the interaction length $L$ must satisfy $L \leq 1.77$m. 
The cylinder diameter can then be deduced geometrically from $L$ and $\alpha$ to be $D = (Z_{max}-L) \tan(\alpha) $. Importantly, the confinement of the Bessel mode within the cylindrical region introduces an additional constraint, namely $Z_{max} < c\tau_p$,  $\tau_p$ being the temporal width of the pulse.

The presence of the constraint on $Z_{max}$ allows for the existence of a period of time during which the conical region containing the Bessel mode can be completely filled. This simplification greatly reduces the computational complexity since the motion of the pulse envelope no longer needs to be considered. Within this particular setup, the duration of the gravitational wave signal $\tau_s$ can be approximated by the time duration over which the laser pulse completely fills the conical region of space generated by the crossed wave fronts of the reflective axicon, such that:

\begin{equation}
    \tau_s \approx \tau_p -\frac{L}{c} ~.
\end{equation}

It is important to acknowledge that Bessel modes are not limited to propagating along null paths in spacetime, as they can exhibit both subluminal and superluminal speeds.
It should be noted that the super-luminal speeds of the Bessel mode do not violate causality, but are instead a consequence of the contact point of the crossing wave fronts created by the axicon optic having an effective velocity that can exceeds the speed of light. This is analogous to the "scissor paradox". This characteristic allows these modes to radiate throughout their entire propagation path \cite{Rätzel_2016}. For the considered setup, effects on the generated gravitational wave signal due to the motion of the temporal envelope of a Bessel mode,  such as the Doppler effect, can not be investigated. Nonetheless, the outcomes of this work are expected to hold true when considering velocities well below the speed of light. 

Specifically, because the strain of gravitational waves is dependant on the energy contained within the source, we expect that the estimated strain amplitude of a gravitational wave generated by a laser pulse whose energy is independant of its frequency to be independent of the velocity of the pulse envelope. The effect of the pulse envelope velocity of a Bessel mode with arbitrary velocity is of theoretical interest and remains to be addressed in future research endeavors.

\subsection{Analytical calculation of spacetime deformations induced by Bessel pulses}

\subsubsection{The stress-energy tensor}

Interestingly, when dealing with relativistic sources, the energy density is not the only relevant quantity for the generation of gravitational waves. All components of the stress-energy tensor can \textit{a priori} contribute equivalently and stresses must also be taken into account. To compute the expected spacetime deformations one first fixes a frame and computes the expression of $T^{\mu \nu}$ in this frame. One chooses the frame in which the expressions of the Bessel modes' $\vec{E}$ and $\vec{B}$ fields given in equations (\ref{Eq. Ex BP} - \ref{Eq. Bz BP}) are expressed, \textit{i.e.} the frame represented in figure  \ref{fig:Region}, whose origin is located at center of the domain $z \in [-L/2,L/2]$ over which the electromagnetic field distribution is defined. The $z$-axis is aligned with the Gaussian pulse direction of propagation anterior to the reflective axicon, and the $x$ and $y$ -axes point respectively vertically and towards the reader. This frame is denoted as the "\textit{laboratory frame (LF)}". In the following, the physical quantities are expressed by default in this frame, unless stated otherwise. This convention will be of particular importance when computing the induced spacetime deformations, as the form of $h_{\mu \nu}$ depends explicitly on the chosen frame.

Plugging equations (\ref{Eq. Ex BP} - \ref{Eq. Bz BP}) into the form of the stress-energy tensor presented in equation (\ref{Eq.TmunuEM}) permits the computation of $T_{\mu\nu}$. To do so the following functions are introduced:

\begin{eqnarray}
        \xi_0 &\equiv&  J_{l}(\beta \rho) \cos\left[ \omega t -k_zz  + l\phi \right]~, \\
        \xi_+ &\equiv& J_{l+1}(\beta \rho) \cos\left[\omega t -k_zz  + (l+1)\phi\right] \\ && +J_{l-1}(\beta \rho) cos\left[\omega t -k_zz + (l-1)\phi\right] ~,  \nonumber \\
     \xi_- &\equiv& J_{l+1}(\beta \rho) \sin\left[\omega t -k_zz+(l+1)\phi \right] \\ && -J_{l-1}(\beta \rho) \sin\left[ \omega t -k_zz + (l-1)\phi \right] ~,\nonumber
\end{eqnarray}

Such that the energy density takes the form

\begin{equation}
        T_{00} = \frac{\epsilon_0 c}{2}E_0^2 \left[ \left( 1+\cos(\alpha)^2 \right) \left( \xi_{+}^2 + \xi_{-}^2 \right) +\sin(\alpha)^2 \xi_0^2 \right] ~.
   \label{Eq. T00 Bessel}
\end{equation}

Due to Bessel functions properties, an increase of the orbital angular momentum parameter $l$ leads to a decrease of the energy density of the individual lobes of $T_{00}$. This does not mean that the total energy is decreased. Indeed the number of the "lobes" increases with $l$, such that the total energy is conserved. 

The previous energy density can be written as the sum of a static and an oscillating component. The static one reads:

\begin{eqnarray}
    T_{00}^{\text{Stat}} =  \frac{\epsilon_0 c}{2}E_0^2  \left[  \left( 1+\cos(\alpha)^2 \right)  \left( J_{l-1}(\beta \rho)^2 \right. \right. \nonumber \\  
        \left. \left. + J_{l+1}(\beta \rho)^2 \right)   + \frac{1}{2}   \sin(\alpha)^2 J_{l}(\beta \rho)^2 \right] ~.
\end{eqnarray}

Once integrated over the source volume, it corresponds to an effective mass on the order of $m_{\text{eff}} \sim \mathcal{E}/c^2 \simeq 10^{-11}$ kg for a 1 MJ laser, $\mathcal{E}$ being the energy of the laser pulse. This is a tiny value, several orders of magnitude below the Planck mass. This is the reason why the chances to detect a gravitational effect due to light directly as a Newtonian force or as tidal effect seem extremely low, which motivates the focus on gravitational waves. 

The oscillating component of the electromagnetic field energy density is given by:

\begin{eqnarray}
        T_{00}^{\text{Osc}} =  \frac{\epsilon_0 c}{2}E_0^2  \left[ \left( 1+\cos(\alpha)^2 \right) J_{l+1}(\beta \rho)J_{l-1}(\beta \rho) \right. \nonumber \\ + \left. \frac{1}{4}\sin^2(\alpha)J_l(\beta \rho)^2 \right] \cos\left[2\omega t -2k_zz  + 2l\phi\right] ~.
        \label{T00 osc}
\end{eqnarray}

The $T_{0i}$ components are not of any significance for this study since, as enhanced by equation (\ref{TT}), they do not contribute to the form of $\bar{h}_{\mu\nu}^{TT}$ expressed in the traceless-transverse gauge. Even so, the expression of the Poynting vector is provided for the interested reader:

\begin{eqnarray}
   T_{0j} = \epsilon_0 c E_0^2 \cos(\alpha)  \begin{pmatrix}
\xi_+\xi_0\\
\xi_-\xi_0 \\
\xi_+^2 + \xi_-^2 
\end{pmatrix} ~.
   \label{Eq. T0j Bessel}
\end{eqnarray}

The $T_{ij}$ components are given by the Maxwell tensor ones $\sigma_{ij}$, that can be expressed as:

\begin{equation}
    T_{ij} = \sigma_{ij} \equiv \sigma_{ij}^{D}+\sigma_{ij}^{ZZ} + \sigma_{ij}^{+} + \sigma_{ij}^{\times} + \sigma_{ij}^{XZ}+\sigma_{ij}^{YZ} ,
\end{equation}

with

\begin{eqnarray}
\hspace{-1cm}
\sigma_{ij}^{D} &=& \frac{\epsilon_0 c E_0^2}{2}  \sin^2(\alpha) \begin{pmatrix}
-\xi_0^2 & 0& 0 \\
0 & -\xi_0^2  & 0 \\
0 & 0& \xi_0^2
\end{pmatrix} ,
\label{eq:maxwellD1} \\
\sigma_{ij}^{ZZ} &=& -\frac{\epsilon_0 c E_0^2}{2} [1+\cos^2(\alpha)] \begin{pmatrix}
0 & 0& 0 \\
0 & 0  & 0 \\
0 & 0& \xi_+^2+\xi_-^2
\end{pmatrix} ,
\label{eq:maxwellD2} \\
    \sigma^{+}_{ij} &=& \frac{\epsilon_0 c E_0^2}{2} \sin^2(\alpha) \begin{pmatrix}
\xi_+^2-\xi_-^2 &0 & 0 \\
0 & -\xi_+^2+\xi_-^2  & 0 \\
0 & 0& 0
\end{pmatrix} ,
\label{eq:maxwellplus} \\
\sigma^{\times}_{ij} &=& \epsilon_0 c E_0^2 \sin^2(\alpha) \begin{pmatrix}
0 & \xi_+\xi_- & 0 \\
\xi_+\xi_- & 0 & 0 \\
0 & 0& 0
\end{pmatrix} ,
\label{eq:maxwellcross} \\
\sigma_{ij}^{XZ} &=& \frac{\epsilon_0 c E_0^2}{2}  \sin^2(\alpha) \begin{pmatrix}
0 & 0& \xi_0 \xi_+ \\
0 & 0  & 0 \\
\xi_0 \xi_+ & 0& 0
\end{pmatrix} ,
\label{eq:maxwellx} \\
\sigma_{ij}^{YZ} &=& \frac{\epsilon_0 c E_0^2}{2}  \sin^2(\alpha) \begin{pmatrix}
0 & 0& 0 \\
0 & 0  & \xi_0 \xi_- \\
0 & \xi_0 \xi_- & 0
\end{pmatrix} .
\label{eq:maxwelly} 
\end{eqnarray}

The Maxwell tensor has been subdivided into three groups:
\begin{itemize}
\item $\left\lbrace \sigma_{ij}^{D},\sigma_{ij}^{ZZ} \right\rbrace$, which coefficients depend on $l$;
\item $\left\lbrace \sigma_{ij}^{+}, \sigma_{ij}^{\times} \right\rbrace$, which coefficients depend on $l\pm1$;
\item $\left\lbrace \sigma_{ij}^{XZ},\sigma_{ij}^{YZ} \right\rbrace$, which coefficients depend on $2l\pm1$.
\end{itemize}

This decomposition is not arbitrary. As it will be made clear in the subsequent analysis, the value of the OAM parameter $l$ dictates which of these groups dominates in $T_{\mu\nu}$. Contributions coming from elements of the two other groups will be neglected. \\

From these expressions it can be ensured that $\partial_\mu T^{\mu \nu} = 0$, thus that the solution equation (\ref{sol}) satisfies the harmonic gauge condition \cite{Hobson}.

\subsubsection{Spacetime deformations}

Considering the symmetries of the system, one writes $\hat{n} \equiv (\sin(\theta),0,\cos(\theta))$.
The variable $\theta$ introduced here denotes the angle in the $z-x$ plane between the optical axis $z$ and the propagation direction of the emitted gravitational waves. \\

Solutions to the Einstein field equations in the far-field region are obtained following equation (\ref{Compactsource}) and performing the integration in cylindrical coordinates over the cylindrical source volume $\mathcal{V}$ depicted in figure  \ref{fig:Region}. Unlike, \textit{e.g.} the seminal paper \cite{Tolman:1931zza}, this analysis does not focus on the total gravitational action of the laser pulse, but only on the associated waves. Any static component can therefore be safely ignored.\\ 

In the following, subsequent definitions will be used:

\begin{eqnarray}
 {h}_0(r) &\equiv& \frac{4\pi \epsilon_0 c E_0^2 G L} {\beta^2 c^5 r} ~, \label{h-amplitude} \\
\psi_q (t,r) &\equiv& 2\omega (t-r/c)+2q(\phi-\pi/2)  ~, \label{Psiq}\\
     \Gamma_{q}(\theta) &\equiv& \int_{0}^{\frac{D \beta}{2}} \tau J_{q}^2(\tau)J_{2q}\left(\frac{\omega \tau }{c \beta} \sin(\theta)\right) d\tau ~,
    \label{GammaintBP} \\ 
    \Lambda^{\pm}_{s}(\theta) &\equiv& \int_{0}^{\frac{D \beta}{2}} \tau J_{l}(\tau) J_{l\pm 1}(\tau)J_{s}\left(\frac{\omega \tau }{ c \beta} \sin(\theta)\right) d\tau 
    \label{Lambdaint} \\ 
    \eta(\theta)&\equiv&\frac{\omega L}{c} \left[ \cos(\theta)- \cos(\alpha) \right] ~,
    \label{Eta}
\end{eqnarray}

With $q \in \{l, Q \}$, where $Q \in \{2, l \pm 1 \}$, and $s \in \{ 1, 2l \pm 1 \}$. The $\Gamma_{q}(\theta)$ and $\Lambda^{\pm}_{s}(\theta)$ represent the radial integrals of the Green's function solutions in the far-field over a planar circular aperture of diameter $D$, for which the substitution $\tau \equiv \beta \rho$ has been used. 

For clarity, only the dependencies upon the coordinates $\left\lbrace t, r, \theta \right\rbrace$ are written down explicitly, and not those related to optical parameters such as $\beta$, $\omega$, etc.
Previous definitions then allow one to introduce the terms:

\begin{eqnarray}
h_{D} &\equiv& \frac{1}{2}{h}_0(r) \sin^2(\alpha)\Gamma_{l}(\theta)\text{Sinc}\left[ \eta(\theta) \right] \sin(\psi_q) ~,
\label{eq:D1}  \\ 
h_{ZZ} &\equiv& 2 {h}_0(r) \left[ 1+\cos^2(\alpha) \right] \Gamma_{l}(\theta)\text{Sinc}\left[ \eta(\theta) \right] \nonumber  \\  && \times \cos(\psi_q) ~,
  \label{eq:D2} \\
   h_{+}^{(Q)} &\equiv&  {h}_0(r) \sin^2(\alpha)\Gamma_{Q}(\theta)\text{Sinc}\left[ \eta(\theta) \right] \cos(\psi_q)~,  \label{eq:hplus}  \\
    {h}_{\times}^{(Q)}  &\equiv&  {h}_0(r) \sin^2(\alpha)\Gamma_{Q}(\theta)\text{Sinc}\left[ \eta(\theta) \right] \sin(\psi_q)~,  \label{eq:hcross} \\ 
    \hspace{-0.5cm}
     {h}_{XZ}^{\pm,(s)}  &\equiv&  \frac{1}{4}{h}_0(r) \sin(2\alpha)\Lambda^{\pm}_{s}(\theta)\text{Sinc}\left[ \eta(\theta) \right] \cos(\psi_q)~,  \label{eq:hx} \\
   \hspace{-0.5cm}
{h}_{YZ}^{\pm,(s)}  &\equiv&  \frac{1}{4} {h}_0(r) \sin(2\alpha)\Lambda^{\pm}_{s}(\theta)\text{Sinc}\left[ \eta(\theta) \right]  \sin(\psi_q)~, 
 \label{eq:hy} \\
    {h}_{N} &\equiv& 2 {h}_0(r) \cos(\alpha)\Gamma_{l}(\theta)\text{Sinc}\left[ \eta(\theta) \right] \cos(\psi_q) ~.
  \label{eq:N} 
\end{eqnarray}

Introducing the following expressions:

\begin{eqnarray}
    {h}_+ &\equiv& 2 {h}_+^{(2)} + {h}_+^{(l+1)} + {h}_+^{(l-1)} ~,
    \label{h_plus decompositon}\\ 
    {h}_{\times} &\equiv& 
    2 {h}_{\times}^{(2)} + {h}_{\times}^{(l+1)} + {h}_{\times}^{(l-1)} ~, \label{h_cross decompositon}\\
    {h}_{XZ} &\equiv& {h}_{XZ}^{+,(1)}+ {h}_{XZ}^{-,(1)} + {h}_{XZ}^{+,(2l+1)} + {h}_{XZ}^{-,(2l-1)} ~, \\
     {h}_{YZ} &\equiv& {h}_{YZ}^{+,(1)}- {h}_{YZ}^{-,(1)} + {h}_{YZ}^{+,(2l+1)} - {h}_{YZ}^{-,(2l-1)} ~,
     \label{h_y decompositon} 
\end{eqnarray}

the $\mu \nu$ components of the oscillating  metric perturbation in the laboratory frame finally read:

\begin{equation}
\small
      {h}_{\mu \nu} =  \begin{pmatrix}
h_{D}-h_{ZZ} & {h}_{XZ}& {h}_{YZ}& {h}_{N} \\  
{h}_{XZ}&{h}_+ - h_{D} & {h}_{\times}& {h}_{XZ} \\
{h}_{YZ}& {h}_{\times}& - {h}_+ - h_{D} & {h}_{YZ} \\
 {h}_{N}  & {h}_{XZ}& {h}_{YZ} &h_{ZZ}+ h_{D} 
\end{pmatrix} ~.
\label{eq:Far-field}
\end{equation}

The expressions (\ref{h_plus decompositon} - \ref{h_y decompositon}) and in turn equation (\ref{eq:Far-field}) can be simplified after a deeper analysis of the $\Gamma_{q}(\theta)$ and $\Lambda^{\pm}_{s}(\theta)$ functions. 
For this purpose the parameter $\zeta_D$ is introduced:

\begin{equation}
\zeta_D \equiv \frac{\omega D}{c} ~.
\end{equation}

This parameter corresponds to the maximal value of the argument of the second Bessel function appearing in equation \ref{GammaintBP}, since $\tau/\beta = \rho \leq D$. It depends on the frequency of the laser pulse $\omega$ as well as on the diameter of the source cylindrical volume $D$. Typical values of $D$ range from 1 m \cite{Dawson} to the diffraction limit (corresponding to half a wavelength), typically of the order of $10^{-7}$m \cite{Brooker}. Considering $w \simeq 10^{15}$ Hz, the associated values of $\zeta_D$ therefore range from unity up to $10^{7}$.\\

The dominant function is always the one with $q=0$, whatever the value of $\zeta_D$. Thus, in the subsequent analysis, only cases where $q=0$ are considered as they correspond to the main sources of gravitational waves (See appendix \ref{Appendix angular distrib l=0} for more information).

Since $q \in \left\lbrace 2, l, l\pm 1 \right\rbrace$, the study is restricted to values of the topological index such that $l \in \left\lbrace 0, \pm 1 \right\rbrace$. Higher values of $l$ lead to Bessel functions of smaller amplitudes and therefore to $\Gamma_{q}(\theta)$ also being smaller. As a consequence, dominant contributions to the ${h}_+$ and ${h}_\times$ functions presented in equations (\ref{h_plus decompositon}) and (\ref{h_cross decompositon}) will come from ${h}_+^{(0)}$ and ${h}_\times^{(0)}$ terms and all the other ${h}_+^{(n)}$ and ${h}_\times^{(n)}$, $n \neq 0$, can be disregarded.\\

Similar deductions can be made about the  $ \Lambda^{\pm}_{s}(\theta)$ functions. Since $ l \in \mathbb{Z}$ it follows that $ s $ is a non-zero integer, whatever the value of $l$. Subsequently there will never be a $s=0$ component in ${h}_{XZ}$ and ${h}_{YZ}$ and they can always be neglected compared to $\left\lbrace {h}_D,h_{ZZ}, {h}_+, {h}_\times \right\rbrace$.\\

After simplifications, spacetime perturbations take the new form:

\begin{equation}
\small
      \bar{h}_{\mu \nu} \simeq  \begin{pmatrix}
h_{D}-h_{ZZ} & 0 & 0 &  {h}_{N} \\  
0 & {h}_+ - h_{D} &  {h}_{\times}& 0 \\
0 &  {h}_{\times} & -  {h}_+ - h_{D} & 0 \\
 \bar{h}_{N}  & 0 & 0 &h_{ZZ}+ h_{D} 
\end{pmatrix}~,
\label{eq:Far-field simplified}
\end{equation}

in which the functions ${h}_+$ and ${h}_\times$ have also been simplified such that:

\begin{eqnarray}
    {h}_+ &\simeq& {h}_+^{(l+1)} + {h}_+^{(l-1)} ~,
    \label{h_plus decompositon simplified}\\ 
    {h}_{\times} &\simeq& 
    {h}_{\times}^{(l+1)} + {h}_{\times}^{(l-1)} ~. \label{h_cross decompositon simplified}
\end{eqnarray}

It is now possible to extract the radiative, or physical, components by projecting ${h}_{ij}$ into the TT gauge associated with any observer in the $\hat{n}$ direction around the source. This projection involves only the spatial components ${h}_{ij}$ of the ${h}_{\mu \nu}$ given equation (\ref{eq:Far-field simplified}). The temporal-temporal $00$ and crossed $0i$ or $i0$ components never contribute to the expressions of the perturbations in the TT gauge and, from now on, the ${h}_N$ function will no longer play any role.

As previously stated, due to the symmetries of the system this study is restricted to the $z-x$ plane, the $z$-axis corresponding to the direction along which the pulse propagates, such that $\hat{n} =(\sin(\theta),0,\cos(\theta) )$. Using the Lambda projector and following (\ref{TT}), the total metric perturbation in the TT gauge is written as the sum of four contributions:

\begin{equation}
  {h}_{\mu \nu}^{TT} = {h}_{\mu \nu}^{D,TT} + {h}_{\mu \nu}^{ZZ,TT} + {h}_{\mu \nu}^{+,TT} + {h}_{\mu \nu}^{\times,TT} ~.
    \label{Eq. hTT BP}
\end{equation}

The respective terms entering formula (\ref{Eq. hTT BP}) are given by:

\begin{widetext}

\begin{equation}
{h}_{\mu \nu}^{D,TT} = \left(
\begin{array}{cccc}
0 & 0 & 0 & 0 \\
 0 & \cos^2(\theta) \left[ 1-\cos(2\theta) \right] &  0 &  \cos(\theta)\sin(\theta) \left[ \cos(2\theta)-1 \right] \\
0 &  0 & \cos(2\theta)-1& 0\\
0 & \cos(\theta) \sin(\theta) \left[ \cos(2\theta)-1 \right] & 0& \sin^2(\theta) \left[ 1-cos(2\theta) \right] \\
\end{array}
\right)
           h_{D} ~,
           \label{BBD:Gauge-transformed}
\end{equation}

\begin{equation}
 {h}_{\mu \nu}^{ZZ,TT} = \left(
\begin{array}{cccc}
0 & 0 & 0 & 0 \\
0 & \cos^2(\theta) \left[ 1-\cos(2\theta) \right] &  0 &  \cos(\theta)\sin(\theta) \left[ \cos(2\theta)-1 \right] \\
0 &  0 & \cos(2\theta)-1& 0\\
0 & \cos(\theta) \sin(\theta) \left[ \cos(2\theta)-1 \right] & 0& \sin^2(\theta) \left[ 1-cos(2\theta) \right] \\
\end{array}
\right)
          h_{ZZ} ~,
           \label{BBZ:Gauge-transformed} 
\end{equation}
\begin{equation}
  {h}_{\mu \nu}^{+,TT} = \left(
\begin{array}{cccc}
0 & 0 & 0 & 0 \\
0 & \frac{1}{2} \cos ^2(\theta ) \left[ \cos ^2(\theta )+1\right] & 0 & -\frac{1}{2} \sin (\theta ) \cos (\theta ) \left[ \cos
   ^2(\theta )+1\right] \\
0 & 0 & -\frac{1}{2} \left[ 1+\cos ^2(\theta )\right] & 0 \\
0 & -\frac{1}{2} \sin (\theta ) \cos (\theta ) \left[ \cos
   ^2(\theta )+1\right] & 0 & \frac{1}{2} \sin ^2(\theta ) \left[ \cos ^2(\theta )+1\right] \\
\end{array}
\right)\bar{h}_{+} ~,
 \label{BB:Gauge-transformed}
\end{equation}
\begin{equation}
{h}_{\mu \nu}^{\times,TT} = \left(
\begin{array}{cccc}
0 & 0 & 0 & 0 \\
0 & 0 &  \cos ^2(\theta ) & 0 \\
0 &  \cos ^2(\theta ) & 0 & - \sin (\theta ) \cos (\theta ) \\
0 & 0 & - \sin (\theta ) \cos (\theta ) & 0 \\
\end{array}
\right) \bar{h}_{\times} ~.  \label{BBx:Gauge-transformed}
\end{equation}
   
\end{widetext}

By straightforward inspection it can be verified that $n^i {H}_{ij}= 0  $ where ${H}_{ij} \in \{{h}_{ij}^{D,TT},{h}_{ij}^{ZZ,TT},{h}_{ij}^{+,TT}, {h}_{ij}^{\times,TT}   \}$. The far-field metric perturbations in the TT gauge are indeed transverse to the propagation direction of the gravitational waves for all values of $\theta$.

The detailed study of those perturbations will be performed in the following sections, for different values of the orbital angular momentum parameter $l$.

It might seem surprising to the reader that the matrices in equations (\ref{BBD:Gauge-transformed}) and (\ref{BBZ:Gauge-transformed}) take the same form, as $h_{D}$ and $h_{ZZ}$ contribute in different ways to ${h}_{\mu \nu}$ in (\ref{eq:Far-field}). However, it has been previously demonstrated that the $\sigma_{ij}^D$ term in the Maxwell tensor gives rise to metric perturbations in which spatial components exhibit the following form:

\begin{equation}
    \bar{h}_{ij}^D = a(t,\vec{x})
    \begin{pmatrix}
1 & 0& 0 \\
0 & 1  & 0 \\
0 & 0& 0
\end{pmatrix}+b(t,\vec{x})
\begin{pmatrix}
0 & 0& 0 \\
0 & 0  & 0 \\
0 & 0& 1
\end{pmatrix} ~,
\label{Breathingmode}
\end{equation}

the functions $a(t,\vec{x})$ and $b(t,\vec{x})$ depending on spatial and temporal coordinates $x^\mu = (ct,\vec{x})$. Neither of the modes exhibited corresponds to the $h^{+}_{\mu \nu}$ and $h^{\times}_{\mu \nu}$ polarizations characteristic of standard General Relativity, thus they are not transverse along the z-axis.  However, both matrices of equation (\ref{Breathingmode}) transform to matrices that have identical form in the TT-gauge and thus ${h}_{ij}^D$ transforms the same way as ${h}_{ij}^{ZZ}$; with scaling factor that is a linear superposition of the functions $a(t,\vec{x})$ and $b(t,\vec{x})$. \footnote{The first term exhibits a form similar to polarization modes known as "breathing" modes, which can be present in alternative theories of gravity in which additional polarisation states of gravitational waves  exist, such as Brans-Dicke or Horndeski theories \cite{Hordeski, PhysRevD.103.104026}.} Therefore, when considering the projection in the TT gauge of both terms entering equation (\ref{Breathingmode}), only the second one matters, and this term is of the same form as the metric perturbation generated by the purely longitudinal source term $\sigma_{ij}^{ZZ}$.\\

\section{Further Experimental Considerations}
\label{section numerical parameters}

\subsubsection{Considered laser facilities}

In the proposed experimental setup, the interaction region remains stationary and thus the interaction length L is a fraction of the laser pulse width. Since from equation.(\ref{h-amplitude}) $h \propto E_0^2 L / r$ it would be expected that the strain correlates solely with the laser pulse energy. However due to the 1/r dependence of the metric perturbation a short laser pulse enables the probing the spacetime perturbation closer to the source, potentially yielding a higher observable strain.  Consequently, we thus introduce two types of laser systems that are currently or expected to become operational in the next decade \cite{Dawson}. 

The National ignition facility (NIF) is a mega-joule class faciliy, designed for inertial confinement fusion (ICF) and nuclear stockpile stewardship research, and allocates a portion of its shot time to the academic community for discovery Science applications \cite{LMJ,NIF1}.
Currently composed of composed of 192 seperate $3 ns$ to $20ns$ pulse width kilo-joule class beams, with total energ of $1.6 MJ$. However recent advancements in plasma optics and beam-combiners \cite{10.1063/1.5016310,10.1063/5.0086068,Kirkwoodnature2022} suggest that NIF may soon be capable of generating a single beamline with an energy up to $0.8 MJ$ in a single 3 ns to 20 ns pulse at 351 nm. 

We consider two different Kilo joule class laser facilities;  The NIF Advanced Radiography Capability (ARC) facility and the Station of Extreme Light (SEL) in Shanghai\cite{NIF2,https://doi.org/10.1002/lpor.202100705}. The NIF ARC facility was developed to enhance the capabilities of high brightness X-ray probes for studying high energy density conditions. NIF ARC employs four of NIF’s 351 nm laser beams (one quad), with each beam split into two halves. This configuration enables the generation of eight peta-Watt class beams, delivering energy ranging from 0.4 kJ to 1.7 kJ, with pulse lengths ranging from 1.3 to 38 ps.  The SEL facility expected to be commissioned later this decade,  primary objective is to generate the most intense laser pulses ever achieved on Earth.  Specifically, the SEL aims to produce a pulse with a power of 100 PW resulting from a 1.5 kJ, 15 fs pulse.

\subsubsection{Power normalisation} 

In order to perform numerical simulations, the Bessel modes power must first be normalised. Since ideal Bessel beams partition their energy equally among the lobes of the beam radial profile, setting the power $P_0 = \epsilon_0 c E_0^2 \beta^{-2}$ adds spurious energy to the system when a Bessel mode beam waist $\beta^{-1}$ is decreased at constant $D$. This is compensated for by equating the power of the incident beam on the axicon optic to the power of the generated Bessel mode, fixing the energy content by conservation of energy \cite{Durnin:88}, such that:

\begin{equation}
   \mathcal{N}_{B}^{-1} P_0 = \frac{\epsilon_0 c E_0^2}{2\beta^2}  ~,
\end{equation}

Where the normalisation factor $\mathcal{N}_{B}$ is defined as:

\begin{equation}
   \mathcal{N}_{B}  = \cos(\alpha) \int_{0}^{\frac{D \beta}{2}} \tau  \left[J_{l-1}^2(\tau)+J_{l+1}^2(\tau) \right] d\tau ~.
\end{equation}

\subsubsection{Axicon optic}

\label{Section: Axicon optic}
The functions $h_{D}$, $h_{ZZ}$, ${h}_{+}^{Q}$ and $h_{\cross}^{Q}$ presented in equations (\ref{eq:D1}  - \ref{eq:hcross}) give the amplitudes of the generated gravitational waves. Their maximal values, corresponding to the highest gravitational wave signal, depend, among other parameters, on the half cone angle $\alpha$. 

One considers a focused Bessel mode associated with large values of $\alpha$ so that all previous but $h_{ZZ}$ functions are maximized. It is important to note that increasing the half cone angle cannot result in arbitrarily small beam waists. This limitation stems from the diffraction-limited nature of Bessel modes. The diffraction limited  spot size $ \Delta_{D}$ is given in terms of the numerical aperture $N_a$ by \cite{Brooker, Simon}:

\begin{equation}
    \Delta_{D} = \frac{\lambda}{2 N_a} ~.
    \label{Eq:Wdiff}
\end{equation}

Equating the beam width $\beta$ to the reciprocal of the diffraction limit beam waist $\Delta_D^{-1}$ leads to $N_a = \pi \sin(\alpha)$. Axicon optics have demonstrated the ability to reach numerical apertures of 0.75 \cite{s21196690}. Therefore, the maximum achievable half cone angle, beyond which the $\beta^{-1}$ cannot decrease further, is $\alpha_{max}$. 

\begin{equation}
    \alpha_{\text{max}} = 0.24106 \simeq \frac{3}{4\pi} ~ \text{rads} ~.
\end{equation}

We evaluate the far-field regime utilizing the Fraunhofer distance \cite{Brooker}. Dependant on the value of $\alpha$ the longitudinal extent of the Bessel mode is either larger or on the same order as the radial extent of the mode. Therefore, we can designate the Fraunhofer distance as:

\begin{equation}
    \epsilon_F = \frac{W_0^2}{2 sin^2(\alpha)\lambda}
\end{equation}

To ensure the validity of the far-field approximation, we constrain the distance from the center of the source by the inequality $r \geq \epsilon_F$.

\section{Numerical Estimates}

\textit{This section is kept brief and numerical results of the amplitude and angular distribution of the Bessel mode generated gravitational wave are presented, for the case of the topological index being zero and unity with minimal comments. Further discussions are provided in section \ref{section discussion}.}\\

\subsection{Bessel pulses with no orbital angular momentum: the $l=0$ case}
\label{Case:l=0}

At first sight one is tempted to claim that, without orbital angular momentum carried by the laser pulse, no gravitational waves are expected to be produced by the system. Neglecting phase noise effects this statement appears to be erroneous because, in addition to the twisting due to the orbital angular momentum, an additional effect is in play.

It is true that when $l=0$, $\xi_{+}^{2} - \xi_{-}^{2} = \xi_+ \xi_- = 0$ and the $\sigma_{ij}^{+}$ equation (\ref{eq:maxwellplus}) and $\sigma_{ij}^{\cross}$ equation (\ref{eq:maxwellcross}) contributions to the Maxwell tensor both vanish. As a consequence ${h}_{+}$ and ${h}_{\times}$ also completely vanish. In addition, it has been demonstrated that the spacetime deformations generated by the $\sigma_{ij}^{XZ}$ and $\sigma_{ij}^{YZ}$ components given in (\ref{eq:maxwellx}) and (\ref{eq:maxwelly}), \textit{i.e} ${h}_{XZ}$ and ${h}_{YZ}$, can be neglected. Nevertheless, the $h_{D}$ and $h_{ZZ}$ deformations, generated by $\sigma_{ij}^{D}$ from (\ref{eq:maxwellD1}) and $\sigma_{ij}^{ZZ}$ from (\ref{eq:maxwellD2}), remain.  

This poses the question: if the pulse carries no orbital angular momentum, what is the source of the gravitational waves? Even when no orbital angular momentum is present, longitudinal energy oscillations along the optical axis within the volume of the cylindrical stress-energy distribution can continue to persist , analogous to the oscillations of a mass in a harmonic oscillator \cite{Maggiore:2007ulw}. However, this is only the case if there exists a mismatch between the group velocity and phase velocity of the electromagnetic field distribution. Which is precisely the case for the proposed experimental scheme for which the group velocity of the interference pattern (the Bessel mode) is zero and the phase velocity is non-zero.

Thus, in the absence of orbital angular momentum, radiative metric perturbations are therefore still expected and their form in the $\hat{n}$ oriented TT gauges are given by:

\begin{equation}
    {h}_{\mu \nu, l=0}^{TT} \simeq {h}_{\mu \nu}^{D,TT} + {h}_{\mu \nu}^{ZZ,TT} ~.
\end{equation}

Of course, it is possible that accumulated phase errors, e.g. those associated with propagation of the laser light in the amplifier media, might negate this effect. However, the low B-integrals normally associated with peta-Watt class laser systems suggest that this is unlikely to manifest itself. Similarly, numerical studies of phase errors of amplified seed beams in Raman amplifiers also indicate that these are also unlikely to play a significant role in pulses associated with beam-combiners \cite{RamanAmplificationRaoul, AdvantageRamanAmp}.

\subsubsection{Frequencies ($l=0$)}

The generated gravitational wave frequency is obviously a key parameter, especially regarding detection aspects. The time dependence of the linearised Einstein field equation solutions ${h}_{\mu \nu}^{D,TT}$ and ${h}_{\mu \nu}^{ZZ,TT}$ presented  in (\ref{BBD:Gauge-transformed}) and (\ref{BBZ:Gauge-transformed}) is located in ${h}_D$ and $h_{ZZ}$. Those functions both oscillate in time with a phase given by $\psi_q(t,r)$, defined in equation (\ref{Psiq}). Gravitational waves generated by oscillations of the electromagnetic energy content therefore oscillate at twice the laser frequency. This result was indeed already obvious from the expression of the energy density shown in (\ref{Eq. T00 Bessel}). For a NIF's 351 nm laser pulse \cite{10.1063/1.5016310} this corresponds to:

\begin{equation}
\omega_g^{l=0} = 2 \omega \simeq 1.07 \times 10^{16} ~ \text{Hz} ~.
\label{freuquency l=0}
\end{equation}

\subsubsection{Amplitudes ($l=0$)}

As the coefficients of the matrices equations (\ref{BBD:Gauge-transformed}) and (\ref{BBZ:Gauge-transformed}) involve only trigonometric functions that will mainly impact the angular distributions of the waves and not their maximal amplitude, the latter property is estimated by the sum of the ${h}_D$ and $h_{ZZ}$ functions established in (\ref{eq:D1}) and (\ref{eq:D2}) respectively. Taking the half-cone angle to take the diffraction limited value $\alpha = 0.24106$ discussed in section (\ref{Section: Axicon optic}) , for $W_0 = \Delta_D = 0.234\; \mu m$ it follows that $\epsilon_F = 1.369 \; \mu m $ for a wavelength of $351 nm$. Thus we can estimate the maximal value of this sum  for a $1 PW$ laser to be: 

\begin{equation}
    h \sim 1.90 \times 10^{-36}\left(\frac{P}{1 PW}\right)  \left(\frac{r}{1.36 \mu m}\right)^{-1}
\end{equation}

\begin{figure}[h]
\hspace*{-0.5cm}
\centering
\includegraphics[scale=0.3]{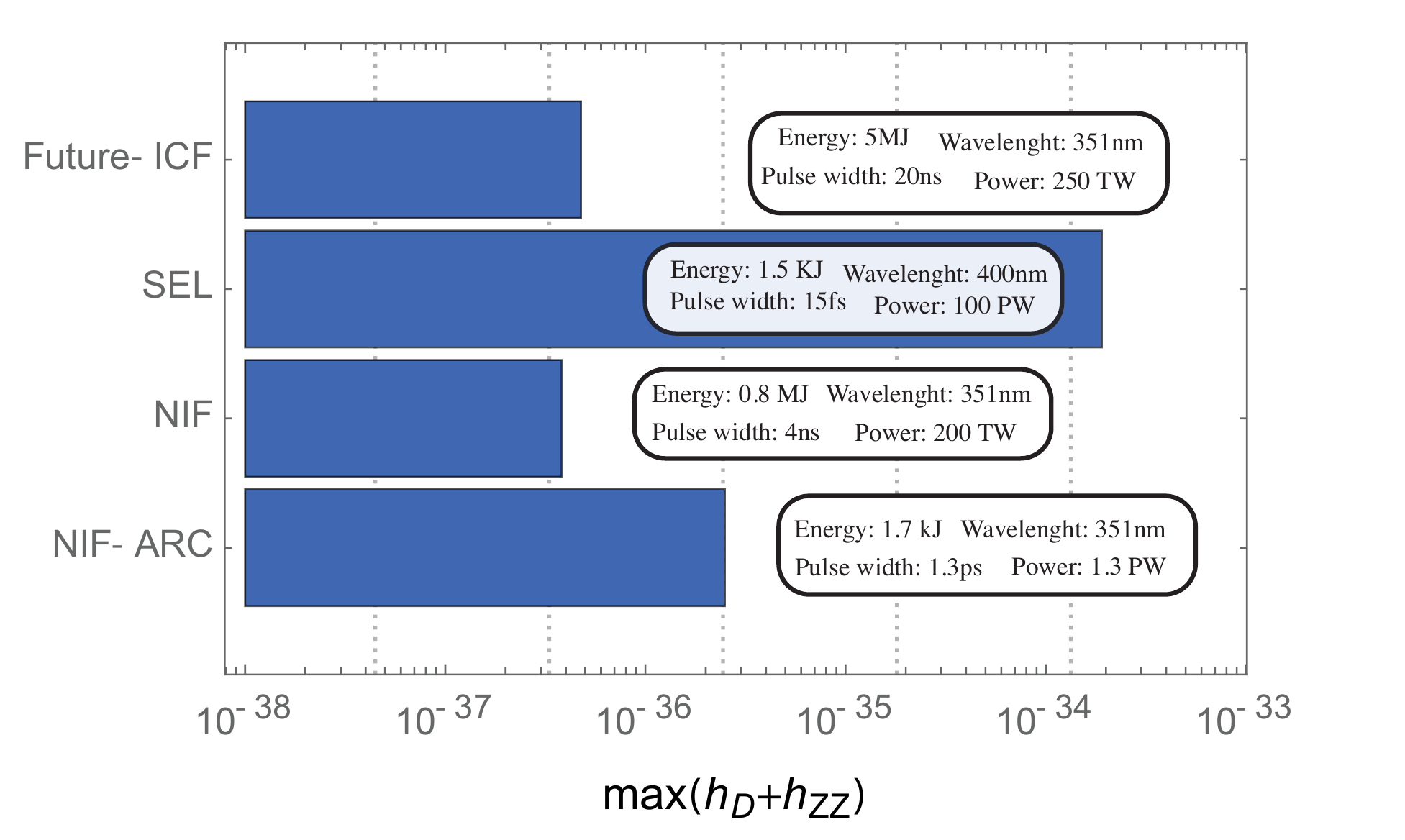}
\caption{Maximum gravitational wave amplitude at distance $r = 1.369 \; \mu m$, for the optical parameters $l=0$, $L= 0.49 \; \mu m$ and $D=0.234 \;\mu m$.}

\label{fig:BBstrain}
\end{figure}

 Strain estimates are explicitly given for the different laser facilities \cite{Dawson,LMJ,France, NIF1,NIF2,https://doi.org/10.1002/lpor.202100705} presented section \ref{section numerical parameters} in figure \ref{fig:BBstrain}. The plot includes data from four laser facilities, including a potential next-generation high-gain ICF laser facility \cite{doi:10.1098/rsta.2020.0005} is denoted "Future-ICF" in the figure.

 It might be that, in the future, one can try to detect gravitational perturbations with detection systems placed at very short distances, using \textit{e.g.} other laser systems such as presented in \cite{Vacalis:2023gdz}. 

\subsubsection{Angular distributions ($l=0$)}

Now consider the angular distribution of the previous metric deformations. In order to better visualise the dependencies to the optical parameters, we consider the laser frequency to be variable and $L$ fixed at $L=1$mm. As long as $\zeta_D < 1$, $\Gamma_0(\theta)$ is maximised in all directions and therefore independent of $\theta$, see Appendix \ref{Appendix angular distrib l=0} for more details. The whole angular dependence is in this case due to the $\text{Sinc}\left[ \eta(\theta) \right]$ function. To better characterise its effect we introduce:

\begin{equation}
   \zeta_{L} \equiv \omega L / c  ~,
\end{equation}

\begin{widetext}

\begin{figure}[h]
     \centering
     \begin{subfigure}[b]{0.24\textwidth}
         \centering
         \includegraphics[width=\textwidth]{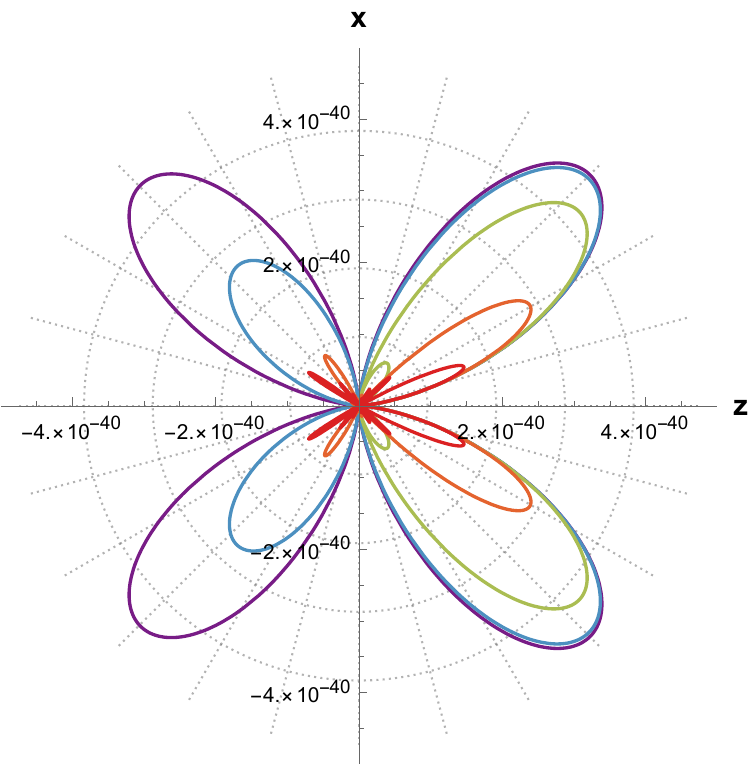}
         \caption{${h}_{11}^{TT}(\theta)$}
         \label{fig:ZD11}
     \end{subfigure}
     \begin{subfigure}[b]{0.24\textwidth}
         \centering
         \includegraphics[width=\textwidth]{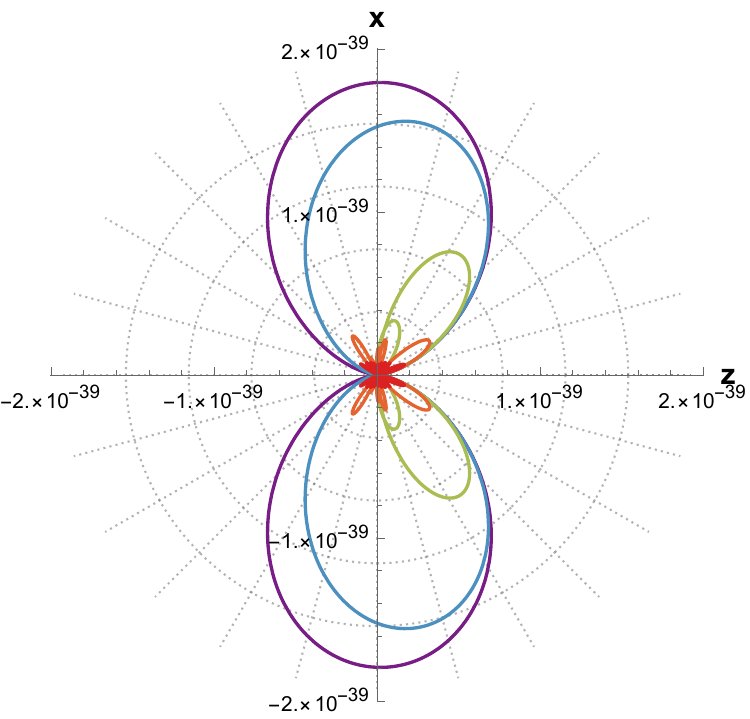}
         \caption{${h}_{22}^{TT}(\theta)$}
         \label{ffig:ZD22}
     \end{subfigure}
      \begin{subfigure}[b]{0.24\textwidth}
         \centering
         \includegraphics[width=\textwidth]{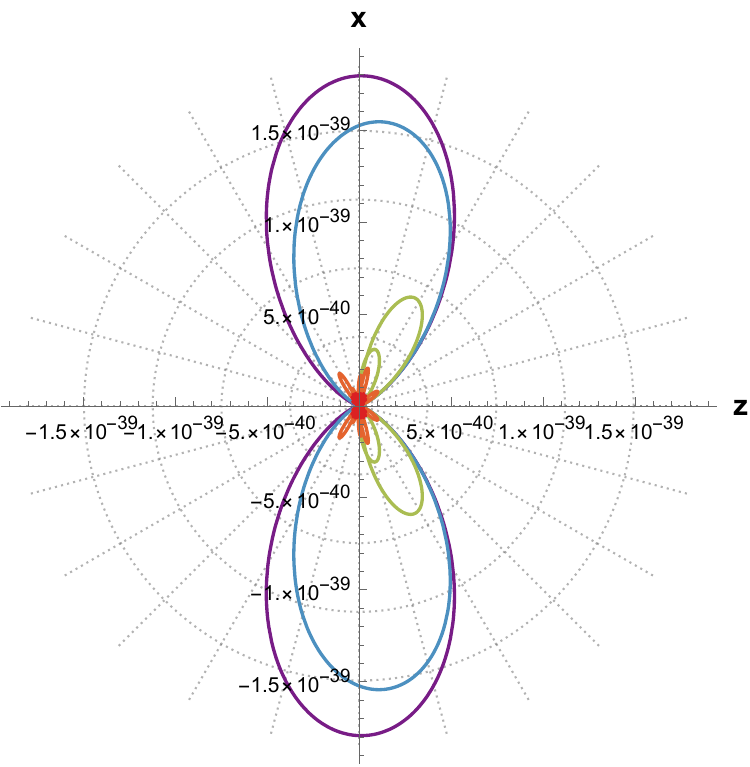}
         \caption{${h}_{33}^{TT}(\theta)$}
         \label{fig:ZD33}
     \end{subfigure}
        \label{fig:three graphs}
    \begin{subfigure}[b]{0.24\textwidth}
         \centering
         \includegraphics[width=\textwidth]{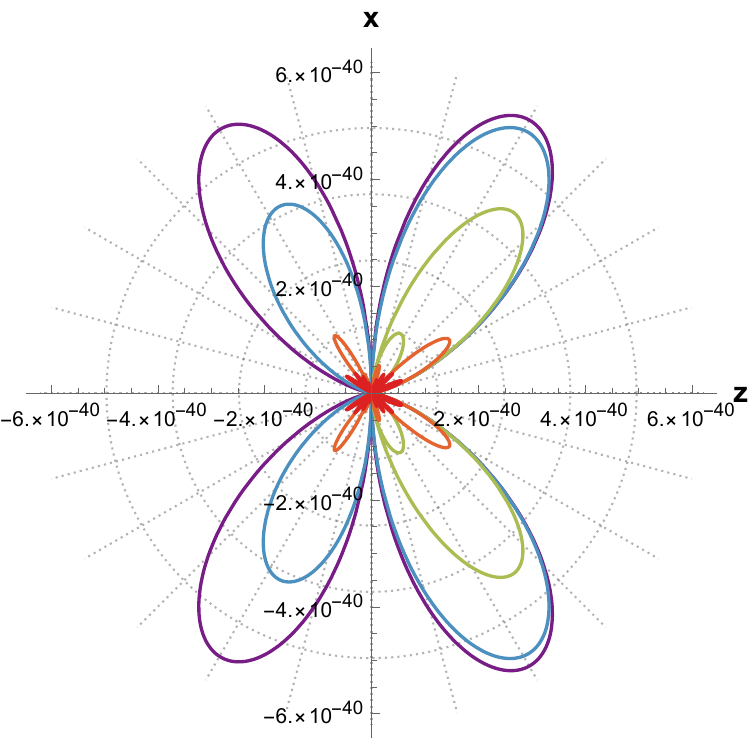}
         \caption{${h}_{13}^{TT}(\theta) = \bar{h}_{31}^{TT}(\theta)$}
         \label{fig:ZD13}
     \end{subfigure}

     \centering
     \vspace{1cm}
     \hspace{-12.2cm}
     \begin{subfigure}[]{0.2\textwidth}
         \centering
         \includegraphics[scale =0.5]{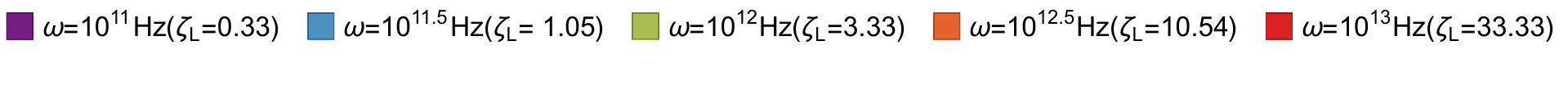}
     \end{subfigure}
     \caption{Angular variation of the z-axis propagating NIF ARC pulse generated $l=0$, TT gauge metric perturbations (${h}_{\mu \nu}^{D,TT} + {h}_{\mu \nu}^{ZZ,TT}$) in the $zx$ plane, at fixed $r=1$m $L=1$mm, $\beta^{-1} =1 \mu$m and $D=10\mu$m, such that $\zeta_D \leq 0.33$.} 
     \label{Angular distrib h l=0}
\end{figure}
\end{widetext}

Such that formula (\ref{Eta}) becomes $\eta(\theta) = \zeta_{L} \left[ \cos(\theta)-\cos(\alpha) \right]$. Similarly to $D$, the length of the source cylinder can vary roughly between $10^{-6}$m (corresponding to a 10 femtosecond laser pulse \cite{10.1063/1.5131174, https://doi.org/10.1002/lpor.202100705}) and 1 m (corresponding to a 10 nanoseconds laser pulse \cite{Dawson, LMJ, NIF1}). Consequently $\zeta_{L}$ varies, as $\zeta_{D}$, between unity and $10^7$ for $w \simeq 10^{15}$ Hz.  The angular distributions of the non-zero, radiative components of ${h}_{\mu \nu}^{TT} = {h}_{\mu \nu}^{D,TT} + {h}_{\mu \nu}^{ZZ,TT}$ for $l=0$ evaluated one meter away from the source are shown in figures \ref{fig:ZD11}-\ref{fig:ZD13}, each time for different values of $\omega$ and hence of $\zeta_{L}$. The values of the optical parameters have been chosen for the purpose of a better visualisation.

Figure \ref{Angular distrib h l=0} shows that, when $l=0$, no gravitational waves are emitted along the optical axis. Indeed, in the absence of orbital angular momentum, the energy distribution displays azimuthal symmetry in the plane perpendicular to the optical axis. Consequently, similarly to how oscillations of a symmetric mass distribution fail to generate gravitational waves, an observer positioned along the optical axis would not detect any gravitational radiation.

Interestingly, when $\zeta_{L} \geq 1$, $\text{Sinc}\left[ \eta(\theta) \right]$ is evaluated outside of its central peak, leading to a noticeable reduction of the ${h}_D +h_{ZZ}$ angular distribution width. To maximise the signal, the condition $\zeta_{L} < 1$ must be ensured. Even more interestingly, those angular distributions exhibit a beaming effect when $\zeta_L$ becomes greater than one towards the angles $\theta_{\text{peak}} = \pm \alpha$, $\alpha$ being the half-cone angle introduced in figure  \ref{fig:BBsetup}. A Bessel beam can be thought as a linear superposition of plane waves which propagation directions are inclined at the half-cone angle $\alpha$ with respect to the $z$-axis. It is therefore not surprising that such plane waves generate metric perturbations whose amplitudes peak at $\theta_{\text{peak}} = \pm \alpha$.
 
\subsubsection{Power emitted by gravitational waves ($l= 0$)}
    
The power per unit solid angle given in equation \ref{Eq:PPUSA} contains all contractions of the radiative contributions and is a suitable observable to study. For $l=0$ it takes the form:

\begin{align}
    \small
   \text{$\frac{dP_{D}}{d\Omega} = g_{D}(\theta) F(\alpha) \mathcal{N}_B^{-2} \left( \omega_g^{l=0} \right)^2 \frac{\pi G L^2 P_0^2 } {c^7} \Gamma^2_0(\theta) \text{Sinc}\left [ \eta(\theta)  \right ]^2 $}~,
    \label{PPUSA:D}
\end{align}

Where $P_{0}$ is the laser power and $g_{D}(\theta)$ and $ F(\alpha) $ are respectively given by:

  \begin{eqnarray}
   g_{D}(\theta) & \equiv& \frac{1}{16} \left[ 3-4 \cos (2 \theta )+\cos (4 \theta ) \right] ~,  \\
  F(\alpha) &\equiv& 2\left[ 1+ \cos^2(\alpha)+\frac{1}{4} \sin^2(\alpha) \right] ~. 
 \end{eqnarray}  

\begin{figure}[h]
        \centering
        \includegraphics[scale=0.65]{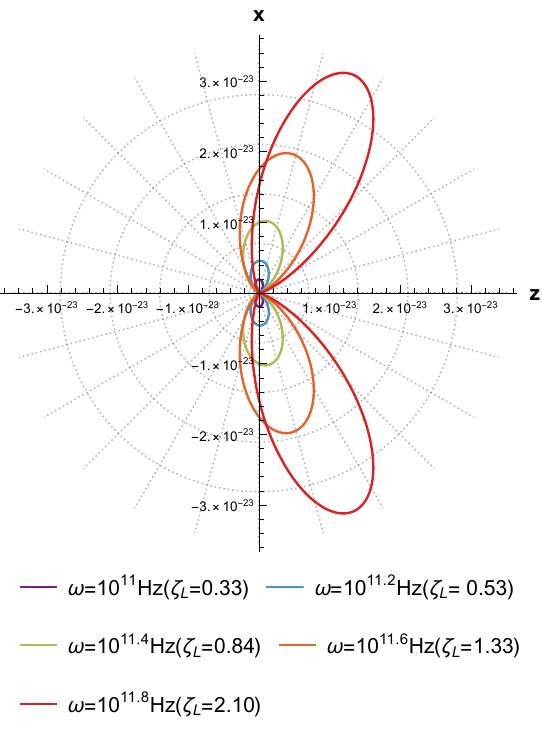}
        \caption{ Angular distribution of the $l=0$, NIF ARC ${h}_D+ {h}_Z$ generated gravitational $dP_D/d\Omega$,  for fixed values of $L=1$mm, $D=100\mu$m and $\beta^{-1}=10 \mu$m. Where the laser pulse propagates along the z-axis.}
        \label{fig:PPUSGB0}
    \end{figure}

This quantity is represented in figure  \ref{fig:PPUSGB0}, for the case of the NIF ARC facility. To improve graphical clarity in figure \ref{fig:PPUSGB0}, different values of $\zeta_L$ were used to that in figure \ref{Angular distrib h l=0}. Indeed, at constant laser power $P_0$, equation (\ref{PPUSA:D}) exhibits that $dP_D / d \Omega \propto \omega_g^2$. The higher the frequency, the bigger the power radiated by gravitational waves (and the sizes of the lobes). In any case this power loss is too tiny to have a substantial impact on the dynamics of the system that would permit an indirect measurement of the gravitational radiation. \\

Also, and similarly to the ${h}_{\mu \nu}^{TT}$ components represented in figure \ref{Angular distrib h l=0}, a beaming effect towards the half-cone angle occurs for values of $\omega$ such that $\zeta_L > 1$. The location at which the greatest signal is expected depends on the pulse frequency.

\subsection{Gravitational waves generated by a Bessel pulse with orbital angular momentum: $l= \pm 1 $}
\label{Case:l=1}

The initial purpose of this study was to characterise the gravitational waves emitted by light carrying orbital angular momentum. To do so we consider from now on non-zero values of the OAM parameter $l$. As discussed earlier one can restrict the study to $l \in \left\lbrace \pm 1 \right\rbrace$ as higher values lead to smaller amplitudes of the $\Gamma_q$ function. Also, using the following Bessel function identity $J_{-n}(x) = (-1)^n J_{n}(x), \; n \in \mathbb{Z}, \; n \geq 0$ it can be shown that $\Gamma_{-q} = \Gamma_q $. Subsequent results are therefore invariant under this sign change of $q$. Similarly to the $l=0$ case, gravitational waves described by the $h_{D}$ and $h_{ZZ}$ functions are still expected due to oscillations of the energy content in the source region. For $l = \pm 1$ this effect is however not dominant when compared to a new one: the twisting of the energy distributions described by the ${h}_+$ and ${h}_\times$ components. Indeed, when $l = \pm 1$, both $h_{D}$ and $h_{ZZ}$ contain $\Gamma_{1}$ functions and are therefore small compared to $\bar{h}_+$ and ${h}_\times$ which contain $\Gamma_{0}$ functions through the ${h}_{+}^{l \pm 1}$ and ${h}_{\times}^{l \pm 1}$ contributions. The metric perturbation in the TT gauge is now well approximated by the sum of two distinct polarisations:

\begin{equation}
    {h}_{\mu\nu, l=\pm 1}^{TT} \simeq {h}_{\mu\nu}^{+,TT} + {h}_{\mu\nu}^{\times,TT} ~.
\end{equation}

\subsubsection{Frequencies ($l=\pm 1$)}

Similarly to the case with no orbital angular momentum, the frequency of the emitted gravitational waves is given by the function $\psi_{q}$ as it is the argument of the $\sin(\psi_q)$ and $\cos(\psi_q)$ functions appearing in equations (\ref{eq:hplus}) and (\ref{eq:hcross}). As for $l=0$, gravitational waves are therefore emitted at twice the frequency of the laser pulse:

\begin{equation}
\omega_g^{l= \pm 1} = 2 \omega \simeq 1.07 \times 10^{16} ~ \text{Hz} ~,
\label{frequency l=1}
\end{equation}

For NIF's 351 nm laser pulse \cite{10.1063/1.5016310}. The non-trivial equivalence between $\omega_g^{l=0}$ and $\omega_g^{l=\pm 1}$ will be further discussed section \ref{section discussion}.

\begin{widetext}

\begin{figure}[H]
     \centering
     \begin{subfigure}[htpb]{0.24\textwidth}
         \centering
         \includegraphics[scale =0.28]{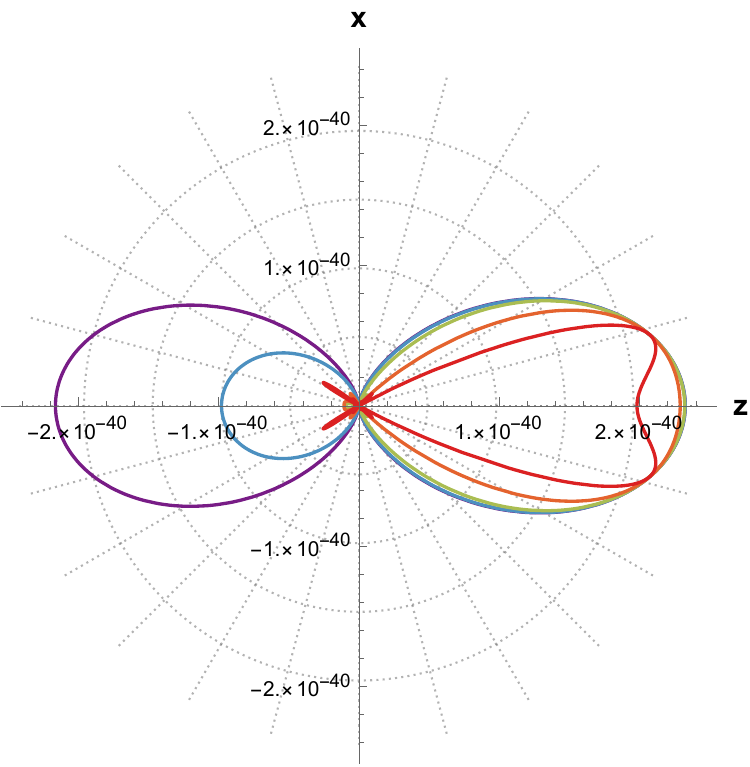}
         \caption{${h}_{11}^{+,TT}(\theta)$}
         \label{fig:PT11}
     \end{subfigure}
     \begin{subfigure}[htpb]{0.24\textwidth}
         \centering
         \includegraphics[scale =0.28]{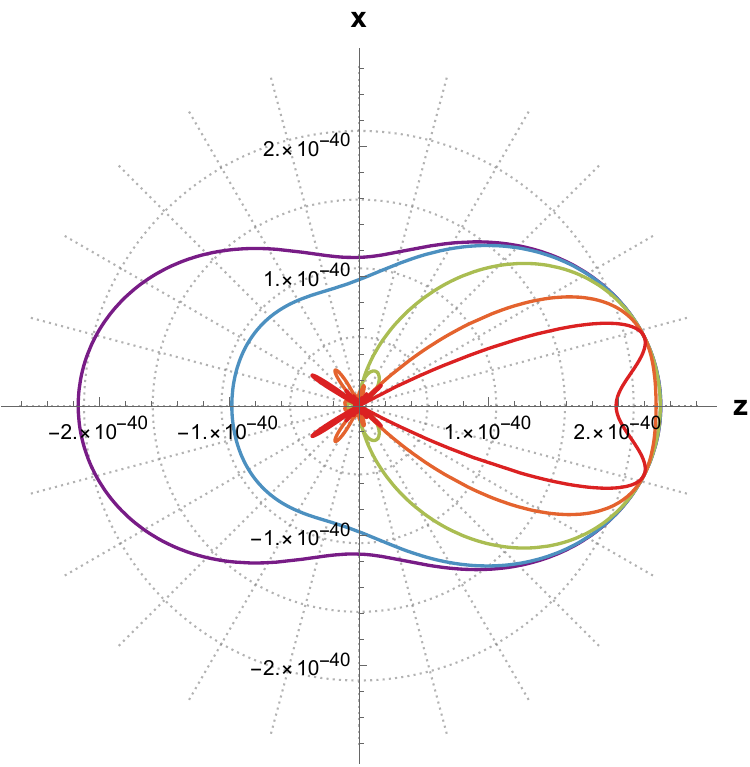}
         \caption{${h}_{22}^{+,TT}(\theta)$}
         \label{ffig:PT22}
     \end{subfigure}
     \begin{subfigure}[htpb]{0.24\textwidth}
         \centering
         \includegraphics[scale =0.28]{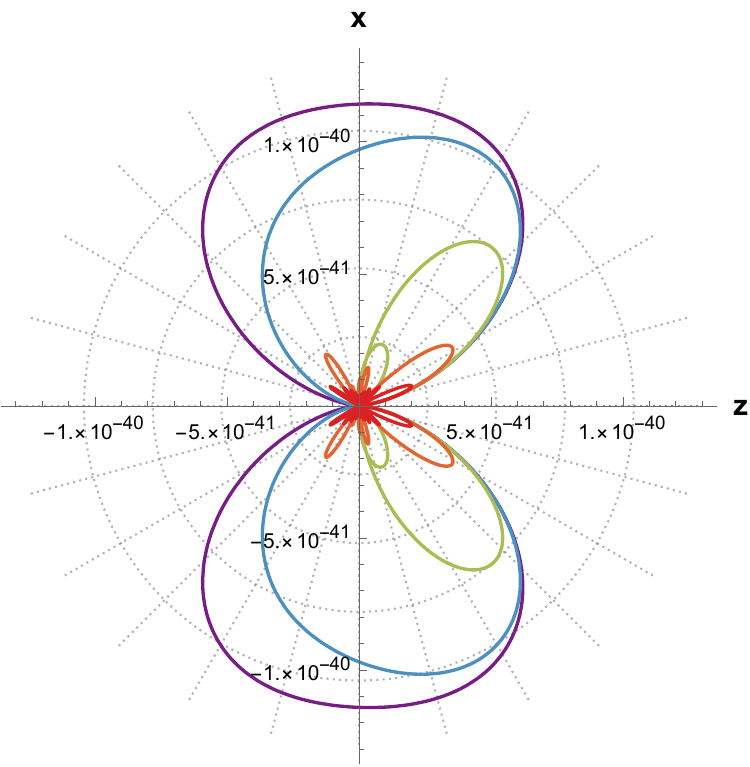}
         \caption{${h}_{33}^{+,TT}(\theta)$}
         \label{fig:PT33}
     \end{subfigure}
        \label{fig:three graphs}
    \begin{subfigure}[htpb]{0.24\textwidth}
         \centering
         \includegraphics[scale =0.28]{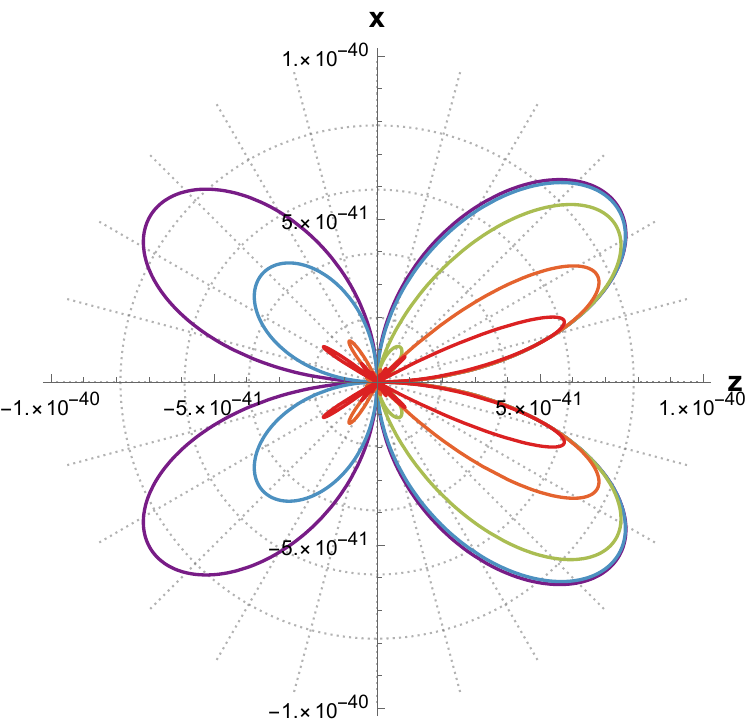}
         \caption{${h}_{13}^{+,TT}(\theta) =  {h}_{31}^{+,TT}(\theta)$}
         \label{fig:PT13}
     \end{subfigure}

     \begin{subfigure}[htpb]{0.24\textwidth}
         \centering
         \includegraphics[scale =0.28]{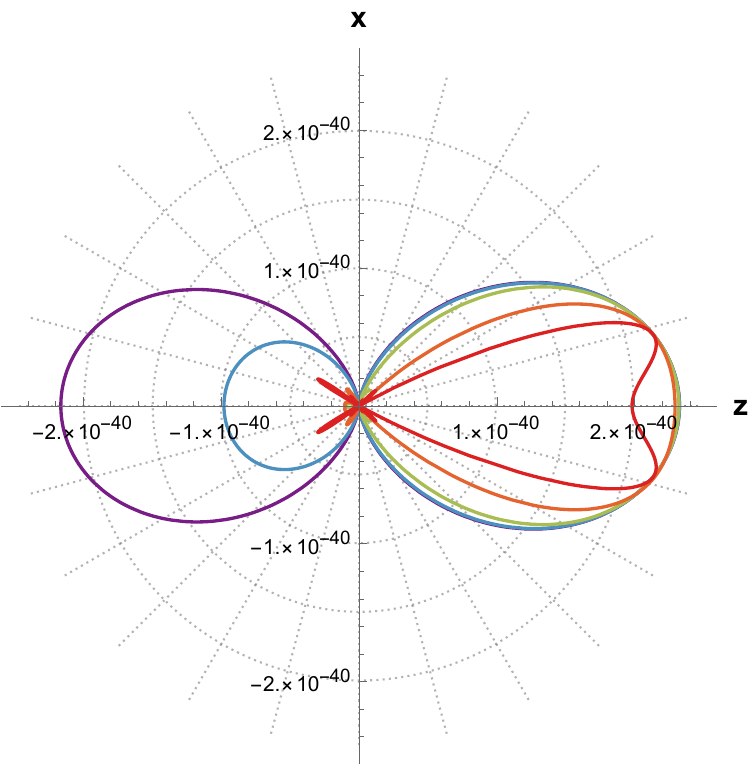}
         \caption{${h}_{12}^{\times,TT}(\theta) = {h}_{21}^{\times,TT}(\theta)$}
         \label{fig:CT12}
     \end{subfigure}
     \begin{subfigure}[htpb]{0.24\textwidth}
         \centering
         \includegraphics[scale =0.28]{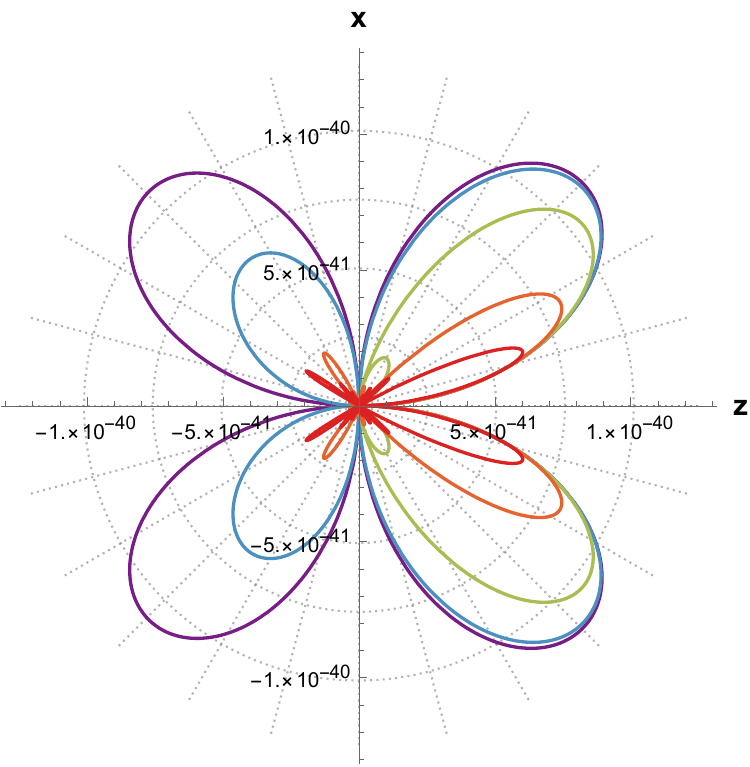}
         \caption{${h}_{23}^{\times,TT}(\theta) = {h}_{32}^{\times,TT}(\theta)$}
         \label{fig:CT32}
     \end{subfigure}
      \begin{subfigure}[htpb]{0.24\textwidth}
         \centering
         \includegraphics[scale =0.56]{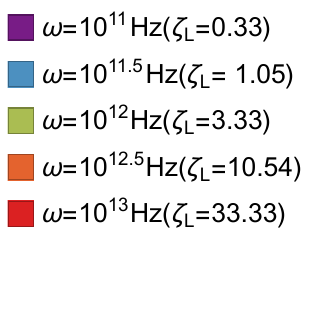}
     
         \label{ffig:Legend2}
     \end{subfigure} 
     \caption{Angular variation of the z-axis propagating NIF ARC pulse generated $l=1$, TT gauge metric perturbations (${h}_{\mu\nu}^{+,TT}$ and ${h}_{\mu\nu}^{\times,TT}$) in the $z-x$ plane, at fixed $r=1$m $L=1$mm, $\beta^{-1} =1 \mu$m and $D=10\mu$m, such that $\zeta_D \leq 0.33$.}
     \label{fig:TTP}
\end{figure}
\end{widetext}
\subsubsection{Amplitudes ($l=\pm 1$)}

\begin{figure}[]
\hspace*{-1cm}
\centering
\includegraphics[scale=0.3]{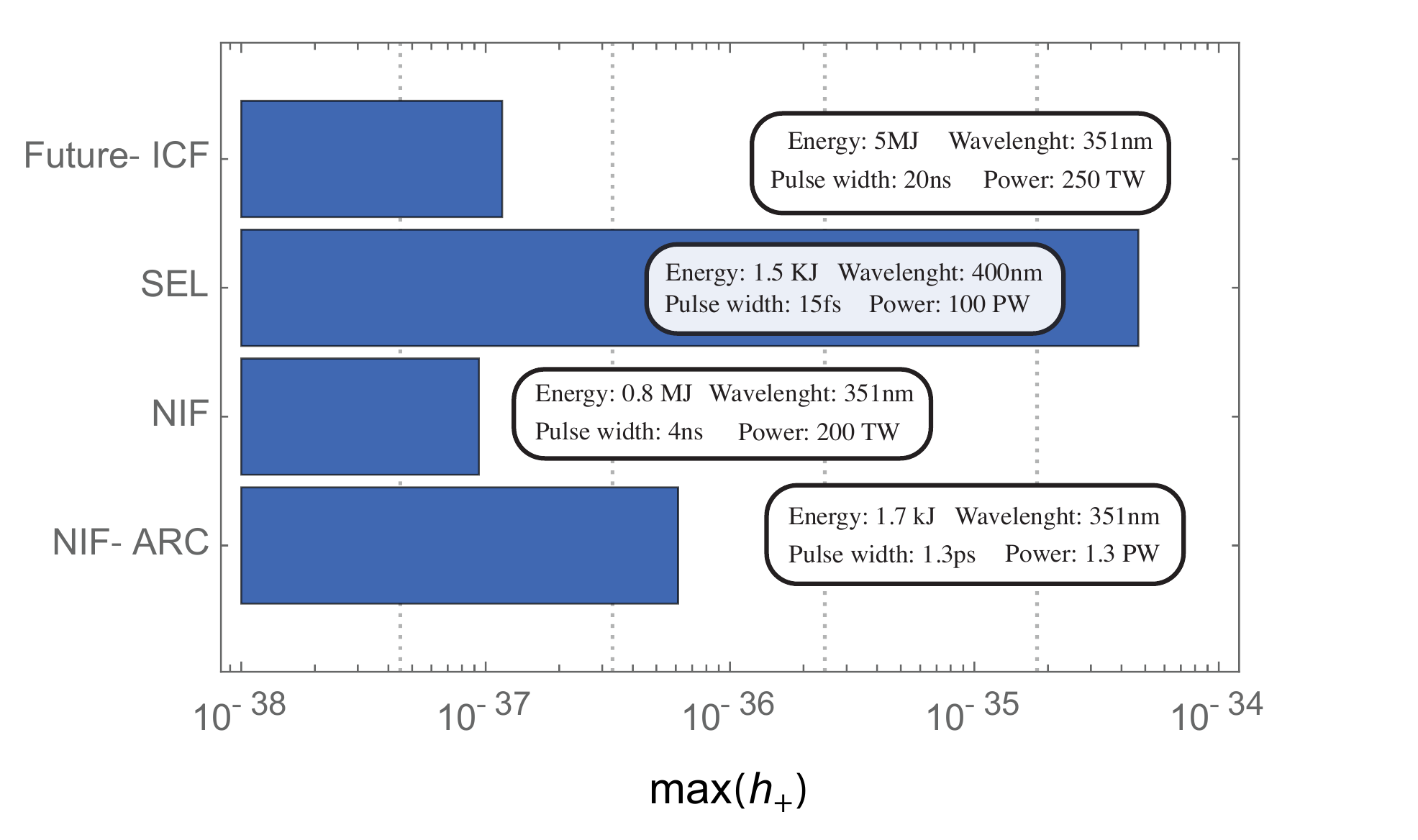}
\caption{ Maximum gravitational wave amplitude at distance $r = 1.369 \; \mu m$, for the optical parameters $l=\pm1$, $L= 0.49 \; \mu m$ and $D=0.234 \;\mu m$.}
\label{fig:BBstrain2}
\end{figure}

The amplitude of the associated gravitational waves is well estimated (once again up to the trigonometric functions entering the matrices in (\ref{BB:Gauge-transformed}) and (\ref{BBx:Gauge-transformed})) by the sum of the ${h}_+$ and ${h}_{\times}$ functions. Their maximal values over the $2 \pi$ region in the $z-x$ plane are identical, and thus we present the estimates for one of them. As previously done, we present estimated strain magnitudes for different facilities in figure {\ref{fig:BBstrain2}}. Those spacetime deformations do not exceed the $l=0$ case already presented in figure  \ref{fig:BBstrain}. Comparing those two figures, it appears that the twisting effect generates weaker spacetime deformations than the energy residual oscillations in the source volume in the absence of OAM. This strangeness is further discussed and justified in Appendix \ref{appendix twisting not dominant}.  Similar to the $l=0$ case, we give the strain estimate for a $1PW$ laser.

\begin{equation}
    h \sim 4.69 \times 10^{-37}\left(\frac{P}{1 PW}\right)  \left(\frac{r}{1.36 \mu m}\right)^{-1}
\end{equation}

Interestingly, naive estimations are recovered as an equivalent system of rotating masses with effective masses $E_L/c^2 \simeq 10^{-11}$ kg (1 MJ laser) separated by $\beta^{-1} = 40$ nm and rotating at a frequency $\omega = 5.37\times 10^{15}$ Hz would produce a strain $h \sim 6.4\times10^{-38}$ one meter away from the source. This estimation is in very good agreement with figure  \ref{fig:BBstrain2}. Obviously, more complex and subtle phenomena cannot be exhibited by this simple equivalence.

\subsubsection{Angular distributions ($l=\pm 1$)}

Figure \ref{fig:TTP} shows the angular distributions of the non-zero ${h}_{\mu\nu}^{+,TT}$ and ${h}_{\mu\nu}^{\times,TT}$ metric components one meter away from the source.

 One noticeable feature that differs from figure \ref{Angular distrib h l=0} is the appearance of gravitational waves emitted along the optical axis $z$.
This discrepancy with respect to the previous $l=0$ case arises due to disruptions of the previous azimuthal symmetry by the presence of a non-zero topological index, giving rise to spiraling patterns in the energy density and gravitational waves along the optical axis. It is stressed here that this does not mean that those gravitational waves are longitudinal, their polarisations remain in the plane transverse to the direction of propagation.\\

 As for all waves emitted in all but the direction corresponding to $\theta = \alpha$, those "optical axis gravitational waves" are suppressed when $\zeta_L > 1$ and the beaming effect towards the half-cone angle takes place.

\subsubsection{Power emitted by gravitational waves ($l=\pm 1$)}

The full metric perturbation being a linear superposition of independent $+$ and $\times$ polarisations, the power per unit solid angle of equation takes the form 

\begin{equation}
   dP/ d\Omega \propto (g_{+}(\theta) \left<h_+^2\right>+g_{\times}(\theta)\left<h_{\times}^2\right> ) ~,
\end{equation}

With

\begin{eqnarray}
       g_{+}(\theta) &\equiv& \frac{\sin(\alpha)^4}{16} \left[ 19+12 \cos(2\theta)+ \cos(4\theta) \right] ~, \\
     g_{\times}(\theta) &\equiv& \sin(\alpha)^4 \left[ 1+ \cos(2\theta) \right] ~.
\end{eqnarray}

When $l=1$, the corresponding angular power distributions are:

\begin{align}
\frac{dP_+}{d\Omega} = g_{+}(\theta)\mathcal{N}_B^{-2} \left( \omega_g^{l=1} \right)^2  \frac{\pi G L^2 P_0^2 } {c^7}\Gamma_{l-1}^2(\theta) \text{Sinc}\left [ \eta(\theta)  \right ]^2 ~,  \\
\frac{dP_{\times}}{d\Omega} = g_{\times}(\theta)\mathcal{N}_B^{-2} \left( \omega_g^{l=1} \right)^2  \frac{\pi G L^2 P_0^2 } {c^7}\Gamma_{l-1}^2(\theta) \text{Sinc}\left [ \eta(\theta)  \right ]^2 ~.
\end{align}

\begin{figure}[htp!]
\centering
\includegraphics[scale=0.65]{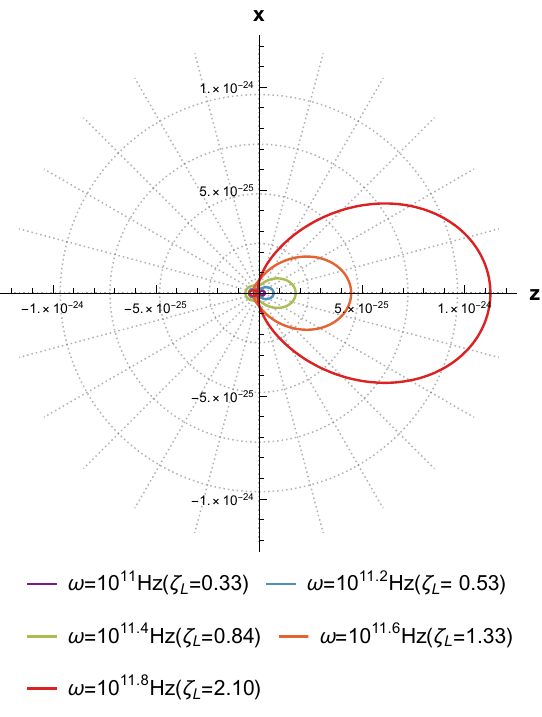}
w\caption{Angular distribution of the $l=1$ NIF ARC $h_{+}$ generated gravitational $dP_D/d\Omega$,  for fixed values of $L=1$mm, $D=100\mu$m and $\beta^{-1}=10$m. Where the laser pulse propagates along the z-axis.}
\label{fig:Pfrequency}
\end{figure}

Figure \ref{fig:Pfrequency} represents the power per unit solid angle of the $+$ polarised gravitational wave radiation. It is sufficient to represent only one polarisation mode as the angular distributions of the gravitational wave radiation flux for the $+$ and $\times$ polarisations are nearly identical, especially when $\zeta_L> 1$. 

\section{Discussion}
\label{section discussion}

What conclusions can be drawn from the results previously obtained?\\

\begin{center}
    \textit{Twisted lasers: the brightest sources of gravitational waves produced by humankind?}
\end{center}

To know whether high energy lasers are the brightest sources of gravitational wave currently generated by humankind (or that could be generated in a near future) or not it is interesting to compare the previous amplitudes to the ones of gravitational waves emitted, \textit{e.g.}, by the orbital accelerator of the Spinlaunch project\footnote{see https://www.spinlaunch.com/}, which we estimated to be $h_{SL} \sim 10^{-36}$ at the optimistic distance of $10m$ to the source. This facility is probably among, if not the biggest, source of terrestrial gravitational waves produced by rotating masses. Even a 5 MJ laser pulse remains two orders of magnitude below (when the gravitational wave amplitude is evaluated at the same distance to the source). If a laser system could be constructed with energy comparable to the kinetic energy contained within the SpinLaunch system mass ($ \sim 4.3 \times 10^{8}$ J), the amplitude of the gravitational waves generated by Bessel modes would however become slightly bigger, $h\sim 10^{-35}$, still at $10m$ away from the source. 

Does it mean that the game is over for high-energy lasers as sources of gravitational waves? Not necessarily, as their interest is manifold:

\begin{enumerate}
    \item Beyond any possible technological application or measurable effect, the question of the gravitational field created by light is, in itself, fascinating and largely unexplored. A lot remains to be understood. This work provides new insights on the subtle properties of spacetime vibrations induced by coherent and monochromatic light.
    \item The source volume is orders of magnitude smaller than the sub-orbital accelerators of the Spinlaunch project. As previously stated in this paper it might be that a detection system as the one proposed \textit{e.g.}, in \cite{Vacalis:2023gdz}, can be put (very) close to the source. The gravitational wave amplitude being $\propto 1/r$, this possibility might compensate the lacking orders of magnitude.
    \item Twisted laser pulses allow for a better control of the gravitational wave properties such as their polarisation states and directions of emission. Thanks to a beam combiner it might even be possible to control their frequencies over a wide range. This possibility will be further explored in a subsequent study. 
    \item Twisted laser pulses may not be the highest sources of gravitational strains $h$ built by humankind, but they definitely are the highest sources of power radiated through gravitational waves, this quantity scaling with the square of the gravitational wave frequency: $dP_D / d\Omega \propto \omega_g^2$. Within the far-field approximation, the peak intensity of the SEL laser system at a distance equal to the laser pulse width and at $10m$ from the source are $1.44 \times 10^{-5} \text{Wm}^{-2}$ and $1.01 \times 10^{-19} \text{Wm}^{-2}  $ respectively, which is many (18!) orders of magnitude greater than the estimated intensity of the Spinlaunch generated gravitational wave at the same distance of 10m from the source: $I_{SL} \sim 1.1 \times 10^{-35} \text{Wm}^{-2} $. Interestingly, the intensity at the pulse boundary is one order of magnitude greater than the  $3.6 \times 10^{-6} \text{Wm}^{-2}$ intensity of the GW15094 signal \cite{Abbott_2016} (measured at the detector location on earth). 
\end{enumerate}
    
\begin{center}
    \textit{Beaming}
\end{center}

This study has revealed that gravitational waves emitted by Bessel mode laser pulses give rise to a beaming effect. The tilts and widths angular distributions presented, \textit{e.g.}, figures \ref{Angular distrib h l=0} or \ref{fig:TTP} depend significantly on the optical parameters, specifically through $\zeta_D \equiv \omega D/c$ and $\zeta_L \equiv \omega L/c$. When those parameters become greater than unity a beaming towards the half-cone angle $\alpha$ arises. This \textit{a priori} non-trivial effect reflects properties of the Bessel pulse energy distributions function. If such waves could be detected one day, varying the optical parameters such as the pulse frequency or its time duration would have direct and important consequences on the signal.

Interestingly, this effect arises due to the non-zero spatial extension of the source. Indeed, when the laser pulse approaches a point source, the $\zeta_{L,D}$ parameters decrease to values for which the beaming does not occur anymore.

\begin{center}
    \textit{Gravitational wave frequencies}
\end{center}

The gravitational waves explored in this paper are emitted at twice the pulse frequency, roughly around $10^{15}$ Hz.

The equivalence between $\omega_g^{l= 0}$ and $\omega_g^{l= \pm 1}$, respectively given equations (\ref{freuquency l=0}) and (\ref{frequency l=1}), might not be obvious at first sight as the sources of gravitational waves differ in both cases. 

Indeed, in the absence of orbital angular momentum, gravitational waves are emitted due to oscillations of the energy content in the source region, oscillations whose frequency is twice the frequency of the fields (the energy simply being linked to the fields squared).

When $l = \pm 1$ the situation is different. Even though the previous effect still exists, the dominant source of gravitational waves is oscillations of the orbital angular momentum induced spirals in the energy density of the electromagnetic field. By examining the locations at which $\sin(\psi_q)$ and $\cos(\psi_q)$ vanish, corresponding to $\psi_q = p\pi$ and $\psi_q = (2p+1)\pi/2, ~ p \in \mathbb{Z}, \; p \geq 0$ respectively, it can be inferred that the number of rotating lobes in the interval $\phi \in [0,2\pi]$ is twice the value of the topological index. Moving on to the temporal evolution of a constant phase point in a plane defined by $z=$constant, it can be demonstrated from $\partial \psi_q / \partial t =0  $ that the rotational frequency of each lobe in a laser with frequency $\omega$ is $\omega/l$.  Nevertheless, the frequency at which the system reproduces identically is dictated by the multiplication of the individual frequency and the number of lobes, yielding an exact frequency of $2\omega$ whatever the value of $\l$. Notably, and interestingly, $\omega_g$ is independent of the topological index.\\ 

 Those gravitational wave frequencies are orders of magnitude above that of standard astrophysical sources and even  those predicted by beyond the standard model physics \cite{Aggarwal:2020olq, Raidal:2017mfl, Franciolini:2022htd, Dong:2015yjs, Ireland:2023avg, Yoshino:2013ofa, Arvanitaki:2014wva, Brito:2014wla, Sun:2020gem,Caprini:2007xq, Caprini:2009fx, Caprini:2009yp,PhysRevD.42.354, PhysRevD.31.3052, ZhouRuiYu:2020bbs, Chang:2021afa}. Barring primordial black holes, those sources hardly go above the $10^{12}-10^{13}$ Hz, even in the most extreme scenarios. Only light black holes of primordial origin could emit gravitational waves at frequencies comparable or even greater than laser-generated gravitational waves\footnote{The upper bound on the frequency of gravitational waves emitted during the inspiralling process can only be avoided in the unlikely event of the observation of a system which is both merging and loosing mass \cite{Blachier:2023ygh}.}

\interfootnotelinepenalty=10000

\begin{center}
    \textit{Detection}
\end{center}

There is not yet a final word on the best way to search for gravitational waves at very high frequencies. The most promising possibility might not be by direct measurement of their strain but through their interaction with (static) electromagnetic fields \cite{zeldovich1973}.

The conversion into photons has been \cite{osti_4070463, osti_4377804, PhysRevD.16.2915} and still is \cite{Aggarwal:2020olq, Domcke:2023qle, Ejlli:2019bqj} studied frequently: in cosmology \cite{PhysRevLett.74.634, Pshirkov:2009sf, Dolgov:2012be, Ringwald:2020ist, Fujita:2020rdx}, and more recently in the context of haloscopes and resonant cavities \cite{Berlin:2021txa, Domcke:2022rgu, Barrau:2023kuv, Domcke:2023bat, Herman:PhysRevD.108.124009,Herman:PhysRevD.104.023524}. This method takes advantage of the high values of the gravitational wave frequencies. As an example, the effective current generated by the conversion of a gravitational wave in a GHz resonant cavity scales with the square of the gravitational wave frequency \cite{Berlin:2021txa}.\\

The experimental exploration of gravitational waves around optical frequencies is mainly twofold. First, based on the graviton-to-photon conversion expected in presence of magnetic fields, reinterpretation of data from already existing experiments initially dedicated to the search for weakly interacting slim particles such as OSQAR \cite{OSQAR:2015qdv} or CAST \cite{CAST:2017uph} has allowed to put constraints on the existence of a stochastic gravitational wave background in the $\left[ 2.7 - 14 \right] \times 10^{14}$ Hz and $\left[ 5 - 12 \right] \times 10^{18}$ Hz windows, with associated characteristic amplitudes of $h_c \approx 6 \times 10^{-26}$ and $h_c \approx 5 \times 10^{-28}$ respectively \cite{Ejlli:2019bqj}. Prospects operated for the JURA follow-up \cite{Beacham:2019nyx} lower this bound to $h_c \approx 10^{-32}$ around $10^{18} Hz$ \cite{Ejlli:2019bqj}. 

The possibility to use high-energy lasers as detectors for $10^{16}$ Hz gravitational waves has also recently been considered \cite{Vacalis:2023gdz}. In this paper the authors have investigated the feasibility of employing high-energy laser pulses as gravitational wave detectors, considering a laser pulse with an energy of 1.8 MJ. In their most optimistic scenario, they determined that a sensitivity level of $h\sim10^{-27}$ could be attained. 

The presented values, already impressive in themselves, remains for the moment orders of magnitude above that required to detect laser-produced gravitational waves. Further explorations of those detection methods will be performed as some advantages might be taken, \textit{e.g}, of the fact that laser pulses can be focused to incredibly small spot sizes.

 \begin{center}
    \textit{How to maximise the signal?}
\end{center}

What optical parameters influence the maximal gravitational wave amplitudes generated by a Bessel pulse? \\

Looking at equations (\ref{eq:D1}) and  (\ref{eq:D2}) and putting the trigonometric functions depending on $\psi_q$ apart, four different functions must \textit{a priori} play a role: $\bar{h}_0(r)$, the trigonometric functions in the half cone angle $\alpha$, $\Gamma_0$ and $\text{Sinc}\left[ \eta (\theta) \right]$. 

At fixed distance $r$ to the source, those functions depend on the following optical parameters: the pulse energy $\mathcal{E}$, its frequency $\omega$, its duration $\tau_P$, its half cone angle $\alpha$, its confinement diameter $D$ and its beam waist $\beta$.\\

\textbf{Pulse energy $\mathcal{E}$:} As shown in figures \ref{fig:BBstrain} and \ref{fig:BBstrain2}, spacetime deformations increase with the pulse energy. The intuitive scaling of the gravitational effect with the energy is recovered: the greater the energy, the bigger the maximal strain. The energy appears in $\bar{h}_{0}$. Indeed, at fixed $r$ and $\beta$, maximising this function is equivalent to maximise $E_0^2 L / \beta^2$, \textit{i.e.} the pulse power times its length, which is the pulse energy.

Higher energy pulses generate larger spacetime deformations. Does this mean that the most energetic laser pulses automatically are the most interesting sources from the point of view of an observer trying to detect them? Not necessarily as, ideally, for lower energy lasers such as the SEL 1.5 kJ pulse, a detection system could be placed closer to the source. Depending on how close one can get to the source, considering lower energy pulses might be favored\footnote{In most situations involving gravitation, the compactness plays a central role. In the sense of \cite{Alho:2022bki}, it is defined as $C=Gm/(Rc^2)$ for a mass $m$ inside a region of size $R$. For black holes, the compactness is $C=1/2$ and it can only be smaller for other sources. In the case of an astrophysical binary system, the radius of the innermost stable orbit scales as $1/C$. In the case considered in this work, the compactness is very small — typically of the order of $10^{-31}$ for a 351
nm beamline on the NIF — which simply reflects the fact that the beam is far from being inside its own gravitational radius. Usually, for binary systems of masses, a smaller compactness prevents the objects from getting very close one from the other and, therefore, decreases the gravitational wave strain ($h \propto C$). The situation considered throughout this work is different as the source here is not self-gravitating. If both $\omega$ and the distance $r$ at which the gravitational wave amplitude is evaluated are fixed, a decrease of the compactness plays in the opposite direction and increases the signal. As stated above, in an ideal scenario, more compact sources could however allow a detection system to be placed closer to the source, making them potentially more relevant.}. 
\\

\textbf{Pulse duration $\tau_P$:} In the expression for the peak gravitational wave strain it is apparent that at fixed L/r that the strain scales with the power of the laser. Thus the minimisation of the pulse duration at constant pulse energy maximises  the peak strain at the cost of the signal time. However, the pulse duration can not be decreased arbitrarily. Fundamentally, it is constrained to a single optical cycle, approximately on the order of $\lambda/c$ \cite{RevModPhys.78.309, Mourou}. For a laser with a central wavelength of $351$ nm, the pulse width constrained by a single optical cycle is $1.17$ fs. However, the current minimum duration of laser pulses in peta-Watt laser systems typically ranges from a few tens to a few hundred femtoseconds \cite{10.1063/1.5131174, https://doi.org/10.1002/lpor.202100705}. To enable the development of lasers operating in the exa-Watt regime \cite{LMJ} various techniques have been devised to compress laser pulses to the limit of a single optical cycle \cite{Li:19, Liu:21, Han}. A comprehensive review of these techniques can be found in \cite{https://doi.org/10.1002/lpor.202100705, 10.1063/1.5131174}.\\

\textbf{Confinement diameter $D$ and beam waist $\beta$:} The beam waist $\beta$ does not exist as an independent parameter. Rather, its value is intricately linked to both the wavelength of light and the desired half-cone angle and is constrained by the diffraction limit. On the other hand, the confinement diameter $D$ maintains its status as an independent parameter. However, it must adhere to the condition $D \geq \Delta_D$. 

As previously discussed and as shown in figure \ref{fig:GammaGraph2}, max$\left( \Gamma_0 \right)$ is independent of $\zeta_D \equiv \omega D /c$. Varying the diameter of the source cylinder has no effect on the maximal value of $\Gamma_0$, nor does a variation in the pulse frequency $\omega$. As shown in Appendix \ref{Appendix angular distrib l=0} the angular distribution of $\Gamma_0$ however depends on $\zeta_D$. Similarly to the Sinc function, for $\Gamma_0$ to take its maximal value whatever $\theta \in \left[ 0 ; 2 \pi \right] $ the condition $\zeta_D <1$ must be fulfilled, \textit{i.e.} $D$ must be lower than $\lambda/(2 \pi) \simeq 55.9$ nm for a NIF's 351 nm beam. For a numerical aperture of $N_a = 0.75$ \cite{s21196690}, it follows that $\zeta_D \geq 4.19$.

The fact that $\beta$ and $D$ must be decreased in order to boost the signal is somehow counter intuitive when making an analogy with massive binary systems, for which the distance between the two bodies increases the lever arm and thus the gravitational strain. The origin of this discrepancy is mainly twofold. First, the expansion of the retarded time to linear order accounts for the non-zero size of the laser pulse. Second, both the amplitude and the phase exhibit inherent variations at each point within the laser pulse. The difference between gravitational waves emitted by laser pulses and massive binary systems becomes more and more pronounced as the values of $\zeta_D$ and $\zeta_L$ increase above one, for which the laser pulse increasingly deviates from resembling a point source.\\

\textbf{Half cone angle $\alpha$:} 
The effect of the half-cone angle is two-fold. Firsly, the value of $\alpha$ cannot be fixed such that all four $h_{D}$, $h_{ZZ}$, ${h}_{+}^{(Q)}$  and ${h}_{x}^{(Q)}$ equations (\ref{eq:D1} - \ref{eq:hcross}) are simultaneously maximised. In order to maximise the highest number of them, the half-cone angle should be maximised. At diffraction limit For a numerical aperture of $N_a = 0.75$ \cite{s21196690}, $\alpha_{\text{max}} =0.24 \simeq \frac{3}{4\pi} ~ \text{rads}$. Secondly, the half-cone angle governs the interaction length and thus $\zeta_L$, which decreases as $\alpha$ increases. Unlike the confinement diameter, the interaction length can continue to decrease. For example, with $\alpha = 0.61$ radians, we find $\zeta_L = 3.65$. \\

\textbf{Pulse frequency $\omega$:} Since for photons $E = \hbar \omega$, we could expect a linear increase of the gravitational strain with the photon frequency\footnote{This differs from usual systems of masses in circular orbits for which the kinetic energy, and hence the strain, is proportional to $\omega^2$.}. However laser facilities deliver a fixed energy. Writing $E_0^2 \sim  n \hbar \omega / \epsilon_0 c$, with $n$ the photon number density, varying the frequency $\omega$ by processes such as second harmonic generation \cite{PhysRev.128.1761} results in diminishing the photon number density $n$ (or \textit{vice versa}) so that the total pulse energy remains unchanged. Therefore, once the pulse energy is fixed and the conditions $L < c / \omega$ and $D < c / \omega$ are satisfied, the pulse frequency plays no role in the amplitudes of the emitted gravitational waves. However the power radiated by gravitational waves scales quickly with $\omega$, as $dP_D / d\Omega \propto \omega^2$. Increasing the frequency is a promising way to increase the gravitational wave power. This effect is highly important as detection through the inverse Gertsensthein effect relies on this power.\\

In summary, to maximise the gravitational waves signal at fixed distance $r$ to the source:

\begin{itemize}
    \item The pulse energy needs to be maximised.
    \item At fixed pulse energy, the pulse frequency does not affect the strain, but its increase is of primary importance for the gravitational wave power. 
    \item Shorter pulse duration maximise the gravitational wave strain at the cost of the signal time. 
    \item The confinement diameter and waist need to be small enough such that $D < \lambda/(2 \pi)$. Once this condition is fulfilled the confinement diameter no longer matters.
    \item The half cone angle $\alpha$ needs to be increased.  
    \item The parameters $\zeta_D$ and $\zeta_L$, which determine the beaming of the generated gravitational wave, need to be minimised. 
\end{itemize}

\section{Conclusion}

The understanding of the detailed structure of the gravitational field produced by light is still in its infancy. In this work, the characteristics of gravitational waves produced by Bessel pulses have been extensively investigated for the first time. Analytical formulations and numerical evaluations of the spacetime deformations and the associated power have been provided. The influence of the different optical parameters on the signal has also been discussed in detail. In addition, the parameters of interest required to boost the signal amplitude have been made explicit. 

Notably, a deeply non-intuitive angular distribution of the signal has been found, revealing fascinating subtleties of spacetime deformations generated by high-power laser pulses. This exemplifies how different laser-generated gravitational waves are from textbook signals, e.g. those generated in astrophysical binary systems of compact objects. Most of the complex patterns found in this study do not exist at all when considering basic rotating masses.\\

Many paths remain to be explored, among which:

\begin{enumerate}

    \item Laguerre-Gaussian pulses are another class of twisted light pulses, often considered in applications. The gravitational waves generated by this second class of pulses with orbital angular momentum might reveal new aspects of gravitational radiation physics.

   \item Multi-beam interactions via beam combiners could facilitate the generation of gravitational waves at the beat frequency of the stress-energy distribution. Theoretically, using modern optical techniques, this could allow for the emission of gravitational waves in the MHz to GHz range. This could be of high interest for other detection methods, such as resonant cavities.  

    \item A laboratory control of gravitational waves may also be useful to probe parametric resonant effects that require high levels of tuning. Notably, in \cite{Brandenberger:2022xbu}, it has been shown that gravitational waves can excite a parametric resonant instability of the electromagnetic field, improving graviton to photon conversion. The inverse process, namely the exponential enhancement of the gravitational wave signal via parametric resonance, can also occur for standing electromagnetic waves in vacuum. This possibility is currently under investigation.
    \item  The group velocity of Bessel modes is not necessarily equal to the speed of light. Subluminal Bessel pulses may exhibit a Doppler effect in the generated gravitational wave signal which has the potential to influence its intensity, angular distribution, and time duration.
\end{enumerate}
All those possibilities will be considered in subsequent studies.\\

In addition, it should be kept in mind that most theories beyond general relativity make specific predictions for gravitational waves. In the usual interferometer setting, generically, signal deviations are degenerated with unknown astrophysical parameters describing the source. Laboratory emission of gravitational waves should lead to new constraints on the deformation parameters, although a quantitative study is far beyond the scope of this work. \\

Beyond any mastering of laboratory controlled gravitational waves, which would undoubtedly lead to a scientific revolution, it is already remarkable in itself that the richness -- and complexity -- of gravitational radiation physics that extends past elementary situations slowly reveals itself in its intricacy.

\section{Acknowledgements} 

The authors gratefully acknowledge useful discussions with Prof. Gianluca Gregori, Prof. Robert Bingham, Dr. Raoul Trines, Professor Steven Balbus, Mr. Georgios Vacalis and Dr Robert Kirkwood. They also thank all of the staff of the Central Laser Facility, Rutherford Appleton Laboratory for their assistance in the development of this work. The work was supported by the Oxford-ShanghaiTech collaboration agreement and UKRI-STFC grant ST/V001655/1.  The work of C.L. was supported by the grants 2018/30/Q/ST9/00795 and 2021/42/E/ST9/00260 from the National Science Centre (Poland).
\
\appendix


\section{Angular distribution of $\Gamma_l(\theta)$}
\label{Appendix angular distrib l=0}

To study the angular distribution of the emitted gravitational waves one first focuses on the study of the two functions that impact \textit{a priori} the direction of emission of gravitational waves in the $zx$ plane: 

\begin{itemize}
    \item $\text{Sinc}\left[ \eta(\theta) \right]$;
    \item  and $\Gamma_{q}^{BP}(\theta)$.
\end{itemize}

\begin{figure}[h]
\centering
\hspace{-0.8cm}
\includegraphics[scale=0.65]{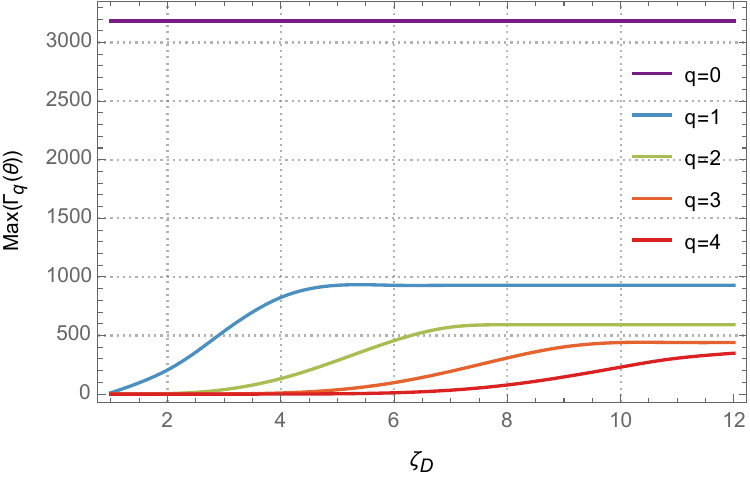}
\caption{Dependencies of $\text{Max}\left[\Gamma_{q}(\theta)\right]$ upon the $\zeta_D$ parameter obtained varying the angle $\theta \in \left[ 0 ; 2 \pi \right]$. The values $L=1$ mm and $D=1 $ mm have been chosen for clarity.}
\label{fig:GammaGraph2}
\end{figure}

We first consider the effect of the topological index. Figure \ref{fig:GammaGraph2} shows, for different values of the index $q$, the maximal values of $\Gamma_{q}(\theta)$ obtained spanning $\theta \in \left[ 0 ; 2 \pi \right]$. Whatever the value of $\zeta_D$, the dominant function is always the one with $q=0$. Thus, in the analysis conducted in the main corpus, only cases where $q=0$ was considered as they correspond to the main sources of gravitational waves.

The angular dependence of $\Gamma_{q}^{BP}(\theta)$ is entirely governed by the $J_{2q} \left[ \omega \tau \sin(\theta) /c \beta \right]$ factor within equation (\ref{GammaintBP}).

\begin{figure}[h]
\centering
\includegraphics[scale=0.72]{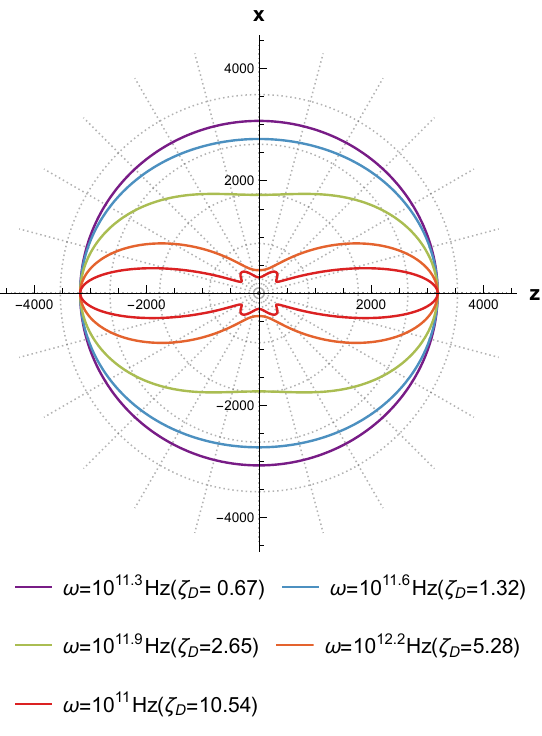}
\caption{Angular variation of $ \Gamma_{0}(\theta)$ in the $zx$ plane, for $\theta \in \left[ 0; 2 \pi \right[$ and for different values of the pulse frequency $\omega$, hence of $\zeta_D$, with fixed value of $D=0.1$mm.}
\label{fig:GammaGraph}
\end{figure}
    
In figure  \ref{fig:GammaGraph2} we have stated that the dominant $\Gamma_q$ function is $\Gamma_{0}(\theta)$ and that its maximal value, occurring along the optical axis when $\theta = 0$, is independent of $\zeta_D$, hence of the cylinder diameter. 
Figure \ref{fig:GammaGraph} represents the variation of $\Gamma_{0}(\theta)$ for $\theta \in \left[ 0 ; 2 \pi \right[$, keeping the size of the cylinder given by $L$ and $D$ fixed, but varying the frequency (which induces variations of $\zeta_D$). For values of $\zeta_D >1$ a beaming effect starts to appear along the direction of propagation of the pulse. This beaming leads to a decrease of $\Gamma_{0}(\theta)$ off z-axis, and thus a decrease of the emitted waves amplitudes off optical axis. To prevent the emergence of this effect it is preferable to work with $\zeta_D < 1$. At fixed frequency, it means considering a cylinder diameter such that $D < c / \omega$. 

\begin{figure}[h]
\centering
\includegraphics[scale=0.75]{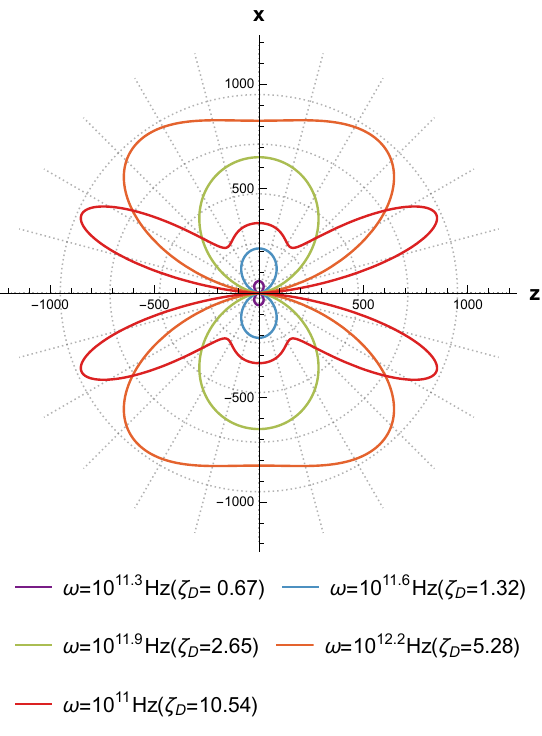}
\caption{Angular variation of $\Gamma_{1}(\theta)$ in the $zx$ plane, for $\theta \in \left[ 0; 2 \pi \right[$ and for different values of the pulse frequency $\omega$, hence of $\zeta_D$, with fixed value of $D=0.1$mm.}
\label{fig:GammaGraph3}
\end{figure}

An increase of $\zeta_D$ induces an increase of $\Gamma_{q \neq 0}$ functions, as depicted in figure  \ref{fig:GammaGraph2}. Unlike $\Gamma_0$ the maximal values of all other $\Gamma_{q \neq 0}$ and the angles $\theta$ at which they occur depend on $\zeta_D$. Indeed, Consider some threshold $\zeta_D^{th}$ which value depends on $q$. For $q=0$, $\zeta_D^{th} \simeq 1$ but for $q \neq 0$ the threshold must be computed numerically. For all values $\zeta_D \leq \zeta_D^{th} $ the maxima of the $\Gamma_{q \neq 0}$ lie along the x-axis. However, when $\zeta_D > \zeta_D^{th} $, the location of the maxima of $\Gamma_{q \neq 0}$ vary and need to be evaluated numerically. As an example, the case of $q=1$ has been depicted in figure  \ref{fig:GammaGraph3}.

Obviously $\Gamma_0$ does not remain the dominant function for all values of $\theta$ and $\zeta_D$. For this function to remain overwhelmingly dominant whatever the direction around the source, rendering all the other generated $\Gamma_q, ~ q \neq 0,$ negligible, one needs to stay in the regime $\zeta_D < \zeta_D^{th}$, for which $\zeta_D^{th} > 1$ if $q \neq 0 $.

This effect arises due to the way the source volume is defined. At fixed values of $\beta$ and $\omega$, increasing the cylinder diameter $D$ increases the number of lobes of the Bessel pulse that act as sources of gravitational waves. Since for Bessel pulses the energy is split equally between the lobes \cite{Durnin:88, Durnin1, Durnin2}, more lobes in the cylinder result in smaller peak intensity in each lobe, thus in smaller gravitational waves amplitudes.

\section{Why is twisting not the dominant effect generating gravitational waves in Bessel pulses?}
\label{appendix twisting not dominant}

How is it that the twisting effect is not dominant in the generation of gravitational waves? To better understand this question, the Maxwell tensor is first discussed in more detail.

The Maxwell tensor serves as a mathematical representation of the "pressure" and "shear stress" exerted on the surrounding spacetime by electromagnetic waves. 

In the case of a sole electromagnetic wave, these stresses are aligned, when acting on a physical object, with the momentum of the wave. For a single plane wave this direction corresponds to its direction of propagation. Without loss of generality and as we did, it is possible to align the z-axis along the propagation direction of the plane wave. In this scenario, the only non-zero component of the Maxwell tensor is $\sigma_{zz}$, indicating the presence of pressure along the z-axis. 

To generate additional pressure and shear stress components in the Maxwell tensor, the interaction of multiple electromagnetic plane waves propagating at angles to each other is necessary. When multiple plane waves interact, the resulting electromagnetic field distribution can give rise to non-zero components of the Maxwell tensor beyond $\sigma_{zz}$. 

The electromagnetic wave field tensors of two plane waves propagating at an angle $2 \Theta$ to each other in the $xz$ plane, described by the 4-wavevectors  $K^{\mu}_{\pm} \equiv \left[ \omega/c, \pm k \sin(\Theta), 0 , k \cos(\Theta) \right]$ are given by:

\begin{eqnarray}
   \frac{\mathcal{F}^{\mu \nu}_{\pm} (\Theta)}{C_\pm}  =  \begin{pmatrix}
0  &-\cos(\Theta) & 0& \pm \sin(\Theta) \\
\cos(\Theta)  & 0  & 0&1 \\
0 & 0  & 0&0 \\
\mp \sin(\Theta) & -1  &0 &   0
\end{pmatrix} ~,
\end{eqnarray}
    
where $C_\pm \equiv (E_0/c) \cos(K^\pm_{\mu}x^{\mu})$. The Maxwell tensor resulting from their interaction is:

\begin{align}
&\sigma_{xx} = \frac{1}{2}\epsilon_0 c^3 \sin(\Theta)^2 (B^2 - A^2) ~,
\label{eq:maxwellxx} \\
&\sigma_{yy} = -\sigma_{xx} ~, \label{eq:maxwellyy} \\
&\sigma_{zz} = -\frac{1}{2}\epsilon_0 c^3 \left((1+\cos(\Theta)^2) A^2 + \sin(\Theta)^2 B^2 \right) ~, \\
&\sigma_{xz} = - \frac{1}{2}\epsilon_0 c^3 \sin(2\Theta)AB ~,\\
&\sigma_{xy} = 0 ~,\\
&\sigma_{yz} = 0 ~.  
\end{align}

Where $A$ and $B$ are  $C_++C_-$ and $C_+-C_-$ respectively. The resulting amplitude of the oscillatory components  $ \Tilde{\sigma}_{xx} $ and $\Tilde{\sigma}_{zz} $ of the Maxwell tensor for this system at $x=0$ are given by $\Tilde{\sigma}_{xx}(\Theta) =  \epsilon_0 E_0^2 \sin^2(\Theta)$ and $ \Tilde{\sigma}_{zz} =\epsilon_0 \left( E_0^2/2 \right) (1-\cos(2\Theta)-4)$ respectively. Within the domain $\Theta \in [0,\pi/2]$, $|\Tilde{\sigma}_{xx}(\Theta)| < |\Tilde{\sigma}_{zz}(\Theta)|$. This reasoning can be generalised to the case of an Bessel mode which is an infinite superposition of plane waves. The difference for Bessel modes is that they can exhibit a twist in the field such that it produces shear stresses on the surounding spacetime that give rise to $\bar{h}_{xy}$, unlike the case of two interacting plane waves which only produce outward pressure. Thus in contrast to the $\sigma_{zz}$ component of the Maxwell tensor, the components $\sigma_{ab}$, $a,b \in \left\lbrace x,y \right\rbrace$ exhibit growth with increasing tilt angle between the corresponding electromagnetic waves. Since a Bessel mode can be regarded as an infinite superposition of plane waves with a characteristic angle of inclination $\alpha$ between them, the amplitude of the generated $+$ and $\times$ polarized gravitational waves is lower than that of the gravitational waves generated by the $\sigma_{ab}^D$ and $\sigma_{ab}^{ZZ}$ terms of the Maxwell tensor.\\


\begin{thebibliography}{0}%
\makeatletter
\providecommand \@ifxundefined [1]{%
 \@ifx{#1\undefined}
}%
\providecommand \@ifnum [1]{%
 \ifnum #1\expandafter \@firstoftwo
 \else \expandafter \@secondoftwo
 \fi
}%
\providecommand \@ifx [1]{%
 \ifx #1\expandafter \@firstoftwo
 \else \expandafter \@secondoftwo
 \fi
}%
\providecommand \natexlab [1]{#1}%
\providecommand \enquote  [1]{``#1''}%
\providecommand \bibnamefont  [1]{#1}%
\providecommand \bibfnamefont [1]{#1}%
\providecommand \citenamefont [1]{#1}%
\providecommand \href@noop [0]{\@secondoftwo}%
\providecommand \href [0]{\begingroup \@sanitize@url \@href}%
\providecommand \@href[1]{\@@startlink{#1}\@@href}%
\providecommand \@@href[1]{\endgroup#1\@@endlink}%
\providecommand \@sanitize@url [0]{\catcode `\\12\catcode `\$12\catcode
  `\&12\catcode `\#12\catcode `\^12\catcode `\_12\catcode `\%12\relax}%
\providecommand \@@startlink[1]{}%
\providecommand \@@endlink[0]{}%
\providecommand \url  [0]{\begingroup\@sanitize@url \@url }%
\providecommand \@url [1]{\endgroup\@href {#1}{\urlprefix }}%
\providecommand \urlprefix  [0]{URL }%
\providecommand \Eprint [0]{\href }%
\providecommand \doibase [0]{https://doi.org/}%
\providecommand \selectlanguage [0]{\@gobble}%
\providecommand \bibinfo  [0]{\@secondoftwo}%
\providecommand \bibfield  [0]{\@secondoftwo}%
\providecommand \translation [1]{[#1]}%
\providecommand \BibitemOpen [0]{}%
\providecommand \bibitemStop [0]{}%
\providecommand \bibitemNoStop [0]{.\EOS\space}%
\providecommand \EOS [0]{\spacefactor3000\relax}%
\providecommand \BibitemShut  [1]{\csname bibitem#1\endcsname}%
\let\auto@bib@innerbib\@empty
\end{thebibliography}%


\begin{thebibliography}{105}%
\makeatletter
\providecommand \@ifxundefined [1]{%
 \@ifx{#1\undefined}
}%
\providecommand \@ifnum [1]{%
 \ifnum #1\expandafter \@firstoftwo
 \else \expandafter \@secondoftwo
 \fi
}%
\providecommand \@ifx [1]{%
 \ifx #1\expandafter \@firstoftwo
 \else \expandafter \@secondoftwo
 \fi
}%
\providecommand \natexlab [1]{#1}%
\providecommand \enquote  [1]{``#1''}%
\providecommand \bibnamefont  [1]{#1}%
\providecommand \bibfnamefont [1]{#1}%
\providecommand \citenamefont [1]{#1}%
\providecommand \href@noop [0]{\@secondoftwo}%
\providecommand \href [0]{\begingroup \@sanitize@url \@href}%
\providecommand \@href[1]{\@@startlink{#1}\@@href}%
\providecommand \@@href[1]{\endgroup#1\@@endlink}%
\providecommand \@sanitize@url [0]{\catcode `\\12\catcode `\$12\catcode
  `\&12\catcode `\#12\catcode `\^12\catcode `\_12\catcode `\%12\relax}%
\providecommand \@@startlink[1]{}%
\providecommand \@@endlink[0]{}%
\providecommand \url  [0]{\begingroup\@sanitize@url \@url }%
\providecommand \@url [1]{\endgroup\@href {#1}{\urlprefix }}%
\providecommand \urlprefix  [0]{URL }%
\providecommand \Eprint [0]{\href }%
\providecommand \doibase [0]{https://doi.org/}%
\providecommand \selectlanguage [0]{\@gobble}%
\providecommand \bibinfo  [0]{\@secondoftwo}%
\providecommand \bibfield  [0]{\@secondoftwo}%
\providecommand \translation [1]{[#1]}%
\providecommand \BibitemOpen [0]{}%
\providecommand \bibitemStop [0]{}%
\providecommand \bibitemNoStop [0]{.\EOS\space}%
\providecommand \EOS [0]{\spacefactor3000\relax}%
\providecommand \BibitemShut  [1]{\csname bibitem#1\endcsname}%
\let\auto@bib@innerbib\@empty
\bibitem [{\citenamefont {Weber}(1960)}]{PhysRev.117.306}%
  \BibitemOpen
  \bibfield  {author} {\bibinfo {author} {\bibfnamefont {J.}~\bibnamefont
  {Weber}},\ }\bibfield  {title} {\bibinfo {title} {Detection and generation of
  gravitational waves},\ }\href {https://doi.org/10.1103/PhysRev.117.306}
  {\bibfield  {journal} {\bibinfo  {journal} {Phys. Rev.}\ }\textbf {\bibinfo
  {volume} {117}},\ \bibinfo {pages} {306} (\bibinfo {year}
  {1960})}\BibitemShut {NoStop}%
\bibitem [{\citenamefont {{Halpern}}\ and\ \citenamefont
  {{Laurent}}(1964)}]{HalpernLaurent}%
  \BibitemOpen
  \bibfield  {author} {\bibinfo {author} {\bibfnamefont {L.}~\bibnamefont
  {{Halpern}}}\ and\ \bibinfo {author} {\bibfnamefont {B.}~\bibnamefont
  {{Laurent}}},\ }\bibfield  {title} {\bibinfo {title} {{On the gravitational
  radiation of microscopic systems}},\ }\href
  {https://doi.org/10.1007/BF02749891} {\bibfield  {journal} {\bibinfo
  {journal} {Il Nuovo Cimento}\ }\textbf {\bibinfo {volume} {33}},\ \bibinfo
  {pages} {728} (\bibinfo {year} {1964})}\BibitemShut {NoStop}%
\bibitem [{\citenamefont {Grishchuk}\ and\ \citenamefont
  {Sazhin}(1974)}]{1974GrishchukSazhin}%
  \BibitemOpen
  \bibfield  {author} {\bibinfo {author} {\bibfnamefont {L.}~\bibnamefont
  {Grishchuk}}\ and\ \bibinfo {author} {\bibfnamefont {M.}~\bibnamefont
  {Sazhin}},\ }\bibfield  {title} {\bibinfo {title} {Emission of gravitational
  waves by an electromagnetic cavity},\ }\href@noop {} {\bibfield  {journal}
  {\bibinfo  {journal} {Journal of Experimental and Theoretical Physics - J EXP
  THEOR PHYS}\ }\textbf {\bibinfo {volume} {38}} (\bibinfo {year}
  {1974})}\BibitemShut {NoStop}%
\bibitem [{\citenamefont {{Grishchuk}}\ and\ \citenamefont
  {{Sazhin}}(1975)}]{1975GrischukSazhin}%
  \BibitemOpen
  \bibfield  {author} {\bibinfo {author} {\bibfnamefont {L.~P.}\ \bibnamefont
  {{Grishchuk}}}\ and\ \bibinfo {author} {\bibfnamefont {M.~V.}\ \bibnamefont
  {{Sazhin}}},\ }\bibfield  {title} {\bibinfo {title} {{Excitation and
  detection of standing gravitational waves}},\ }\href@noop {} {\bibfield
  {journal} {\bibinfo  {journal} {Zhurnal Eksperimentalnoi i Teoreticheskoi
  Fiziki}\ }\textbf {\bibinfo {volume} {68}},\ \bibinfo {pages} {1569}
  (\bibinfo {year} {1975})}\BibitemShut {NoStop}%
\bibitem [{\citenamefont {Grishchuk}(1977)}]{LeonidPGrishchuk_1977}%
  \BibitemOpen
  \bibfield  {author} {\bibinfo {author} {\bibfnamefont {L.~P.}\ \bibnamefont
  {Grishchuk}},\ }\bibfield  {title} {\bibinfo {title} {Gravitational waves in
  the cosmos and the laboratory},\ }\href
  {https://doi.org/10.1070/PU1977v020n04ABEH005327} {\bibfield  {journal}
  {\bibinfo  {journal} {Soviet Physics Uspekhi}\ }\textbf {\bibinfo {volume}
  {20}},\ \bibinfo {pages} {319} (\bibinfo {year} {1977})}\BibitemShut
  {NoStop}%
\bibitem [{\citenamefont {Grishchuk}(2003)}]{Grishchuk:2003un}%
  \BibitemOpen
  \bibfield  {author} {\bibinfo {author} {\bibfnamefont {L.~P.}\ \bibnamefont
  {Grishchuk}},\ }\bibfield  {title} {\bibinfo {title} {{Electromagnetic
  generators and detectors of gravitational waves}},\ }in\ \href@noop {} {\emph
  {\bibinfo {booktitle} {{1st Conference on High Frequency Gravitational
  Waves}}}}\ (\bibinfo {year} {2003})\ \Eprint
  {https://arxiv.org/abs/gr-qc/0306013} {arXiv:gr-qc/0306013} \BibitemShut
  {NoStop}%
 \bibitem [{\citenamefont {Bulanov}\ \emph {et~al.}(2023)\citenamefont
  {Esirkepov}, \citenamefont {Thomas}, \citenamefont {Koga}, and\ \citenamefont {Bulanov}}]{PhysRevLett.105.220407}%
  \BibitemOpen
  \bibfield  {author} {\bibinfo {author} {\bibfnamefont {S.}~\bibnamefont
  {Bulanov}}, \bibinfo {author} {\bibfnamefont {T.}\ \bibnamefont
  {Esirkepov}}, \bibinfo {author} {\bibfnamefont {A.}\ \bibnamefont
  {Thomas}},\bibinfo {author} {\bibfnamefont {J.}\ \bibnamefont
  {Koga}},\ and\ \bibinfo {author} {\bibfnamefont {S.}~\bibnamefont {Bulanov}},\ }\bibfield
  {title} {\bibinfo {title} {{Schwinger Limit Attainability with Extreme Power Lasers}},\ }\href {10.1103/PhysRevLett.105.220407} {\bibfield
  {journal} {\bibinfo  {journal} {Phys. Rev. Lett.}\ }\textbf {\bibinfo
  {volume} {105}},\ \bibinfo {pages} {220407} (\bibinfo {year} {2010})},\
  \Eprint {https://link.aps.org/doi/10.1103/PhysRevLett.105.220407} 
  \BibitemShut %
\bibitem [{\citenamefont {Kolosnitsyn}\ and\ \citenamefont
  {Rudenko}(2015)}]{Kolosnitsyn:2015zua}%
  \BibitemOpen
  \bibfield  {author} {\bibinfo {author} {\bibfnamefont {N.~I.}\ \bibnamefont
  {Kolosnitsyn}}\ and\ \bibinfo {author} {\bibfnamefont {V.~N.}\ \bibnamefont
  {Rudenko}},\ }\bibfield  {title} {\bibinfo {title} {{Gravitational Hertz
  experiment with electromagnetic radiation in a strong magnetic field}},\
  }\href {https://doi.org/10.1088/0031-8949/90/7/074059} {\bibfield  {journal}
  {\bibinfo  {journal} {Phys. Scripta}\ }\textbf {\bibinfo {volume} {90}},\
  \bibinfo {pages} {074059} (\bibinfo {year} {2015})},\ \Eprint
  {https://arxiv.org/abs/1504.06548} {arXiv:1504.06548 [gr-qc]} \BibitemShut
  {NoStop}%
\bibitem [{\citenamefont {{Faccio}}\ \emph {et~al.}(2006)\citenamefont
  {{Faccio}}, \citenamefont {{Clerici}},\ and\ \citenamefont
  {{Tambuchi}}}]{Hertz}%
  \BibitemOpen
  \bibfield  {author} {\bibinfo {author} {\bibfnamefont {D.}~\bibnamefont
  {{Faccio}}}, \bibinfo {author} {\bibfnamefont {M.}~\bibnamefont
  {{Clerici}}},\ and\ \bibinfo {author} {\bibfnamefont {D.}~\bibnamefont
  {{Tambuchi}}},\ }\bibfield  {title} {\bibinfo {title} {{Revisiting the 1888
  Hertz experiment}},\ }\href {https://doi.org/10.1119/1.2238886} {\bibfield
  {journal} {\bibinfo  {journal} {American Journal of Physics}\ }\textbf
  {\bibinfo {volume} {74}},\ \bibinfo {pages} {992} (\bibinfo {year} {2006})},\
  \Eprint {https://arxiv.org/abs/physics/0602073} {arXiv:physics/0602073
  [physics.ed-ph]} \BibitemShut {NoStop}%
\bibitem [{\citenamefont {{Morozov}}\ \emph {et~al.}(2021)\citenamefont
  {{Morozov}}, \citenamefont {{Pustovoit}},\ and\ \citenamefont
  {{Fomin}}}]{2021Morozov}%
  \BibitemOpen
  \bibfield  {author} {\bibinfo {author} {\bibfnamefont {A.~N.}\ \bibnamefont
  {{Morozov}}}, \bibinfo {author} {\bibfnamefont {V.~I.}\ \bibnamefont
  {{Pustovoit}}},\ and\ \bibinfo {author} {\bibfnamefont {I.~V.}\ \bibnamefont
  {{Fomin}}},\ }\bibfield  {title} {\bibinfo {title} {{Generation of
  Gravitational Waves by a Standing Electromagnetic Wave}},\ }\href
  {https://doi.org/10.1134/S020228932101014X} {\bibfield  {journal} {\bibinfo
  {journal} {Gravitation and Cosmology}\ }\textbf {\bibinfo {volume} {27}},\
  \bibinfo {pages} {24} (\bibinfo {year} {2021})}\BibitemShut {NoStop}%
\bibitem [{\citenamefont {Tolman}\ \emph {et~al.}(1931)\citenamefont {Tolman},
  \citenamefont {Ehrenfest},\ and\ \citenamefont {Podolsky}}]{Tolman:1931zza}%
  \BibitemOpen
  \bibfield  {author} {\bibinfo {author} {\bibfnamefont {R.~C.}\ \bibnamefont
  {Tolman}}, \bibinfo {author} {\bibfnamefont {P.}~\bibnamefont {Ehrenfest}},\
  and\ \bibinfo {author} {\bibfnamefont {B.}~\bibnamefont {Podolsky}},\
  }\bibfield  {title} {\bibinfo {title} {On the gravitational field produced by
  light},\ }\href {https://doi.org/10.1103/PhysRev.37.602} {\bibfield
  {journal} {\bibinfo  {journal} {Phys. Rev.}\ }\textbf {\bibinfo {volume}
  {37}},\ \bibinfo {pages} {602} (\bibinfo {year} {1931})}\BibitemShut
  {NoStop}%
\bibitem [{\citenamefont {{M. E. Gertsenshtein}}(1962)}]{Gert}%
  \BibitemOpen
  \bibfield  {author} {\bibinfo {author} {\bibnamefont {{M. E.
  Gertsenshtein}}},\ }\bibfield  {title} {\bibinfo {title} {{Wave resonance of
  light and gravitional waves}},\ }\href@noop {} {\bibfield  {journal}
  {\bibinfo  {journal} {{Journal of Experimental and Theoretical Physics}}\
  }\textbf {\bibinfo {volume} {41}},\ \bibinfo {pages} {113} (\bibinfo {year}
  {1962})}\BibitemShut {NoStop}%
\bibitem [{\citenamefont {Danson}\ \emph {et~al.}(2019)\citenamefont {Danson},
  \citenamefont {Haefner}, \citenamefont {Bromage}, \citenamefont {Butcher},
  \citenamefont {Chanteloup}, \citenamefont {Chowdhury}, \citenamefont
  {Galvanauskas}, \citenamefont {Gizzi}, \citenamefont {Hein}, \citenamefont
  {Hillier}, \citenamefont {Hopps}, \citenamefont {Kato}, \citenamefont
  {Khazanov}, \citenamefont {Kodama}, \citenamefont {Korn}, \citenamefont {Li},
  \citenamefont {Li}, \citenamefont {Limpert}, \citenamefont {Ma},
  \citenamefont {Nam}, \citenamefont {Neely}, \citenamefont {Papadopoulos},
  \citenamefont {Penman}, \citenamefont {Qian}, \citenamefont {Rocca},
  \citenamefont {Shaykin}, \citenamefont {Siders}, \citenamefont {Spindloe},
  \citenamefont {Szatmári}, \citenamefont {Trines}, \citenamefont {Zhu},
  \citenamefont {Zhu},\ and\ \citenamefont {Zuegel}}]{Dawson}%
  \BibitemOpen
  \bibfield  {author} {\bibinfo {author} {\bibfnamefont {C.~N.}\ \bibnamefont
  {Danson}}, \bibinfo {author} {\bibfnamefont {C.}~\bibnamefont {Haefner}},
  \bibinfo {author} {\bibfnamefont {J.}~\bibnamefont {Bromage}}, \bibinfo
  {author} {\bibfnamefont {T.}~\bibnamefont {Butcher}}, \bibinfo {author}
  {\bibfnamefont {J.-C.~F.}\ \bibnamefont {Chanteloup}}, \bibinfo {author}
  {\bibfnamefont {E.~A.}\ \bibnamefont {Chowdhury}}, \bibinfo {author}
  {\bibfnamefont {A.}~\bibnamefont {Galvanauskas}}, \bibinfo {author}
  {\bibfnamefont {L.~A.}\ \bibnamefont {Gizzi}}, \bibinfo {author}
  {\bibfnamefont {J.}~\bibnamefont {Hein}}, \bibinfo {author} {\bibfnamefont
  {D.~I.}\ \bibnamefont {Hillier}}, \bibinfo {author} {\bibfnamefont {N.~W.}\
  \bibnamefont {Hopps}}, \bibinfo {author} {\bibfnamefont {Y.}~\bibnamefont
  {Kato}}, \bibinfo {author} {\bibfnamefont {E.~A.}\ \bibnamefont {Khazanov}},
  \bibinfo {author} {\bibfnamefont {R.}~\bibnamefont {Kodama}}, \bibinfo
  {author} {\bibfnamefont {G.}~\bibnamefont {Korn}}, \bibinfo {author}
  {\bibfnamefont {R.}~\bibnamefont {Li}}, \bibinfo {author} {\bibfnamefont
  {Y.}~\bibnamefont {Li}}, \bibinfo {author} {\bibfnamefont {J.}~\bibnamefont
  {Limpert}}, \bibinfo {author} {\bibfnamefont {J.}~\bibnamefont {Ma}},
  \bibinfo {author} {\bibfnamefont {C.~H.}\ \bibnamefont {Nam}}, \bibinfo
  {author} {\bibfnamefont {D.}~\bibnamefont {Neely}}, \bibinfo {author}
  {\bibfnamefont {D.}~\bibnamefont {Papadopoulos}}, \bibinfo {author}
  {\bibfnamefont {R.~R.}\ \bibnamefont {Penman}}, \bibinfo {author}
  {\bibfnamefont {L.}~\bibnamefont {Qian}}, \bibinfo {author} {\bibfnamefont
  {J.~J.}\ \bibnamefont {Rocca}}, \bibinfo {author} {\bibfnamefont {A.~A.}\
  \bibnamefont {Shaykin}}, \bibinfo {author} {\bibfnamefont {C.~W.}\
  \bibnamefont {Siders}}, \bibinfo {author} {\bibfnamefont {C.}~\bibnamefont
  {Spindloe}}, \bibinfo {author} {\bibfnamefont {S.}~\bibnamefont {Szatmári}},
  \bibinfo {author} {\bibfnamefont {R.~M. G.~M.}\ \bibnamefont {Trines}},
  \bibinfo {author} {\bibfnamefont {J.}~\bibnamefont {Zhu}}, \bibinfo {author}
  {\bibfnamefont {P.}~\bibnamefont {Zhu}},\ and\ \bibinfo {author}
  {\bibfnamefont {J.~D.}\ \bibnamefont {Zuegel}},\ }\bibfield  {title}
  {\bibinfo {title} {Petawatt and exawatt class lasers worldwide},\ }\href
  {https://doi.org/10.1017/hpl.2019.36} {\bibfield  {journal} {\bibinfo
  {journal} {High power laser science and engineering}\ }\textbf {\bibinfo
  {volume} {7}},\ \bibinfo {pages} {e54} (\bibinfo {year} {2019})}\BibitemShut
  {NoStop}%
\bibitem [{\citenamefont {Rätzel}\ \emph {et~al.}(2016)\citenamefont
  {Rätzel}, \citenamefont {Wilkens},\ and\ \citenamefont
  {Menzel}}]{Rätzel_2016}%
  \BibitemOpen
  \bibfield  {author} {\bibinfo {author} {\bibfnamefont {D.}~\bibnamefont
  {Rätzel}}, \bibinfo {author} {\bibfnamefont {M.}~\bibnamefont {Wilkens}},\
  and\ \bibinfo {author} {\bibfnamefont {R.}~\bibnamefont {Menzel}},\
  }\bibfield  {title} {\bibinfo {title} {Gravitational properties of
  light—the gravitational field of a laser pulse},\ }\href
  {https://doi.org/10.1088/1367-2630/18/2/023009} {\bibfield  {journal}
  {\bibinfo  {journal} {New Journal of Physics}\ }\textbf {\bibinfo {volume}
  {18}},\ \bibinfo {pages} {023009} (\bibinfo {year} {2016})}\BibitemShut
  {NoStop}%
\bibitem [{\citenamefont {Scully}(1979)}]{PhysRevD.19.3582}%
  \BibitemOpen
  \bibfield  {author} {\bibinfo {author} {\bibfnamefont {M.~O.}\ \bibnamefont
  {Scully}},\ }\bibfield  {title} {\bibinfo {title} {General-relativistic
  treatment of the gravitational coupling between laser beams},\ }\href
  {https://doi.org/10.1103/PhysRevD.19.3582} {\bibfield  {journal} {\bibinfo
  {journal} {Phys. Rev. D}\ }\textbf {\bibinfo {volume} {19}},\ \bibinfo
  {pages} {3582} (\bibinfo {year} {1979})}\BibitemShut {NoStop}%
\bibitem [{\citenamefont {Füzfa}(2018)}]{füzfa2018electromagnetic}%
  \BibitemOpen
  \bibfield  {author} {\bibinfo {author} {\bibfnamefont {A. }\ \bibnamefont
  {Füzfa}},\ }\bibfield  {title} {\bibinfo {title} {Electromagnetic Gravitational Waves Antennas for Directional Emission and Reception},\ }\href
  {https://doi.org/10.48550/arXiv.1702.06052} {\bibfield  {journal} {\bibinfo
  {journal} {}\ }\textbf {\bibinfo {volume} {}},\ \bibinfo
  {pages} {} (\bibinfo {year} {2018})}\BibitemShut {NoStop}%
\bibitem [{\citenamefont {Lageyre}\ \emph {et~al.}(2022)\citenamefont
  {Lageyre}, \citenamefont {d'Humi\`eres},\ and\ \citenamefont
  {Ribeyre}}]{PhysRevD.105.104052}%
  \BibitemOpen
  \bibfield  {author} {\bibinfo {author} {\bibfnamefont {P.}~\bibnamefont
  {Lageyre}}, \bibinfo {author} {\bibfnamefont {E.}~\bibnamefont
  {d'Humi\`eres}},\ and\ \bibinfo {author} {\bibfnamefont {X.}~\bibnamefont
  {Ribeyre}},\ }\bibfield  {title} {\bibinfo {title} {Gravitational influence
  of high power laser pulses},\ }\href
  {https://doi.org/10.1103/PhysRevD.105.104052} {\bibfield  {journal} {\bibinfo
   {journal} {Phys. Rev. D}\ }\textbf {\bibinfo {volume} {105}},\ \bibinfo
  {pages} {104052} (\bibinfo {year} {2022})}\BibitemShut {NoStop}%
\bibitem [{\citenamefont {Ng}\ and\ \citenamefont
  {Ooi}(2013)}]{10.1063/1.4803644}%
  \BibitemOpen
  \bibfield  {author} {\bibinfo {author} {\bibfnamefont {K.~S.}\ \bibnamefont
  {Ng}}\ and\ \bibinfo {author} {\bibfnamefont {C.~H.~R.}\ \bibnamefont
  {Ooi}},\ }\bibfield  {title} {\bibinfo {title} {{Weak gravitational field of
  Bessel beam}},\ }\href {https://doi.org/10.1063/1.4803644} {\bibfield
  {journal} {\bibinfo  {journal} {AIP Conference Proceedings}\ }\textbf
  {\bibinfo {volume} {1528}},\ \bibinfo {pages} {456} (\bibinfo {year}
  {2013})},\ \Eprint
  {https://arxiv.org/abs/https://pubs.aip.org/aip/acp/article-pdf/1528/1/456/12204645/456\_1\_online.pdf}
  {https://pubs.aip.org/aip/acp/article-pdf/1528/1/456/12204645/456\_1\_online.pdf}
  \BibitemShut {NoStop}%
\bibitem [{\citenamefont {Schneiter}\ \emph {et~al.}(2018)\citenamefont
  {Schneiter}, \citenamefont {R{\"a}tzel},\ and\ \citenamefont
  {Braun}}]{Schneiter_2018}%
  \BibitemOpen
  \bibfield  {author} {\bibinfo {author} {\bibfnamefont {F.}~\bibnamefont
  {Schneiter}}, \bibinfo {author} {\bibfnamefont {D.}~\bibnamefont
  {R{\"a}tzel}},\ and\ \bibinfo {author} {\bibfnamefont {D.}~\bibnamefont
  {Braun}},\ }\bibfield  {title} {\bibinfo {title} {The gravitational field of
  a laser beam beyond the short wavelength approximation},\ }\href
  {https://doi.org/10.1088/1361-6382/aadc81} {\bibfield  {journal} {\bibinfo
  {journal} {Classical and Quantum Gravity}\ }\textbf {\bibinfo {volume}
  {35}},\ \bibinfo {pages} {195007} (\bibinfo {year} {2018})}\BibitemShut
  {NoStop}%
\bibitem [{\citenamefont {McGloin}\ and\ \citenamefont
  {Dholakia}(2005)}]{doi:10.1080/0010751042000275259}%
  \BibitemOpen
  \bibfield  {author} {\bibinfo {author} {\bibfnamefont {D.}~\bibnamefont
  {McGloin}}\ and\ \bibinfo {author} {\bibfnamefont {K.}~\bibnamefont
  {Dholakia}},\ }\bibfield  {title} {\bibinfo {title} {Bessel beams:
  Diffraction in a new light},\ }\href
  {https://doi.org/10.1080/0010751042000275259} {\bibfield  {journal} {\bibinfo
   {journal} {Contemporary Physics}\ }\textbf {\bibinfo {volume} {46}},\
  \bibinfo {pages} {15} (\bibinfo {year} {2005})},\ \Eprint
  {https://arxiv.org/abs/https://doi.org/10.1080/0010751042000275259}
  {https://doi.org/10.1080/0010751042000275259} \BibitemShut {NoStop}%
\bibitem [{\citenamefont {Volke-Sepulveda}\ \emph {et~al.}(2002)\citenamefont
  {Volke-Sepulveda}, \citenamefont {Garcés-Chávez}, \citenamefont
  {Chávez-Cerda}, \citenamefont {Arlt},\ and\ \citenamefont
  {Dholakia}}]{VolkeSepulveda_2002}%
  \BibitemOpen
  \bibfield  {author} {\bibinfo {author} {\bibfnamefont {K.}~\bibnamefont
  {Volke-Sepulveda}}, \bibinfo {author} {\bibfnamefont {V.}~\bibnamefont
  {Garcés-Chávez}}, \bibinfo {author} {\bibfnamefont {S.}~\bibnamefont
  {Chávez-Cerda}}, \bibinfo {author} {\bibfnamefont {J.}~\bibnamefont
  {Arlt}},\ and\ \bibinfo {author} {\bibfnamefont {K.}~\bibnamefont
  {Dholakia}},\ }\bibfield  {title} {\bibinfo {title} {Orbital angular momentum
  of a high-order bessel light beam},\ }\href
  {https://doi.org/10.1088/1464-4266/4/2/373} {\bibfield  {journal} {\bibinfo
  {journal} {Journal of Optics B: Quantum and Semiclassical Optics}\ }\textbf
  {\bibinfo {volume} {4}},\ \bibinfo {pages} {S82} (\bibinfo {year}
  {2002})}\BibitemShut {NoStop}%
\bibitem [{\citenamefont {Gomes}\ and\ \citenamefont
  {Rovelli}(2023)}]{Gomes:2023xda}%
  \BibitemOpen
  \bibfield  {author} {\bibinfo {author} {\bibfnamefont {H.}~\bibnamefont
  {Gomes}}\ and\ \bibinfo {author} {\bibfnamefont {C.}~\bibnamefont
  {Rovelli}},\ }\bibfield  {title} {\bibinfo {title} {{On the analogies between
  gravitational and electromagnetic radiative energy}},\ }\href@noop {} {\
  (\bibinfo {year} {2023})},\ \Eprint {https://arxiv.org/abs/2303.14064}
  {arXiv:2303.14064 [physics.hist-ph]} \BibitemShut {NoStop}%
\bibitem [{\citenamefont {Buonanno}\ and\ \citenamefont
  {Sathyaprakash}(2015)}]{buonanno_sathyaprakash_2015}%
  \BibitemOpen
  \bibfield  {author} {\bibinfo {author} {\bibfnamefont {A.}~\bibnamefont
  {Buonanno}}\ and\ \bibinfo {author} {\bibfnamefont {B.~S.}\ \bibnamefont
  {Sathyaprakash}},\ }\bibinfo {title} {Sources of gravitational waves: Theory
  and observations},\ in\ \href {https://doi.org/10.1017/CBO9781139583961.009}
  {\emph {\bibinfo {booktitle} {General Relativity and Gravitation: A
  Centennial Perspective}}},\ \bibinfo {editor} {edited by\ \bibinfo {editor}
  {\bibfnamefont {A.}~\bibnamefont {Ashtekar}}, \bibinfo {editor}
  {\bibfnamefont {B.~K.}\ \bibnamefont {Berger}}, \bibinfo {editor}
  {\bibfnamefont {J.}~\bibnamefont {Isenberg}},\ and\ \bibinfo {editor}
  {\bibfnamefont {M.}~\bibnamefont {MacCallum}}}\ (\bibinfo  {publisher}
  {Cambridge University Press},\ \bibinfo {year} {2015})\ p.\ \bibinfo {pages}
  {287–346}\BibitemShut {NoStop}%
\bibitem [{\citenamefont {Aggarwal}\ \emph {et~al.}(2021)\citenamefont
  {Aggarwal} \emph {et~al.}}]{Aggarwal:2020olq}%
  \BibitemOpen
  \bibfield  {author} {\bibinfo {author} {\bibfnamefont {N.}~\bibnamefont
  {Aggarwal}} \emph {et~al.},\ }\bibfield  {title} {\bibinfo {title}
  {{Challenges and opportunities of gravitational-wave searches at MHz to GHz
  frequencies}},\ }\href {https://doi.org/10.1007/s41114-021-00032-5}
  {\bibfield  {journal} {\bibinfo  {journal} {Living Rev. Rel.}\ }\textbf
  {\bibinfo {volume} {24}},\ \bibinfo {pages} {4} (\bibinfo {year} {2021})},\
  \Eprint {https://arxiv.org/abs/2011.12414} {arXiv:2011.12414 [gr-qc]}
  \BibitemShut {NoStop}%
\bibitem [{\citenamefont {Maggiore}(2007)}]{Maggiore:2007ulw}%
  \BibitemOpen
  \bibfield  {author} {\bibinfo {author} {\bibfnamefont {M.}~\bibnamefont
  {Maggiore}},\ }\href@noop {} {\emph {\bibinfo {title} {{Gravitational Waves.
  Vol. 1: Theory and Experiments}}}},\ Oxford Master Series in Physics\
  (\bibinfo  {publisher} {Oxford University Press},\ \bibinfo {year}
  {2007})\BibitemShut {NoStop}%
\bibitem [{\citenamefont {Eddington}(1922)}]{doi:10.1098/rspa.1922.0085}%
  \BibitemOpen
  \bibfield  {author} {\bibinfo {author} {\bibfnamefont {A.~S.}\ \bibnamefont
  {Eddington}},\ }\bibfield  {title} {\bibinfo {title} {The propagation of
  gravitational waves},\ }\href {https://doi.org/10.1098/rspa.1922.0085}
  {\bibfield  {journal} {\bibinfo  {journal} {Proceedings of the Royal Society
  of London. Series A, Containing Papers of a Mathematical and Physical
  Character}\ }\textbf {\bibinfo {volume} {102}},\ \bibinfo {pages} {268}
  (\bibinfo {year} {1922})}\BibitemShut {NoStop}%
\bibitem [{\citenamefont {Éanna É~Flanagan}\ and\ \citenamefont
  {Hughes}(2005)}]{Flanagan_2005}%
  \BibitemOpen
  \bibfield  {author} {\bibinfo {author} {\bibnamefont {Éanna É~Flanagan}}\
  and\ \bibinfo {author} {\bibfnamefont {S.~A.}\ \bibnamefont {Hughes}},\
  }\bibfield  {title} {\bibinfo {title} {The basics of gravitational wave
  theory},\ }\href {https://doi.org/10.1088/1367-2630/7/1/204} {\bibfield
  {journal} {\bibinfo  {journal} {New Journal of Physics}\ }\textbf {\bibinfo
  {volume} {7}},\ \bibinfo {pages} {204} (\bibinfo {year} {2005})}\BibitemShut
  {NoStop}%
\bibitem [{\citenamefont {Lifshitz}(2017)}]{lifshitz_2017}%
  \BibitemOpen
  \bibfield  {author} {\bibinfo {author} {\bibfnamefont {E.}~\bibnamefont
  {Lifshitz}},\ }\bibfield  {title} {\bibinfo {title} {Republication of: On the
  gravitational stability of the expanding universe},\ }\bibfield  {journal}
  {\bibinfo  {journal} {General Relativity and Gravitation}\ }\textbf {\bibinfo
  {volume} {49}},\ \href {https://doi.org/10.1007/s10714-016-2165-8}
  {10.1007/s10714-016-2165-8} (\bibinfo {year} {2017})\BibitemShut {NoStop}%
\bibitem [{\citenamefont {Bertschinger}(2000)}]{bertschinger2000cosmological}%
  \BibitemOpen
  \bibfield  {author} {\bibinfo {author} {\bibfnamefont {E.}~\bibnamefont
  {Bertschinger}},\ }\href@noop {} {\bibinfo {title} {Cosmological perturbation
  theory and structure formation}} (\bibinfo {year} {2000}),\ \Eprint
  {https://arxiv.org/abs/astro-ph/0101009} {arXiv:astro-ph/0101009 [astro-ph]}
  \BibitemShut {NoStop}%
\bibitem [{\citenamefont {Hobson}\ \emph {et~al.}(2006)\citenamefont {Hobson},
  \citenamefont {Efstathiou},\ and\ \citenamefont {Lasenby}}]{Hobson}%
  \BibitemOpen
  \bibfield  {author} {\bibinfo {author} {\bibfnamefont {M.~P.}\ \bibnamefont
  {Hobson}}, \bibinfo {author} {\bibfnamefont {G.}~\bibnamefont {Efstathiou}},\
  and\ \bibinfo {author} {\bibfnamefont {A.~N.}\ \bibnamefont {Lasenby}},\
  }\href@noop {} {\emph {\bibinfo {title} {{General relativity: An introduction
  for physicists}}}}\ (\bibinfo  {publisher} {Cambridge University Press},\
  \bibinfo {year} {2006})\BibitemShut {NoStop}%
\bibitem [{\citenamefont {Allen}\ and\ \citenamefont {Padgett}(2011)}]{Torres}%
  \BibitemOpen
  \bibfield  {author} {\bibinfo {author} {\bibfnamefont {L.}~\bibnamefont
  {Allen}}\ and\ \bibinfo {author} {\bibfnamefont {M.}~\bibnamefont
  {Padgett}},\ }\bibinfo {title} {The orbital angular momentum of light: An
  introduction},\ in\ \href
  {https://doi.org/https://doi.org/10.1002/9783527635368.ch1} {\emph {\bibinfo
  {booktitle} {Twisted Photons}}}\ (\bibinfo  {publisher} {John Wiley and Sons,
  Ltd},\ \bibinfo {year} {2011})\ Chap.~\bibinfo {chapter} {1}, pp.\ \bibinfo
  {pages} {1--12}\BibitemShut {NoStop}%
\bibitem [{\citenamefont {Allen}\ \emph
  {et~al.}(1992{\natexlab{a}})\citenamefont {Allen}, \citenamefont
  {Beijersbergen}, \citenamefont {Spreeuw},\ and\ \citenamefont
  {Woerdman}}]{PhysRevA.45.8185}%
  \BibitemOpen
  \bibfield  {author} {\bibinfo {author} {\bibfnamefont {L.}~\bibnamefont
  {Allen}}, \bibinfo {author} {\bibfnamefont {M.~W.}\ \bibnamefont
  {Beijersbergen}}, \bibinfo {author} {\bibfnamefont {R.~J.~C.}\ \bibnamefont
  {Spreeuw}},\ and\ \bibinfo {author} {\bibfnamefont {J.~P.}\ \bibnamefont
  {Woerdman}},\ }\bibfield  {title} {\bibinfo {title} {Orbital angular momentum
  of light and the transformation of laguerre-gaussian laser modes},\ }\href
  {https://doi.org/10.1103/PhysRevA.45.8185} {\bibfield  {journal} {\bibinfo
  {journal} {Phys. Rev. A}\ }\textbf {\bibinfo {volume} {45}},\ \bibinfo
  {pages} {8185} (\bibinfo {year} {1992}{\natexlab{a}})}\BibitemShut {NoStop}%
\bibitem [{\citenamefont {Barnett}\ and\ \citenamefont
  {Allen}(1994)}]{BARNETT1994670}%
  \BibitemOpen
  \bibfield  {author} {\bibinfo {author} {\bibfnamefont {S.~M.}\ \bibnamefont
  {Barnett}}\ and\ \bibinfo {author} {\bibfnamefont {L.}~\bibnamefont
  {Allen}},\ }\bibfield  {title} {\bibinfo {title} {Orbital angular momentum
  and nonparaxial light beams},\ }\href
  {https://doi.org/https://doi.org/10.1016/0030-4018(94)90269-0} {\bibfield
  {journal} {\bibinfo  {journal} {Optics Communications}\ }\textbf {\bibinfo
  {volume} {110}},\ \bibinfo {pages} {670} (\bibinfo {year}
  {1994})}\BibitemShut {NoStop}%
\bibitem [{\citenamefont {Durnin}(1987)}]{Durnin1}%
  \BibitemOpen
  \bibfield  {author} {\bibinfo {author} {\bibfnamefont {J.}~\bibnamefont
  {Durnin}},\ }\bibfield  {title} {\bibinfo {title} {Exact solutions for
  nondiffracting beams. i. the scalar theory},\ }\href
  {https://doi.org/10.1364/JOSAA.4.000651} {\bibfield  {journal} {\bibinfo
  {journal} {J. Opt. Soc. Am. A}\ }\textbf {\bibinfo {volume} {4}},\ \bibinfo
  {pages} {651} (\bibinfo {year} {1987})}\BibitemShut {NoStop}%
\bibitem [{\citenamefont {Durnin}\ \emph {et~al.}(1987)\citenamefont {Durnin},
  \citenamefont {Miceli},\ and\ \citenamefont {Eberly}}]{Durnin2}%
  \BibitemOpen
  \bibfield  {author} {\bibinfo {author} {\bibfnamefont {J.}~\bibnamefont
  {Durnin}}, \bibinfo {author} {\bibfnamefont {J.~J.}\ \bibnamefont {Miceli}},\
  and\ \bibinfo {author} {\bibfnamefont {J.~H.}\ \bibnamefont {Eberly}},\
  }\bibfield  {title} {\bibinfo {title} {Diffraction-free beams},\ }\href
  {https://doi.org/10.1103/PhysRevLett.58.1499} {\bibfield  {journal} {\bibinfo
   {journal} {Phys. Rev. Lett.}\ }\textbf {\bibinfo {volume} {58}},\ \bibinfo
  {pages} {1499} (\bibinfo {year} {1987})}\BibitemShut {NoStop}%
\bibitem [{\citenamefont {Quinteiro}\ \emph {et~al.}(2015)\citenamefont
  {Quinteiro}, \citenamefont {Reiter},\ and\ \citenamefont {Kuhn}}]{Quin}%
  \BibitemOpen
  \bibfield  {author} {\bibinfo {author} {\bibfnamefont {G.~F.}\ \bibnamefont
  {Quinteiro}}, \bibinfo {author} {\bibfnamefont {D.~E.}\ \bibnamefont
  {Reiter}},\ and\ \bibinfo {author} {\bibfnamefont {T.}~\bibnamefont {Kuhn}},\
  }\bibfield  {title} {\bibinfo {title} {Formulation of the
  twisted-light--matter interaction at the phase singularity: The twisted-light
  gauge},\ }\href {https://doi.org/10.1103/PhysRevA.91.033808} {\bibfield
  {journal} {\bibinfo  {journal} {Phys. Rev. A}\ }\textbf {\bibinfo {volume}
  {91}},\ \bibinfo {pages} {033808} (\bibinfo {year} {2015})}\BibitemShut
  {NoStop}%
\bibitem [{\citenamefont {Jackson}(1999)}]{Jackson}%
  \BibitemOpen
  \bibfield  {author} {\bibinfo {author} {\bibfnamefont {J.~D.}\ \bibnamefont
  {Jackson}},\ }\href {http://cdsweb.cern.ch/record/490457} {\emph {\bibinfo
  {title} {Classical electrodynamics}}},\ \bibinfo {edition} {3rd}\ ed.\
  (\bibinfo  {publisher} {Wiley},\ \bibinfo {address} {New York, {NY}},\
  \bibinfo {year} {1999})\BibitemShut {NoStop}%
\bibitem [{\citenamefont {Allen}\ \emph
  {et~al.}(1992{\natexlab{b}})\citenamefont {Allen}, \citenamefont
  {Beijersbergen}, \citenamefont {Spreeuw},\ and\ \citenamefont
  {Woerdman}}]{allen_1992}%
  \BibitemOpen
  \bibfield  {author} {\bibinfo {author} {\bibfnamefont {L.}~\bibnamefont
  {Allen}}, \bibinfo {author} {\bibfnamefont {M.~W.}\ \bibnamefont
  {Beijersbergen}}, \bibinfo {author} {\bibfnamefont {R.~J.~C.}\ \bibnamefont
  {Spreeuw}},\ and\ \bibinfo {author} {\bibfnamefont {J.~P.}\ \bibnamefont
  {Woerdman}},\ }\bibfield  {title} {\bibinfo {title} {Orbital angular momentum
  of light and the transformation of laguerre-gaussian laser modes},\ }\href
  {https://doi.org/10.1103/PhysRevA.45.8185} {\bibfield  {journal} {\bibinfo
  {journal} {Phys. Rev. A}\ }\textbf {\bibinfo {volume} {45}},\ \bibinfo
  {pages} {8185} (\bibinfo {year} {1992}{\natexlab{b}})}\BibitemShut {NoStop}%
\bibitem [{\citenamefont {Aboushelbaya}\ \emph {et~al.}(2019)\citenamefont
  {Aboushelbaya}, \citenamefont {Glize}, \citenamefont {Savin}, \citenamefont
  {Mayr}, \citenamefont {Spiers}, \citenamefont {Wang}, \citenamefont
  {Collier}, \citenamefont {Marklund}, \citenamefont {Trines}, \citenamefont
  {Bingham},\ and\ \citenamefont {Norreys}}]{PhysRevLett.123.113604}%
  \BibitemOpen
  \bibfield  {author} {\bibinfo {author} {\bibfnamefont {R.}~\bibnamefont
  {Aboushelbaya}}, \bibinfo {author} {\bibfnamefont {K.}~\bibnamefont {Glize}},
  \bibinfo {author} {\bibfnamefont {A.~F.}\ \bibnamefont {Savin}}, \bibinfo
  {author} {\bibfnamefont {M.}~\bibnamefont {Mayr}}, \bibinfo {author}
  {\bibfnamefont {B.}~\bibnamefont {Spiers}}, \bibinfo {author} {\bibfnamefont
  {R.}~\bibnamefont {Wang}}, \bibinfo {author} {\bibfnamefont {J.}~\bibnamefont
  {Collier}}, \bibinfo {author} {\bibfnamefont {M.}~\bibnamefont {Marklund}},
  \bibinfo {author} {\bibfnamefont {R.~M. G.~M.}\ \bibnamefont {Trines}},
  \bibinfo {author} {\bibfnamefont {R.}~\bibnamefont {Bingham}},\ and\ \bibinfo
  {author} {\bibfnamefont {P.~A.}\ \bibnamefont {Norreys}},\ }\bibfield
  {title} {\bibinfo {title} {Orbital angular momentum coupling in elastic
  photon-photon scattering},\ }\href
  {https://doi.org/10.1103/PhysRevLett.123.113604} {\bibfield  {journal}
  {\bibinfo  {journal} {Phys. Rev. Lett.}\ }\textbf {\bibinfo {volume} {123}},\
  \bibinfo {pages} {113604} (\bibinfo {year} {2019})}\BibitemShut {NoStop}%
\bibitem [{\citenamefont {Denoeud}\ \emph {et~al.}(2017)\citenamefont
  {Denoeud}, \citenamefont {Chopineau}, \citenamefont {Leblanc},\ and\
  \citenamefont {Qu\'er\'e}}]{PhysRevLett.118.033902}%
  \BibitemOpen
  \bibfield  {author} {\bibinfo {author} {\bibfnamefont {A.}~\bibnamefont
  {Denoeud}}, \bibinfo {author} {\bibfnamefont {L.}~\bibnamefont {Chopineau}},
  \bibinfo {author} {\bibfnamefont {A.}~\bibnamefont {Leblanc}},\ and\ \bibinfo
  {author} {\bibfnamefont {F.}~\bibnamefont {Qu\'er\'e}},\ }\bibfield  {title}
  {\bibinfo {title} {Interaction of ultraintense laser vortices with plasma
  mirrors},\ }\href {https://doi.org/10.1103/PhysRevLett.118.033902} {\bibfield
   {journal} {\bibinfo  {journal} {Phys. Rev. Lett.}\ }\textbf {\bibinfo
  {volume} {118}},\ \bibinfo {pages} {033902} (\bibinfo {year}
  {2017})}\BibitemShut {NoStop}%
\bibitem [{\citenamefont {Shi}\ \emph {et~al.}(2022)\citenamefont {Shi},
  \citenamefont {Blackman}, \citenamefont {Zhu},\ and\ \citenamefont
  {Arefiev}}]{shi_blackman_zhu_arefiev_2022}%
  \BibitemOpen
  \bibfield  {author} {\bibinfo {author} {\bibfnamefont {Y.}~\bibnamefont
  {Shi}}, \bibinfo {author} {\bibfnamefont {D.~R.}\ \bibnamefont {Blackman}},
  \bibinfo {author} {\bibfnamefont {P.}~\bibnamefont {Zhu}},\ and\ \bibinfo
  {author} {\bibfnamefont {A.}~\bibnamefont {Arefiev}},\ }\bibfield  {title}
  {\bibinfo {title} {Electron pulse train accelerated by a linearly polarized
  laguerre–gaussian laser beam},\ }\href
  {https://doi.org/10.1017/hpl.2022.37} {\bibfield  {journal} {\bibinfo
  {journal} {High Power Laser Science and Engineering}\ }\textbf {\bibinfo
  {volume} {10}},\ \bibinfo {pages} {e45} (\bibinfo {year} {2022})}\BibitemShut
  {NoStop}%
\bibitem [{\citenamefont {Zhang}\ \emph {et~al.}(2022)\citenamefont {Zhang},
  \citenamefont {Ji},\ and\ \citenamefont {Shen}}]{zhang_ji_shen_2022}%
  \BibitemOpen
  \bibfield  {author} {\bibinfo {author} {\bibfnamefont {L.}~\bibnamefont
  {Zhang}}, \bibinfo {author} {\bibfnamefont {L.}~\bibnamefont {Ji}},\ and\
  \bibinfo {author} {\bibfnamefont {B.}~\bibnamefont {Shen}},\ }\bibfield
  {title} {\bibinfo {title} {Intense harmonic generation driven by a
  relativistic spatiotemporal vortex beam},\ }\href
  {https://doi.org/10.1017/hpl.2022.38} {\bibfield  {journal} {\bibinfo
  {journal} {High Power Laser Science and Engineering}\ }\textbf {\bibinfo
  {volume} {10}},\ \bibinfo {pages} {e46} (\bibinfo {year} {2022})}\BibitemShut
  {NoStop}%
\bibitem [{\citenamefont {Shi}\ \emph {et~al.}(2018)\citenamefont {Shi},
  \citenamefont {Vieira}, \citenamefont {Trines}, \citenamefont {Bingham},
  \citenamefont {Shen},\ and\ \citenamefont
  {Kingham}}]{PhysRevLett.121.145002}%
  \BibitemOpen
  \bibfield  {author} {\bibinfo {author} {\bibfnamefont {Y.}~\bibnamefont
  {Shi}}, \bibinfo {author} {\bibfnamefont {J.}~\bibnamefont {Vieira}},
  \bibinfo {author} {\bibfnamefont {R.~M. G.~M.}\ \bibnamefont {Trines}},
  \bibinfo {author} {\bibfnamefont {R.}~\bibnamefont {Bingham}}, \bibinfo
  {author} {\bibfnamefont {B.~F.}\ \bibnamefont {Shen}},\ and\ \bibinfo
  {author} {\bibfnamefont {R.~J.}\ \bibnamefont {Kingham}},\ }\bibfield
  {title} {\bibinfo {title} {Magnetic field generation in plasma waves driven
  by copropagating intense twisted lasers},\ }\href
  {https://doi.org/10.1103/PhysRevLett.121.145002} {\bibfield  {journal}
  {\bibinfo  {journal} {Phys. Rev. Lett.}\ }\textbf {\bibinfo {volume} {121}},\
  \bibinfo {pages} {145002} (\bibinfo {year} {2018})}\BibitemShut {NoStop}%
\bibitem [{\citenamefont {Mirhosseini}\ \emph {et~al.}(2013)\citenamefont
  {Mirhosseini}, \citenamefont {Malik}, \citenamefont {Shi},\ and\
  \citenamefont {Boyd}}]{Mirhosseini}%
  \BibitemOpen
  \bibfield  {author} {\bibinfo {author} {\bibfnamefont {M.}~\bibnamefont
  {Mirhosseini}}, \bibinfo {author} {\bibfnamefont {M.}~\bibnamefont {Malik}},
  \bibinfo {author} {\bibfnamefont {Z.}~\bibnamefont {Shi}},\ and\ \bibinfo
  {author} {\bibfnamefont {R.~W.}\ \bibnamefont {Boyd}},\ }\bibfield  {title}
  {\bibinfo {title} {Efficient separation of the orbital angular momentum
  eigenstates of light},\ }\href {https://doi.org/10.1038/ncomms3781}
  {\bibfield  {journal} {\bibinfo  {journal} {Nature Communications}\ }\textbf
  {\bibinfo {volume} {4}},\ \bibinfo {pages} {2781} (\bibinfo {year}
  {2013})}\BibitemShut {NoStop}%
\bibitem [{\citenamefont {Krenn}\ \emph {et~al.}(2014)\citenamefont {Krenn},
  \citenamefont {Fickler}, \citenamefont {Fink}, \citenamefont {Handsteiner},
  \citenamefont {Malik}, \citenamefont {Scheidl}, \citenamefont {Ursin},\ and\
  \citenamefont {Zeilinger}}]{Krenn_2014}%
  \BibitemOpen
  \bibfield  {author} {\bibinfo {author} {\bibfnamefont {M.}~\bibnamefont
  {Krenn}}, \bibinfo {author} {\bibfnamefont {R.}~\bibnamefont {Fickler}},
  \bibinfo {author} {\bibfnamefont {M.}~\bibnamefont {Fink}}, \bibinfo {author}
  {\bibfnamefont {J.}~\bibnamefont {Handsteiner}}, \bibinfo {author}
  {\bibfnamefont {M.}~\bibnamefont {Malik}}, \bibinfo {author} {\bibfnamefont
  {T.}~\bibnamefont {Scheidl}}, \bibinfo {author} {\bibfnamefont
  {R.}~\bibnamefont {Ursin}},\ and\ \bibinfo {author} {\bibfnamefont
  {A.}~\bibnamefont {Zeilinger}},\ }\bibfield  {title} {\bibinfo {title}
  {Communication with spatially modulated light through turbulent air across
  vienna},\ }\href {https://doi.org/10.1088/1367-2630/16/11/113028} {\bibfield
  {journal} {\bibinfo  {journal} {New Journal of Physics}\ }\textbf {\bibinfo
  {volume} {16}},\ \bibinfo {pages} {113028} (\bibinfo {year}
  {2014})}\BibitemShut {NoStop}%
\bibitem [{\citenamefont {Lei}\ \emph {et~al.}(2017)\citenamefont {Lei},
  \citenamefont {Gao}, \citenamefont {Li}, \citenamefont {Yuan}, \citenamefont
  {Li}, \citenamefont {Zhang}, \citenamefont {Liu}, \citenamefont {Xu},
  \citenamefont {Tian},\ and\ \citenamefont {Yuan}}]{7835197}%
  \BibitemOpen
  \bibfield  {author} {\bibinfo {author} {\bibfnamefont {T.}~\bibnamefont
  {Lei}}, \bibinfo {author} {\bibfnamefont {S.}~\bibnamefont {Gao}}, \bibinfo
  {author} {\bibfnamefont {Z.}~\bibnamefont {Li}}, \bibinfo {author}
  {\bibfnamefont {Y.}~\bibnamefont {Yuan}}, \bibinfo {author} {\bibfnamefont
  {Y.}~\bibnamefont {Li}}, \bibinfo {author} {\bibfnamefont {M.}~\bibnamefont
  {Zhang}}, \bibinfo {author} {\bibfnamefont {G.~N.}\ \bibnamefont {Liu}},
  \bibinfo {author} {\bibfnamefont {X.}~\bibnamefont {Xu}}, \bibinfo {author}
  {\bibfnamefont {J.}~\bibnamefont {Tian}},\ and\ \bibinfo {author}
  {\bibfnamefont {X.}~\bibnamefont {Yuan}},\ }\bibfield  {title} {\bibinfo
  {title} {Fast-switchable oam-based high capacity density optical router},\
  }\href {https://doi.org/10.1109/JPHOT.2017.2659639} {\bibfield  {journal}
  {\bibinfo  {journal} {IEEE Photonics Journal}\ }\textbf {\bibinfo {volume}
  {9}},\ \bibinfo {pages} {1} (\bibinfo {year} {2017})}\BibitemShut {NoStop}%
\bibitem [{\citenamefont {Khonina}\ \emph {et~al.}(2021)\citenamefont
  {Khonina}, \citenamefont {Kazanskiy}, \citenamefont {Khorin},\ and\
  \citenamefont {Butt}}]{s21196690}%
  \BibitemOpen
  \bibfield  {author} {\bibinfo {author} {\bibfnamefont {S.~N.}\ \bibnamefont
  {Khonina}}, \bibinfo {author} {\bibfnamefont {N.~L.}\ \bibnamefont
  {Kazanskiy}}, \bibinfo {author} {\bibfnamefont {P.~A.}\ \bibnamefont
  {Khorin}},\ and\ \bibinfo {author} {\bibfnamefont {M.~A.}\ \bibnamefont
  {Butt}},\ }\bibfield  {title} {\bibinfo {title} {Modern types of axicons: New
  functions and applications},\ }\bibfield  {journal} {\bibinfo  {journal}
  {Sensors}\ }\textbf {\bibinfo {volume} {21}},\ \href
  {https://doi.org/10.3390/s21196690} {10.3390/s21196690} (\bibinfo {year}
  {2021})\BibitemShut {NoStop}%
\bibitem [{\citenamefont {Chen}\ \emph {et~al.}(2020)\citenamefont {Chen},
  \citenamefont {Zhou}, \citenamefont {Moretti}, \citenamefont {Wang},\ and\
  \citenamefont {Li}}]{8894467}%
  \BibitemOpen
  \bibfield  {author} {\bibinfo {author} {\bibfnamefont {R.}~\bibnamefont
  {Chen}}, \bibinfo {author} {\bibfnamefont {H.}~\bibnamefont {Zhou}}, \bibinfo
  {author} {\bibfnamefont {M.}~\bibnamefont {Moretti}}, \bibinfo {author}
  {\bibfnamefont {X.}~\bibnamefont {Wang}},\ and\ \bibinfo {author}
  {\bibfnamefont {J.}~\bibnamefont {Li}},\ }\bibfield  {title} {\bibinfo
  {title} {Orbital angular momentum waves: Generation, detection, and emerging
  applications},\ }\href {https://doi.org/10.1109/COMST.2019.2952453}
  {\bibfield  {journal} {\bibinfo  {journal} {IEEE Communications Surveys \&
  Tutorials}\ }\textbf {\bibinfo {volume} {22}},\ \bibinfo {pages} {840}
  (\bibinfo {year} {2020})}\BibitemShut {NoStop}%
\bibitem [{\citenamefont {Wei}\ \emph {et~al.}(2015)\citenamefont {Wei},
  \citenamefont {Liu}, \citenamefont {Niu}, \citenamefont {Zhang},
  \citenamefont {Wang}, \citenamefont {Yang},\ and\ \citenamefont
  {Liu}}]{Wei:15}%
  \BibitemOpen
  \bibfield  {author} {\bibinfo {author} {\bibfnamefont {X.}~\bibnamefont
  {Wei}}, \bibinfo {author} {\bibfnamefont {C.}~\bibnamefont {Liu}}, \bibinfo
  {author} {\bibfnamefont {L.}~\bibnamefont {Niu}}, \bibinfo {author}
  {\bibfnamefont {Z.}~\bibnamefont {Zhang}}, \bibinfo {author} {\bibfnamefont
  {K.}~\bibnamefont {Wang}}, \bibinfo {author} {\bibfnamefont {Z.}~\bibnamefont
  {Yang}},\ and\ \bibinfo {author} {\bibfnamefont {J.}~\bibnamefont {Liu}},\
  }\bibfield  {title} {\bibinfo {title} {Generation of arbitrary order bessel
  beams via 3d printed axicons at the terahertz frequency range},\ }\href
  {https://doi.org/10.1364/AO.54.010641} {\bibfield  {journal} {\bibinfo
  {journal} {Appl. Opt.}\ }\textbf {\bibinfo {volume} {54}},\ \bibinfo {pages}
  {10641} (\bibinfo {year} {2015})}\BibitemShut {NoStop}%
\bibitem [{\citenamefont {Boucher}\ \emph {et~al.}(2018)\citenamefont
  {Boucher}, \citenamefont {Hoyo}, \citenamefont {Billet}, \citenamefont
  {Pinel}, \citenamefont {Labroille},\ and\ \citenamefont
  {Courvoisier}}]{Boucher:18}%
  \BibitemOpen
  \bibfield  {author} {\bibinfo {author} {\bibfnamefont {P.}~\bibnamefont
  {Boucher}}, \bibinfo {author} {\bibfnamefont {J.~D.}\ \bibnamefont {Hoyo}},
  \bibinfo {author} {\bibfnamefont {C.}~\bibnamefont {Billet}}, \bibinfo
  {author} {\bibfnamefont {O.}~\bibnamefont {Pinel}}, \bibinfo {author}
  {\bibfnamefont {G.}~\bibnamefont {Labroille}},\ and\ \bibinfo {author}
  {\bibfnamefont {F.}~\bibnamefont {Courvoisier}},\ }\bibfield  {title}
  {\bibinfo {title} {Generation of high conical angle Bessel–Gauss beams with reflective axicons},\ }\href
  {https://doi.org/10.1364/AO.57.006725} {\bibfield  {journal} {\bibinfo
  {journal} {Appl. Opt.}\ }\textbf {\bibinfo {volume} {57}},\ \bibinfo {pages}
  {6725} (\bibinfo {year} {2018})}\BibitemShut {NoStop}%
\bibitem [{\citenamefont {McLeod}(1954)}]{McLeod:54}%
  \BibitemOpen
  \bibfield  {author} {\bibinfo {author} {\bibfnamefont {J.~H.}\ \bibnamefont
  {McLeod}},\ }\bibfield  {title} {\bibinfo {title} {The axicon: A new type of
  optical element},\ }\href {https://doi.org/10.1364/JOSA.44.000592} {\bibfield
   {journal} {\bibinfo  {journal} {J. Opt. Soc. Am.}\ }\textbf {\bibinfo
  {volume} {44}},\ \bibinfo {pages} {592} (\bibinfo {year} {1954})}\BibitemShut
  {NoStop}%
\bibitem [{\citenamefont {Durnin}\ \emph {et~al.}(1988)\citenamefont {Durnin},
  \citenamefont {Miceli},\ and\ \citenamefont {Eberly}}]{Durnin:88}%
  \BibitemOpen
  \bibfield  {author} {\bibinfo {author} {\bibfnamefont {J.}~\bibnamefont
  {Durnin}}, \bibinfo {author} {\bibfnamefont {J.~J.}\ \bibnamefont {Miceli}},\
  and\ \bibinfo {author} {\bibfnamefont {J.~H.}\ \bibnamefont {Eberly}},\
  }\bibfield  {title} {\bibinfo {title} {Comparison of bessel and gaussian
  beams},\ }\href {https://doi.org/10.1364/OL.13.000079} {\bibfield  {journal}
  {\bibinfo  {journal} {Opt. Lett.}\ }\textbf {\bibinfo {volume} {13}},\
  \bibinfo {pages} {79} (\bibinfo {year} {1988})}\BibitemShut {NoStop}%
\bibitem [{\citenamefont {Nowack}(2012)}]{Nowack}%
  \BibitemOpen
  \bibfield  {author} {\bibinfo {author} {\bibfnamefont {Robert L.}\ \bibnamefont {Nowack}},\ }\bibfield  {title} {\bibinfo {title} {A tale of two beams: an elementary overview of Gaussian beams and Bessel beams},\ }\href {https://doi.org/10.1007/s11200-011-0002-1} {\bibfield  {journal} {\bibinfo  {journal} {Studia Geophysica et Geodaetica}\ }\textbf {\bibinfo {volume} {56}},\ \bibinfo {pages} {355} (\bibinfo {year} {2012})}\BibitemShut {NoStop}%
\bibitem [{\citenamefont {McDonald}(2000)}]{mcdonald2000bessel}%
  \BibitemOpen
  \bibfield  {author} {\bibinfo {author} {\bibfnamefont {Kirk T.}\ \bibnamefont {McDonald}},\ }\bibfield  {title} {\bibinfo {title} {Bessel Beams},\ }\href {https://arxiv.org/abs/physics/0006046} {\bibfield  {journal} {\bibinfo  {journal} {arXiv:physics.optics}\ } (\bibinfo {year} {2000})}\BibitemShut {NoStop}%
\bibitem [{\citenamefont {Brooker}(2003)}]{Brooker}%
  \BibitemOpen
  \bibfield  {author} {\bibinfo {author} {\bibfnamefont {G.}~\bibnamefont
  {Brooker}},\ }\href@noop {} {\emph {\bibinfo {title} {{Modern classical
  optics}}}},\ Vol.~\bibinfo {volume} {8}\ (\bibinfo  {publisher} {Oxford
  University Press},\ \bibinfo {year} {2003})\BibitemShut {NoStop}%
\bibitem [{\citenamefont {Tiwari}\ \emph {et~al.}(2011)\citenamefont {Tiwari},
  \citenamefont {Mishra},\ and\ \citenamefont {Ram}}]{10.1117/1.3530080}%
  \BibitemOpen
  \bibfield  {author} {\bibinfo {author} {\bibfnamefont {S.}~\bibnamefont
  {Tiwari}}, \bibinfo {author} {\bibfnamefont {S.~R.}\ \bibnamefont {Mishra}},\
  and\ \bibinfo {author} {\bibfnamefont {S.~P.}\ \bibnamefont {Ram}},\
  }\bibfield  {title} {\bibinfo {title} {{Generation of a variable-diameter
  collimated hollow laser beam using metal axicon mirrors}},\ }\href
  {https://doi.org/10.1117/1.3530080} {\bibfield  {journal} {\bibinfo
  {journal} {Optical Engineering}\ }\textbf {\bibinfo {volume} {50}},\ \bibinfo
  {pages} {014001} (\bibinfo {year} {2011})}\BibitemShut {NoStop}%
\bibitem [{\citenamefont {Hou}\ \emph {et~al.}(2018)\citenamefont {Hou},
  \citenamefont {Gong},\ and\ \citenamefont {Liu}}]{Hordeski}%
  \BibitemOpen
  \bibfield  {author} {\bibinfo {author} {\bibfnamefont {S.}~\bibnamefont
  {Hou}}, \bibinfo {author} {\bibfnamefont {Y.}~\bibnamefont {Gong}},\ and\
  \bibinfo {author} {\bibfnamefont {Y.}~\bibnamefont {Liu}},\ }\bibfield
  {title} {\bibinfo {title} {Polarizations of gravitational waves in horndeski
  theory},\ }\href {https://doi.org/10.1140/epjc/s10052-018-5869-y} {\bibfield
  {journal} {\bibinfo  {journal} {The European Physical Journal C}\ }\textbf
  {\bibinfo {volume} {78}},\ \bibinfo {pages} {378} (\bibinfo {year}
  {2018})}\BibitemShut {NoStop}%
\bibitem [{\citenamefont {Tahura}\ \emph {et~al.}(2021)\citenamefont {Tahura},
  \citenamefont {Nichols}, \citenamefont {Saffer}, \citenamefont {Stein},\ and\
  \citenamefont {Yagi}}]{PhysRevD.103.104026}%
  \BibitemOpen
  \bibfield  {author} {\bibinfo {author} {\bibfnamefont {S.}~\bibnamefont
  {Tahura}}, \bibinfo {author} {\bibfnamefont {D.~A.}\ \bibnamefont {Nichols}},
  \bibinfo {author} {\bibfnamefont {A.}~\bibnamefont {Saffer}}, \bibinfo
  {author} {\bibfnamefont {L.~C.}\ \bibnamefont {Stein}},\ and\ \bibinfo
  {author} {\bibfnamefont {K.}~\bibnamefont {Yagi}},\ }\bibfield  {title}
  {\bibinfo {title} {Brans-dicke theory in bondi-sachs form: Asymptotically
  flat solutions, asymptotic symmetries, and gravitational-wave memory
  effects},\ }\href {https://doi.org/10.1103/PhysRevD.103.104026} {\bibfield
  {journal} {\bibinfo  {journal} {Phys. Rev. D}\ }\textbf {\bibinfo {volume}
  {103}},\ \bibinfo {pages} {104026} (\bibinfo {year} {2021})}\BibitemShut
  {NoStop}%
\bibitem [{\citenamefont {Tajima}\ and\ \citenamefont {Mourou}(2002)}]{LMJ}%
  \BibitemOpen
  \bibfield  {author} {\bibinfo {author} {\bibfnamefont {T.}~\bibnamefont
  {Tajima}}\ and\ \bibinfo {author} {\bibfnamefont {G.}~\bibnamefont
  {Mourou}},\ }\bibfield  {title} {\bibinfo {title} {Zettawatt-exawatt lasers
  and their applications in ultrastrong-field physics},\ }\href
  {https://doi.org/10.1103/PhysRevSTAB.5.031301} {\bibfield  {journal}
  {\bibinfo  {journal} {Phys. Rev. ST Accel. Beams}\ }\textbf {\bibinfo
  {volume} {5}},\ \bibinfo {pages} {031301} (\bibinfo {year}
  {2002})}\BibitemShut {NoStop}%
\bibitem [{\citenamefont {Andre}(1999)}]{France}%
  \BibitemOpen
  \bibfield  {author} {\bibinfo {author} {\bibfnamefont {M.}~\bibnamefont
  {Andre}},\ }\bibfield  {title} {\bibinfo {title} {{The French megajoule laser
  project (LMJ)}},\ }\href@noop {} {\bibfield  {journal} {\bibinfo  {journal}
  {Fusion Engineering and Design}\ ,\ \bibinfo {pages} {43}} (\bibinfo {year}
  {1999})}\BibitemShut {NoStop}%
\bibitem [{NIF(2017)}]{NIF1}%
  \BibitemOpen
  \href {https://doi.org/10.1117/12.2257571} {\emph {\bibinfo {title}
  {Proc.SPIE}}},\ Vol.\ \bibinfo {volume} {10084}\ (\bibinfo {year}
  {2017})\BibitemShut {NoStop}%
\bibitem [{\citenamefont {Spaeth}\ \emph {et~al.}(2016)\citenamefont {Spaeth},
  \citenamefont {Manes}, \citenamefont {Kalantar}, \citenamefont {Miller},
  \citenamefont {Heebner}, \citenamefont {Bliss}, \citenamefont {Spec},
  \citenamefont {Parham}, \citenamefont {Whitman}, \citenamefont {Wegner},
  \citenamefont {Baisden}, \citenamefont {Menapace}, \citenamefont {Bowers},
  \citenamefont {Cohen}, \citenamefont {Suratwala}, \citenamefont {Di~Nicola},
  \citenamefont {Newton}, \citenamefont {Adams}, \citenamefont {Trenholme},
  \citenamefont {Finucane}, \citenamefont {Bonanno}, \citenamefont {Rardin},
  \citenamefont {Arnold}, \citenamefont {Dixit}, \citenamefont {Erbert},
  \citenamefont {Erlandson}, \citenamefont {Fair}, \citenamefont {Feigenbaum},
  \citenamefont {Gourdin}, \citenamefont {Hawley}, \citenamefont {Honig},
  \citenamefont {House}, \citenamefont {Jancaitis}, \citenamefont {LaFortune},
  \citenamefont {Larson}, \citenamefont {Le~Galloudec}, \citenamefont {Lindl},
  \citenamefont {MacGowan}, \citenamefont {Marshall}, \citenamefont
  {McCandless}, \citenamefont {McCracken}, \citenamefont {Montesanti},
  \citenamefont {Moses}, \citenamefont {Nostrand}, \citenamefont {Pryatel},
  \citenamefont {Roberts}, \citenamefont {Rodriguez}, \citenamefont {Rowe},
  \citenamefont {Sacks}, \citenamefont {Salmon}, \citenamefont {Shaw},
  \citenamefont {Sommer}, \citenamefont {Stolz}, \citenamefont {Tietbohl},
  \citenamefont {Widmayer},\ and\ \citenamefont {Zacharias}}]{NIF2}%
  \BibitemOpen
  \bibfield  {author} {\bibinfo {author} {\bibfnamefont {M.~L.}\ \bibnamefont
  {Spaeth}}, \bibinfo {author} {\bibfnamefont {K.~R.}\ \bibnamefont {Manes}},
  \bibinfo {author} {\bibfnamefont {D.~H.}\ \bibnamefont {Kalantar}}, \bibinfo
  {author} {\bibfnamefont {P.~E.}\ \bibnamefont {Miller}}, \bibinfo {author}
  {\bibfnamefont {J.~E.}\ \bibnamefont {Heebner}}, \bibinfo {author}
  {\bibfnamefont {E.~S.}\ \bibnamefont {Bliss}}, \bibinfo {author}
  {\bibfnamefont {D.~R.}\ \bibnamefont {Spec}}, \bibinfo {author}
  {\bibfnamefont {T.~G.}\ \bibnamefont {Parham}}, \bibinfo {author}
  {\bibfnamefont {P.~K.}\ \bibnamefont {Whitman}}, \bibinfo {author}
  {\bibfnamefont {P.~J.}\ \bibnamefont {Wegner}}, \bibinfo {author}
  {\bibfnamefont {P.~A.}\ \bibnamefont {Baisden}}, \bibinfo {author}
  {\bibfnamefont {J.~A.}\ \bibnamefont {Menapace}}, \bibinfo {author}
  {\bibfnamefont {M.~W.}\ \bibnamefont {Bowers}}, \bibinfo {author}
  {\bibfnamefont {S.~J.}\ \bibnamefont {Cohen}}, \bibinfo {author}
  {\bibfnamefont {T.~I.}\ \bibnamefont {Suratwala}}, \bibinfo {author}
  {\bibfnamefont {J.~M.}\ \bibnamefont {Di~Nicola}}, \bibinfo {author}
  {\bibfnamefont {M.~A.}\ \bibnamefont {Newton}}, \bibinfo {author}
  {\bibfnamefont {J.~J.}\ \bibnamefont {Adams}}, \bibinfo {author}
  {\bibfnamefont {J.~B.}\ \bibnamefont {Trenholme}}, \bibinfo {author}
  {\bibfnamefont {R.~G.}\ \bibnamefont {Finucane}}, \bibinfo {author}
  {\bibfnamefont {R.~E.}\ \bibnamefont {Bonanno}}, \bibinfo {author}
  {\bibfnamefont {D.~C.}\ \bibnamefont {Rardin}}, \bibinfo {author}
  {\bibfnamefont {P.~A.}\ \bibnamefont {Arnold}}, \bibinfo {author}
  {\bibfnamefont {S.~N.}\ \bibnamefont {Dixit}}, \bibinfo {author}
  {\bibfnamefont {G.~V.}\ \bibnamefont {Erbert}}, \bibinfo {author}
  {\bibfnamefont {A.~C.}\ \bibnamefont {Erlandson}}, \bibinfo {author}
  {\bibfnamefont {J.~E.}\ \bibnamefont {Fair}}, \bibinfo {author}
  {\bibfnamefont {E.}~\bibnamefont {Feigenbaum}}, \bibinfo {author}
  {\bibfnamefont {W.~H.}\ \bibnamefont {Gourdin}}, \bibinfo {author}
  {\bibfnamefont {R.~A.}\ \bibnamefont {Hawley}}, \bibinfo {author}
  {\bibfnamefont {J.}~\bibnamefont {Honig}}, \bibinfo {author} {\bibfnamefont
  {R.~K.}\ \bibnamefont {House}}, \bibinfo {author} {\bibfnamefont {K.~S.}\
  \bibnamefont {Jancaitis}}, \bibinfo {author} {\bibfnamefont {K.~N.}\
  \bibnamefont {LaFortune}}, \bibinfo {author} {\bibfnamefont {D.~W.}\
  \bibnamefont {Larson}}, \bibinfo {author} {\bibfnamefont {B.~J.}\
  \bibnamefont {Le~Galloudec}}, \bibinfo {author} {\bibfnamefont {J.~D.}\
  \bibnamefont {Lindl}}, \bibinfo {author} {\bibfnamefont {B.~J.}\ \bibnamefont
  {MacGowan}}, \bibinfo {author} {\bibfnamefont {C.~D.}\ \bibnamefont
  {Marshall}}, \bibinfo {author} {\bibfnamefont {K.~P.}\ \bibnamefont
  {McCandless}}, \bibinfo {author} {\bibfnamefont {R.~W.}\ \bibnamefont
  {McCracken}}, \bibinfo {author} {\bibfnamefont {R.~C.}\ \bibnamefont
  {Montesanti}}, \bibinfo {author} {\bibfnamefont {E.~I.}\ \bibnamefont
  {Moses}}, \bibinfo {author} {\bibfnamefont {M.~C.}\ \bibnamefont {Nostrand}},
  \bibinfo {author} {\bibfnamefont {J.~A.}\ \bibnamefont {Pryatel}}, \bibinfo
  {author} {\bibfnamefont {V.~S.}\ \bibnamefont {Roberts}}, \bibinfo {author}
  {\bibfnamefont {S.~B.}\ \bibnamefont {Rodriguez}}, \bibinfo {author}
  {\bibfnamefont {A.~W.}\ \bibnamefont {Rowe}}, \bibinfo {author}
  {\bibfnamefont {R.~A.}\ \bibnamefont {Sacks}}, \bibinfo {author}
  {\bibfnamefont {J.~T.}\ \bibnamefont {Salmon}}, \bibinfo {author}
  {\bibfnamefont {M.~J.}\ \bibnamefont {Shaw}}, \bibinfo {author}
  {\bibfnamefont {S.}~\bibnamefont {Sommer}}, \bibinfo {author} {\bibfnamefont
  {C.~J.}\ \bibnamefont {Stolz}}, \bibinfo {author} {\bibfnamefont {G.~L.}\
  \bibnamefont {Tietbohl}}, \bibinfo {author} {\bibfnamefont {C.~C.}\
  \bibnamefont {Widmayer}},\ and\ \bibinfo {author} {\bibfnamefont
  {R.}~\bibnamefont {Zacharias}},\ }\bibfield  {title} {\bibinfo {title}
  {Description of the nif laser},\ }\href {https://doi.org/10.13182/FST15-144}
  {\bibfield  {journal} {\bibinfo  {journal} {Fusion Science and Technology}\
  }\textbf {\bibinfo {volume} {69}},\ \bibinfo {pages} {25} (\bibinfo {year}
  {2016})}\BibitemShut {NoStop}%
\bibitem [{\citenamefont {Li}\ \emph {et~al.}(2023)\citenamefont {Li},
  \citenamefont {Leng},\ and\ \citenamefont
  {Li}}]{https://doi.org/10.1002/lpor.202100705}%
  \BibitemOpen
  \bibfield  {author} {\bibinfo {author} {\bibfnamefont {Z.}~\bibnamefont
  {Li}}, \bibinfo {author} {\bibfnamefont {Y.}~\bibnamefont {Leng}},\ and\
  \bibinfo {author} {\bibfnamefont {R.}~\bibnamefont {Li}},\ }\bibfield
  {title} {\bibinfo {title} {Further development of the short-pulse petawatt
  laser: Trends, technologies, and bottlenecks},\ }\href
  {https://doi.org/https://doi.org/10.1002/lpor.202100705} {\bibfield
  {journal} {\bibinfo  {journal} {Laser \& Photonics Reviews}\ }\textbf
  {\bibinfo {volume} {17}},\ \bibinfo {pages} {2100705} (\bibinfo {year}
  {2023})}\BibitemShut {NoStop}%
\bibitem [{\citenamefont {Norreys~et~al.}(2021)}]{doi:10.1098/rsta.2020.0005}%
  \BibitemOpen
  \bibfield  {author} {\bibinfo {author} {\bibnamefont {Norreys, P. A.}}\ and
  \bibinfo {author} {\bibnamefont {Ceurvorst, L.}}\ and \bibinfo {author}
  {\bibnamefont {Sadler, J. D.}}\ and \bibinfo {author} {\bibnamefont {Spiers,
  B. T.}}\ and \bibinfo {author} {\bibnamefont {Aboushelbaya, R.}}\ and
  \bibinfo {author} {\bibnamefont {Mayr, M. W.}}\ and \bibinfo {author}
  {\bibnamefont {Paddock, R.}}\ and \bibinfo {author} {\bibnamefont {Ratan,
  N.}}\ and \bibinfo {author} {\bibnamefont {Savin, A. F.}}\ and \bibinfo
  {author} {\bibnamefont {Wang, R. H. W.}}\ and \bibinfo {author} {\bibnamefont
  {Glize, K.}}\ and \bibinfo {author} {\bibnamefont {Trines, R. M. G. M.}}\ and
  \bibinfo {author} {\bibnamefont {Bingham, R.}}\ and \bibinfo {author}
  {\bibnamefont {Hill, M. P.}}\ and \bibinfo {author} {\bibnamefont {Sircombe,
  N.}}\ and \bibinfo {author} {\bibnamefont {Ramsay, M.}}\ and \bibinfo
  {author} {\bibnamefont {Allan, P.}}\ and \bibinfo {author} {\bibnamefont
  {Hobbs, L.}}\ and \bibinfo {author} {\bibnamefont {James, S.}}\ and \bibinfo
  {author} {\bibnamefont {Skidmore, J.}}\ and \bibinfo {author} {\bibnamefont
  {Fyrth, J.}}\ and \bibinfo {author} {\bibnamefont {Luis, J.}}\ and \bibinfo
  {author} {\bibnamefont {Floyd, E.}}\ and \bibinfo {author} {\bibnamefont
  {Brown, C.}}\ and \bibinfo {author} {\bibnamefont {Haines, B. M.}}\ and
  \bibinfo {author} {\bibnamefont {Olson, R. E.}}\ and \bibinfo {author}
  {\bibnamefont {Yi, S. A.}}\ and \bibinfo {author} {\bibnamefont {Zylstra, A.
  B.}}\ and \bibinfo {author} {\bibnamefont {Flippo, K.}}\ and \bibinfo {author}
  {\bibnamefont {Bradley, P. A.}}\ and \bibinfo {author} {\bibnamefont
  {Peterson, R. R.}}\ and \bibinfo {author} {\bibnamefont {Kline, J. L.}}\ and
  \bibinfo {author} {\bibnamefont {Leeper, R. J.}}}\bibfield  {title} {\bibinfo
  {title} {Preparations for a European R\&D roadmap for an inertial fusion demo
  reactor},\ }\href {https://doi.org/10.1098/rsta.2020.0005} {\bibfield
  {journal} {\bibinfo  {journal} {Philosophical Transactions of the Royal
  Society A: Mathematical, Physical and Engineering Sciences}\ }\textbf
  {\bibinfo {volume} {379}},\ \bibinfo {pages} {20200005} (\bibinfo {year}
  {2021})}\BibitemShut {NoStop}%
\bibitem [{\citenamefont {Kirkwood}\ \emph {et~al.}(2018)\citenamefont
  {Kirkwood}, \citenamefont {Turnbull}, \citenamefont {Chapman}, \citenamefont
  {Wilks}, \citenamefont {Rosen}, \citenamefont {London}, \citenamefont
  {Pickworth}, \citenamefont {Colaitis}, \citenamefont {Dunlop}, \citenamefont
  {Poole}, \citenamefont {Moody}, \citenamefont {Strozzi}, \citenamefont
  {Michel}, \citenamefont {Divol}, \citenamefont {Landen}, \citenamefont
  {MacGowan}, \citenamefont {Van~Wonterghem}, \citenamefont {Fournier},\ and\
  \citenamefont {Blue}}]{10.1063/1.5016310}%
  \BibitemOpen
  \bibfield  {author} {\bibinfo {author} {\bibfnamefont {R.~K.}\ \bibnamefont
  {Kirkwood}}, \bibinfo {author} {\bibfnamefont {D.~P.}\ \bibnamefont
  {Turnbull}}, \bibinfo {author} {\bibfnamefont {T.}~\bibnamefont {Chapman}},
  \bibinfo {author} {\bibfnamefont {S.~C.}\ \bibnamefont {Wilks}}, \bibinfo
  {author} {\bibfnamefont {M.~D.}\ \bibnamefont {Rosen}}, \bibinfo {author}
  {\bibfnamefont {R.~A.}\ \bibnamefont {London}}, \bibinfo {author}
  {\bibfnamefont {L.~A.}\ \bibnamefont {Pickworth}}, \bibinfo {author}
  {\bibfnamefont {A.}~\bibnamefont {Colaitis}}, \bibinfo {author}
  {\bibfnamefont {W.~H.}\ \bibnamefont {Dunlop}}, \bibinfo {author}
  {\bibfnamefont {P.}~\bibnamefont {Poole}}, \bibinfo {author} {\bibfnamefont
  {J.~D.}\ \bibnamefont {Moody}}, \bibinfo {author} {\bibfnamefont {D.~J.}\
  \bibnamefont {Strozzi}}, \bibinfo {author} {\bibfnamefont {P.~A.}\
  \bibnamefont {Michel}}, \bibinfo {author} {\bibfnamefont {L.}~\bibnamefont
  {Divol}}, \bibinfo {author} {\bibfnamefont {O.~L.}\ \bibnamefont {Landen}},
  \bibinfo {author} {\bibfnamefont {B.~J.}\ \bibnamefont {MacGowan}}, \bibinfo
  {author} {\bibfnamefont {B.~M.}\ \bibnamefont {Van~Wonterghem}}, \bibinfo
  {author} {\bibfnamefont {K.~B.}\ \bibnamefont {Fournier}},\ and\ \bibinfo
  {author} {\bibfnamefont {B.~E.}\ \bibnamefont {Blue}},\ }\bibfield  {title}
  {\bibinfo {title} {{A plasma amplifier to combine multiple beams at NIF}},\
  }\bibfield  {journal} {\bibinfo  {journal} {Physics of Plasmas}\ }\textbf
  {\bibinfo {volume} {25}},\ \href {https://doi.org/10.1063/1.5016310}
  {10.1063/1.5016310} (\bibinfo {year} {2018}),\ \bibinfo {note}
  {056701}\BibitemShut {NoStop}%
\bibitem [{\citenamefont {Kirkwood}\ \emph {et~al.}(2022)\citenamefont {Kirkwood},
  \citenamefont {Poole}, \citenamefont {Kalantar}, \citenamefont {Chapman},
  \citenamefont {Wilks}, \citenamefont {Edwards}, \citenamefont {Turnbull},
  \citenamefont {Michel}, \citenamefont {Divol}, \citenamefont {Fisch},
  \citenamefont {Norreys}, \citenamefont {Rozmus}, \citenamefont {Bude},
  \citenamefont {Blue}, \citenamefont {Fournier}, \citenamefont {Van
  Wonterghem},\ and\ \citenamefont {MacKinnon}}]{10.1063/5.0086068}%
  \BibitemOpen
  \bibfield  {author} {\bibinfo {author} {\bibfnamefont {R.~K.}\ \bibnamefont
  {Kirkwood}}, \bibinfo {author} {\bibfnamefont {P.~L.}\ \bibnamefont
  {Poole}}, \bibinfo {author} {\bibfnamefont {D.~H.}\ \bibnamefont
  {Kalantar}}, \bibinfo {author} {\bibfnamefont {T.~D.}\ \bibnamefont
  {Chapman}}, \bibinfo {author} {\bibfnamefont {S.~C.}\ \bibnamefont {Wilks}},
  \bibinfo {author} {\bibfnamefont {M.~R.}\ \bibnamefont {Edwards}}, \bibinfo
  {author} {\bibfnamefont {D.~P.}\ \bibnamefont {Turnbull}}, \bibinfo {author}
  {\bibfnamefont {P.}\ \bibnamefont {Michel}}, \bibinfo {author} {\bibfnamefont
  {L.}\ \bibnamefont {Divol}}, \bibinfo {author} {\bibfnamefont {N.~J.}\
  \bibnamefont {Fisch}}, \bibinfo {author} {\bibfnamefont {P.}\ \bibnamefont
  {Norreys}}, \bibinfo {author} {\bibfnamefont {W.}\ \bibnamefont {Rozmus}},
  \bibinfo {author} {\bibfnamefont {J.}\ \bibnamefont {Bude}}, \bibinfo
  {author} {\bibfnamefont {B.~E.}\ \bibnamefont {Blue}}, \bibinfo {author}
  {\bibfnamefont {K.~B.}\ \bibnamefont {Fournier}}, \bibinfo {author}
  {\bibfnamefont {B.~M.}\ \bibnamefont {Van Wonterghem}}, \ and\ \bibinfo
  {author} {\bibfnamefont {A.}\ \bibnamefont {MacKinnon}},\ }\bibfield  {title}
  {\bibinfo {title} {Production of high fluence laser beams using ion wave
  plasma optics},\ }\href
  {https://doi.org/10.1063/5.0086068} {\bibfield  {journal} {\bibinfo
  {journal} {Applied Physics Letters}\ }\textbf {\bibinfo {volume} {120}},\
  \bibinfo {pages} {200501} (\bibinfo {year} {2022})}\BibitemShut {NoStop}%
\bibitem [{\citenamefont {Kirkwood}\ \emph {et~al.}(2018)\citenamefont
  {Kirkwood}, \citenamefont {Turnbull}, \citenamefont {Chapman}, \citenamefont
  {Wilks}, \citenamefont {Rosen}, \citenamefont {London}, \citenamefont
  {Pickworth}, \citenamefont {Dunlop}, \citenamefont {Moody}, \citenamefont
  {Strozzi}, \citenamefont {Michel}, \citenamefont {Divol}, \citenamefont
  {Landen}, \citenamefont {MacGowan}, \citenamefont {Van Wonterghem},
  \citenamefont {Fournier},\ and\ \citenamefont {Blue}}]{Kirkwoodnature2022}%
  \BibitemOpen
  \bibfield  {author} {\bibinfo {author} {\bibfnamefont {R.~K.}\ \bibnamefont
  {Kirkwood}}, \bibinfo {author} {\bibfnamefont {D.~P.}\ \bibnamefont
  {Turnbull}}, \bibinfo {author} {\bibfnamefont {T.}\ \bibnamefont {Chapman}},
  \bibinfo {author} {\bibfnamefont {S.~C.}\ \bibnamefont {Wilks}}, \bibinfo
  {author} {\bibfnamefont {M.~D.}\ \bibnamefont {Rosen}}, \bibinfo {author}
  {\bibfnamefont {R.~A.}\ \bibnamefont {London}}, \bibinfo {author}
  {\bibfnamefont {L.~A.}\ \bibnamefont {Pickworth}}, \bibinfo {author}
  {\bibfnamefont {W.~H.}\ \bibnamefont {Dunlop}}, \bibinfo {author}
  {\bibfnamefont {J.~D.}\ \bibnamefont {Moody}}, \bibinfo {author}
  {\bibfnamefont {D.~J.}\ \bibnamefont {Strozzi}}, \bibinfo {author}
  {\bibfnamefont {P.~A.}\ \bibnamefont {Michel}}, \bibinfo {author}
  {\bibfnamefont {L.}\ \bibnamefont {Divol}}, \bibinfo {author} {\bibfnamefont
  {O.~L.}\ \bibnamefont {Landen}}, \bibinfo {author} {\bibfnamefont {B.~J.}\
  \bibnamefont {MacGowan}}, \bibinfo {author} {\bibfnamefont {B.~M.}\
  \bibnamefont {Van Wonterghem}}, \bibinfo {author} {\bibfnamefont {K.~B.}\
  \bibnamefont {Fournier}}, \ and\ \bibinfo {author} {\bibfnamefont {B.~E.}\
  \bibnamefont {Blue}},\ }\bibfield  {title} {\bibinfo {title} {Plasma-based
  beam combiner for very high fluence and energy},\ }\href
  {https://doi.org/10.1038/nphys4312} {\bibfield  {journal} {\bibinfo
  {journal} {Nature Physics}\ }\textbf {\bibinfo {volume} {14}},\ \bibinfo
  {pages} {80--84} (\bibinfo {year} {2018})}\BibitemShut {NoStop}%
\bibitem [{\citenamefont {Hooker}\ and\ \citenamefont {Webb}(2010)}]{Simon}%
  \BibitemOpen
  \bibfield  {author} {\bibinfo {author} {\bibfnamefont {S.}~\bibnamefont
  {Hooker}}\ and\ \bibinfo {author} {\bibfnamefont {C.}~\bibnamefont {Webb}},\
  }\href@noop {} {\emph {\bibinfo {title} {{Laser physics}}}},\ Vol.~\bibinfo
  {volume} {9}\ (\bibinfo  {publisher} {Oxford University Press},\ \bibinfo
  {year} {2010})\BibitemShut {NoStop}%
\bibitem [{\citenamefont {Trines}\ \emph {et~al.}(2010)\citenamefont {Trines},
  \citenamefont {Fluza}, \citenamefont {Bingham}, \citenamefont {Fonseca},
  \citenamefont {Silva}, \citenamefont {Cairns},\ and\ \citenamefont
  {Norreys}}]{RamanAmplificationRaoul}%
  \BibitemOpen
  \bibfield  {author} {\bibinfo {author} {\bibfnamefont {R.}~\bibnamefont
  {Trines}}, \bibinfo {author} {\bibfnamefont {F.}~\bibnamefont {Fluza}},
  \bibinfo {author} {\bibfnamefont {R.}~\bibnamefont {Bingham}}, \bibinfo
  {author} {\bibfnamefont {R.}~\bibnamefont {Fonseca}}, \bibinfo {author}
  {\bibfnamefont {L.}~\bibnamefont {Silva}}, \bibinfo {author} {\bibfnamefont
  {R.}~\bibnamefont {Cairns}},\ and\ \bibinfo {author} {\bibfnamefont
  {P.}~\bibnamefont {Norreys}},\ }\bibfield  {title} {\bibinfo {title}
  {Simulations of efficient raman amplification into the multipetawatt
  regime},\ }\href {https://doi.org/10.1038/nphys1793} {\bibfield  {journal}
  {\bibinfo  {journal} {Nature Physics}\ }\textbf {\bibinfo {volume} {7}}
  (\bibinfo {year} {2010})}\BibitemShut {NoStop}%
\bibitem [{\citenamefont {Sadler}\ \emph {et~al.}(2018)\citenamefont {Sadler},
  \citenamefont {Silva}, \citenamefont {Fonseca}, \citenamefont {Glize},
  \citenamefont {Kasim}, \citenamefont {Savin}, \citenamefont {Aboushelbaya},
  \citenamefont {Mayr}, \citenamefont {Spiers}, \citenamefont {Wang},
  \citenamefont {Bingham}, \citenamefont {Trines},\ and\ \citenamefont
  {Norreys}}]{AdvantageRamanAmp}%
  \BibitemOpen
  \bibfield  {author} {\bibinfo {author} {\bibfnamefont {J.}~\bibnamefont
  {Sadler}}, \bibinfo {author} {\bibfnamefont {L.}~\bibnamefont {Silva}},
  \bibinfo {author} {\bibfnamefont {R.}~\bibnamefont {Fonseca}}, \bibinfo
  {author} {\bibfnamefont {K.}~\bibnamefont {Glize}}, \bibinfo {author}
  {\bibfnamefont {M.}~\bibnamefont {Kasim}}, \bibinfo {author} {\bibfnamefont
  {A.}~\bibnamefont {Savin}}, \bibinfo {author} {\bibfnamefont
  {R.}~\bibnamefont {Aboushelbaya}}, \bibinfo {author} {\bibfnamefont
  {M.}~\bibnamefont {Mayr}}, \bibinfo {author} {\bibfnamefont {B.}~\bibnamefont
  {Spiers}}, \bibinfo {author} {\bibfnamefont {R.}~\bibnamefont {Wang}},
  \bibinfo {author} {\bibfnamefont {R.}~\bibnamefont {Bingham}}, \bibinfo
  {author} {\bibfnamefont {R.}~\bibnamefont {Trines}},\ and\ \bibinfo {author}
  {\bibfnamefont {P.}~\bibnamefont {Norreys}},\ }\bibfield  {title}
  {{\bibinfo {title} {Advantages to a diverging raman
  amplifier}},\ }\bibfield  {journal} {\bibinfo  {journal} {Communications
  Physics}\ }\textbf {\bibinfo {volume} {1}},\ \href
  {https://doi.org/10.1038/s42005-018-0021-8} {10.1038/s42005-018-0021-8}
  (\bibinfo {year} {2018})\BibitemShut {NoStop}%
\bibitem [{\citenamefont {Vacalis}\ \emph {et~al.}(2023)\citenamefont
  {Vacalis}, \citenamefont {Marocco}, \citenamefont {Bamber}, \citenamefont
  {Bingham},\ and\ \citenamefont {Gregori}}]{Vacalis:2023gdz}%
  \BibitemOpen
  \bibfield  {author} {\bibinfo {author} {\bibfnamefont {G.}~\bibnamefont
  {Vacalis}}, \bibinfo {author} {\bibfnamefont {G.}~\bibnamefont {Marocco}},
  \bibinfo {author} {\bibfnamefont {J.}~\bibnamefont {Bamber}}, \bibinfo
  {author} {\bibfnamefont {R.}~\bibnamefont {Bingham}},\ and\ \bibinfo {author}
  {\bibfnamefont {G.}~\bibnamefont {Gregori}},\ }\bibfield  {title} {\bibinfo
  {title} {Detection of high-frequency gravitational waves using high-energy
  pulsed lasers},\ }\href {https://doi.org/10.1088/1361-6382/acd517} {\bibfield
   {journal} {\bibinfo  {journal} {Classical and Quantum Gravity}\ }\textbf
  {\bibinfo {volume} {40}},\ \bibinfo {pages} {155006} (\bibinfo {year}
  {2023})}\BibitemShut {NoStop}%
\bibitem [{\citenamefont {Laurence}\ \emph {et~al.}(2020)\citenamefont
  {Laurence}, \citenamefont {Alessi}, \citenamefont {Feigenbaum}, \citenamefont
  {Negres}, \citenamefont {Qiu}, \citenamefont {Siders}, \citenamefont
  {Spinka},\ and\ \citenamefont {Stolz}}]{10.1063/1.5131174}%
  \BibitemOpen
  \bibfield  {author} {\bibinfo {author} {\bibfnamefont {T.~A.}\ \bibnamefont
  {Laurence}}, \bibinfo {author} {\bibfnamefont {D.~A.}\ \bibnamefont
  {Alessi}}, \bibinfo {author} {\bibfnamefont {E.}~\bibnamefont {Feigenbaum}},
  \bibinfo {author} {\bibfnamefont {R.~A.}\ \bibnamefont {Negres}}, \bibinfo
  {author} {\bibfnamefont {S.~R.}\ \bibnamefont {Qiu}}, \bibinfo {author}
  {\bibfnamefont {C.~W.}\ \bibnamefont {Siders}}, \bibinfo {author}
  {\bibfnamefont {T.~M.}\ \bibnamefont {Spinka}},\ and\ \bibinfo {author}
  {\bibfnamefont {C.~J.}\ \bibnamefont {Stolz}},\ }\bibfield  {title} {\bibinfo
  {title} {{Mirrors for petawatt lasers: Design principles, limitations, and
  solutions}},\ }\bibfield  {journal} {\bibinfo  {journal} {Journal of Applied
  Physics}\ }\textbf {\bibinfo {volume} {128}},\ \href
  {https://doi.org/10.1063/1.5131174} {10.1063/1.5131174} (\bibinfo {year}
  {2020}),\ \bibinfo {note} {071101}\BibitemShut {NoStop}%
\bibitem [{\citenamefont {Abbott}\ \emph {et~al.}(2016)\citenamefont {Abbott}
  \emph {et~al.}}]{Abbott_2016}%
  \BibitemOpen
  \bibfield  {author} {\bibinfo {author} {\bibfnamefont {B.~P.}\ \bibnamefont
  {Abbott}} \emph {et~al.} (\bibinfo {collaboration} {LIGO Scientific
  Collaboration and Virgo Collaboration}),\ }\bibfield  {title} {\bibinfo
  {title} {Properties of the binary black hole merger gw150914},\ }\href
  {https://doi.org/10.1103/PhysRevLett.116.241102} {\bibfield  {journal}
  {\bibinfo  {journal} {Phys. Rev. Lett.}\ }\textbf {\bibinfo {volume} {116}},\
  \bibinfo {pages} {241102} (\bibinfo {year} {2016})}\BibitemShut {NoStop}%
\bibitem [{\citenamefont {Bauswein}\ \emph {et~al.}(2016)\citenamefont
  {Bauswein}, \citenamefont {Stergioulas},\ and\ \citenamefont
  {Janka}}]{Bauswein:2015vxa}%
  \BibitemOpen
  \bibfield  {author} {\bibinfo {author} {\bibfnamefont {A.}~\bibnamefont
  {Bauswein}}, \bibinfo {author} {\bibfnamefont {N.}~\bibnamefont
  {Stergioulas}},\ and\ \bibinfo {author} {\bibfnamefont {H.-T.}\ \bibnamefont
  {Janka}},\ }\bibfield  {title} {\bibinfo {title} {{Exploring properties of
  high-density matter through remnants of neutron-star mergers}},\ }\href
  {https://doi.org/10.1140/epja/i2016-16056-7} {\bibfield  {journal} {\bibinfo
  {journal} {Eur. Phys. J. A}\ }\textbf {\bibinfo {volume} {52}},\ \bibinfo
  {pages} {56} (\bibinfo {year} {2016})},\ \Eprint
  {https://arxiv.org/abs/1508.05493} {arXiv:1508.05493 [astro-ph.HE]}
  \BibitemShut {NoStop}%
\bibitem [{\citenamefont {Raidal}\ \emph {et~al.}(2017)\citenamefont {Raidal},
  \citenamefont {Vaskonen},\ and\ \citenamefont {Veerm\"ae}}]{Raidal:2017mfl}%
  \BibitemOpen
  \bibfield  {author} {\bibinfo {author} {\bibfnamefont {M.}~\bibnamefont
  {Raidal}}, \bibinfo {author} {\bibfnamefont {V.}~\bibnamefont {Vaskonen}},\
  and\ \bibinfo {author} {\bibfnamefont {H.}~\bibnamefont {Veerm\"ae}},\
  }\bibfield  {title} {\bibinfo {title} {{Gravitational Waves from Primordial
  Black Hole Mergers}},\ }\href {https://doi.org/10.1088/1475-7516/2017/09/037}
  {\bibfield  {journal} {\bibinfo  {journal} {JCAP}\ }\textbf {\bibinfo
  {volume} {09}},\ \bibinfo {pages} {037}},\ \Eprint
  {https://arxiv.org/abs/1707.01480} {arXiv:1707.01480 [astro-ph.CO]}
  \BibitemShut {NoStop}%
\bibitem [{\citenamefont {Franciolini}\ \emph {et~al.}(2022)\citenamefont
  {Franciolini}, \citenamefont {Maharana},\ and\ \citenamefont
  {Muia}}]{Franciolini:2022htd}%
  \BibitemOpen
  \bibfield  {author} {\bibinfo {author} {\bibfnamefont {G.}~\bibnamefont
  {Franciolini}}, \bibinfo {author} {\bibfnamefont {A.}~\bibnamefont
  {Maharana}},\ and\ \bibinfo {author} {\bibfnamefont {F.}~\bibnamefont
  {Muia}},\ }\bibfield  {title} {\bibinfo {title} {{Hunt for light primordial
  black hole dark matter with ultrahigh-frequency gravitational waves}},\
  }\href {https://doi.org/10.1103/PhysRevD.106.103520} {\bibfield  {journal}
  {\bibinfo  {journal} {Phys. Rev. D}\ }\textbf {\bibinfo {volume} {106}},\
  \bibinfo {pages} {103520} (\bibinfo {year} {2022})},\ \Eprint
  {https://arxiv.org/abs/2205.02153} {arXiv:2205.02153 [astro-ph.CO]}
  \BibitemShut {NoStop}%
\bibitem [{\citenamefont {Dong}\ \emph {et~al.}(2016)\citenamefont {Dong},
  \citenamefont {Kinney},\ and\ \citenamefont {Stojkovic}}]{Dong:2015yjs}%
  \BibitemOpen
  \bibfield  {author} {\bibinfo {author} {\bibfnamefont {R.}~\bibnamefont
  {Dong}}, \bibinfo {author} {\bibfnamefont {W.~H.}\ \bibnamefont {Kinney}},\
  and\ \bibinfo {author} {\bibfnamefont {D.}~\bibnamefont {Stojkovic}},\
  }\bibfield  {title} {\bibinfo {title} {{Gravitational wave production by
  Hawking radiation from rotating primordial black holes}},\ }\href
  {https://doi.org/10.1088/1475-7516/2016/10/034} {\bibfield  {journal}
  {\bibinfo  {journal} {JCAP}\ }\textbf {\bibinfo {volume} {10}},\ \bibinfo
  {pages} {034}},\ \Eprint {https://arxiv.org/abs/1511.05642} {arXiv:1511.05642
  [astro-ph.CO]} \BibitemShut {NoStop}%
\bibitem [{\citenamefont {Ireland}\ \emph {et~al.}(2023)\citenamefont
  {Ireland}, \citenamefont {Profumo},\ and\ \citenamefont
  {Scharnhorst}}]{Ireland:2023avg}%
  \BibitemOpen
  \bibfield  {author} {\bibinfo {author} {\bibfnamefont {A.}~\bibnamefont
  {Ireland}}, \bibinfo {author} {\bibfnamefont {S.}~\bibnamefont {Profumo}},\
  and\ \bibinfo {author} {\bibfnamefont {J.}~\bibnamefont {Scharnhorst}},\
  }\bibfield  {title} {\bibinfo {title} {{Primordial gravitational waves from
  black hole evaporation in standard and nonstandard cosmologies}},\ }\href
  {https://doi.org/10.1103/PhysRevD.107.104021} {\bibfield  {journal} {\bibinfo
   {journal} {Phys. Rev. D}\ }\textbf {\bibinfo {volume} {107}},\ \bibinfo
  {pages} {104021} (\bibinfo {year} {2023})},\ \Eprint
  {https://arxiv.org/abs/2302.10188} {arXiv:2302.10188 [gr-qc]} \BibitemShut
  {NoStop}%
\bibitem [{\citenamefont {Yoshino}\ and\ \citenamefont
  {Kodama}(2014)}]{Yoshino:2013ofa}%
  \BibitemOpen
  \bibfield  {author} {\bibinfo {author} {\bibfnamefont {H.}~\bibnamefont
  {Yoshino}}\ and\ \bibinfo {author} {\bibfnamefont {H.}~\bibnamefont
  {Kodama}},\ }\bibfield  {title} {\bibinfo {title} {{Gravitational radiation
  from an axion cloud around a black hole: Superradiant phase}},\ }\href
  {https://doi.org/10.1093/ptep/ptu029} {\bibfield  {journal} {\bibinfo
  {journal} {PTEP}\ }\textbf {\bibinfo {volume} {2014}},\ \bibinfo {pages}
  {043E02} (\bibinfo {year} {2014})},\ \Eprint
  {https://arxiv.org/abs/1312.2326} {arXiv:1312.2326 [gr-qc]} \BibitemShut
  {NoStop}%
\bibitem [{\citenamefont {Arvanitaki}\ \emph {et~al.}(2015)\citenamefont
  {Arvanitaki}, \citenamefont {Baryakhtar},\ and\ \citenamefont
  {Huang}}]{Arvanitaki:2014wva}%
  \BibitemOpen
  \bibfield  {author} {\bibinfo {author} {\bibfnamefont {A.}~\bibnamefont
  {Arvanitaki}}, \bibinfo {author} {\bibfnamefont {M.}~\bibnamefont
  {Baryakhtar}},\ and\ \bibinfo {author} {\bibfnamefont {X.}~\bibnamefont
  {Huang}},\ }\bibfield  {title} {\bibinfo {title} {{Discovering the QCD Axion
  with Black Holes and Gravitational Waves}},\ }\href
  {https://doi.org/10.1103/PhysRevD.91.084011} {\bibfield  {journal} {\bibinfo
  {journal} {Phys. Rev. D}\ }\textbf {\bibinfo {volume} {91}},\ \bibinfo
  {pages} {084011} (\bibinfo {year} {2015})},\ \Eprint
  {https://arxiv.org/abs/1411.2263} {arXiv:1411.2263 [hep-ph]} \BibitemShut
  {NoStop}%
\bibitem [{\citenamefont {Brito}\ \emph {et~al.}(2015)\citenamefont {Brito},
  \citenamefont {Cardoso},\ and\ \citenamefont {Pani}}]{Brito:2014wla}%
  \BibitemOpen
  \bibfield  {author} {\bibinfo {author} {\bibfnamefont {R.}~\bibnamefont
  {Brito}}, \bibinfo {author} {\bibfnamefont {V.}~\bibnamefont {Cardoso}},\
  and\ \bibinfo {author} {\bibfnamefont {P.}~\bibnamefont {Pani}},\ }\bibfield
  {title} {\bibinfo {title} {{Black holes as particle detectors: evolution of
  superradiant instabilities}},\ }\href
  {https://doi.org/10.1088/0264-9381/32/13/134001} {\bibfield  {journal}
  {\bibinfo  {journal} {Class. Quant. Grav.}\ }\textbf {\bibinfo {volume}
  {32}},\ \bibinfo {pages} {134001} (\bibinfo {year} {2015})},\ \Eprint
  {https://arxiv.org/abs/1411.0686} {arXiv:1411.0686 [gr-qc]} \BibitemShut
  {NoStop}%
\bibitem [{\citenamefont {Sun}\ and\ \citenamefont
  {Zhang}(2021)}]{Sun:2020gem}%
  \BibitemOpen
  \bibfield  {author} {\bibinfo {author} {\bibfnamefont {S.}~\bibnamefont
  {Sun}}\ and\ \bibinfo {author} {\bibfnamefont {Y.-L.}\ \bibnamefont
  {Zhang}},\ }\bibfield  {title} {\bibinfo {title} {{Fast gravitational wave
  bursts from axion clumps}},\ }\href
  {https://doi.org/10.1103/PhysRevD.104.103009} {\bibfield  {journal} {\bibinfo
   {journal} {Phys. Rev. D}\ }\textbf {\bibinfo {volume} {104}},\ \bibinfo
  {pages} {103009} (\bibinfo {year} {2021})},\ \Eprint
  {https://arxiv.org/abs/2003.10527} {arXiv:2003.10527 [hep-ph]} \BibitemShut
  {NoStop}%
\bibitem [{\citenamefont {Caprini}\ \emph {et~al.}(2008)\citenamefont
  {Caprini}, \citenamefont {Durrer},\ and\ \citenamefont
  {Servant}}]{Caprini:2007xq}%
  \BibitemOpen
  \bibfield  {author} {\bibinfo {author} {\bibfnamefont {C.}~\bibnamefont
  {Caprini}}, \bibinfo {author} {\bibfnamefont {R.}~\bibnamefont {Durrer}},\
  and\ \bibinfo {author} {\bibfnamefont {G.}~\bibnamefont {Servant}},\
  }\bibfield  {title} {\bibinfo {title} {{Gravitational wave generation from
  bubble collisions in first-order phase transitions: An analytic approach}},\
  }\href {https://doi.org/10.1103/PhysRevD.77.124015} {\bibfield  {journal}
  {\bibinfo  {journal} {Phys. Rev. D}\ }\textbf {\bibinfo {volume} {77}},\
  \bibinfo {pages} {124015} (\bibinfo {year} {2008})},\ \Eprint
  {https://arxiv.org/abs/0711.2593} {arXiv:0711.2593 [astro-ph]} \BibitemShut
  {NoStop}%
\bibitem [{\citenamefont {Caprini}\ \emph
  {et~al.}(2009{\natexlab{a}})\citenamefont {Caprini}, \citenamefont {Durrer},
  \citenamefont {Konstandin},\ and\ \citenamefont {Servant}}]{Caprini:2009fx}%
  \BibitemOpen
  \bibfield  {author} {\bibinfo {author} {\bibfnamefont {C.}~\bibnamefont
  {Caprini}}, \bibinfo {author} {\bibfnamefont {R.}~\bibnamefont {Durrer}},
  \bibinfo {author} {\bibfnamefont {T.}~\bibnamefont {Konstandin}},\ and\
  \bibinfo {author} {\bibfnamefont {G.}~\bibnamefont {Servant}},\ }\bibfield
  {title} {\bibinfo {title} {{General Properties of the Gravitational Wave
  Spectrum from Phase Transitions}},\ }\href
  {https://doi.org/10.1103/PhysRevD.79.083519} {\bibfield  {journal} {\bibinfo
  {journal} {Phys. Rev. D}\ }\textbf {\bibinfo {volume} {79}},\ \bibinfo
  {pages} {083519} (\bibinfo {year} {2009}{\natexlab{a}})},\ \Eprint
  {https://arxiv.org/abs/0901.1661} {arXiv:0901.1661 [astro-ph.CO]}
  \BibitemShut {NoStop}%
\bibitem [{\citenamefont {Caprini}\ \emph
  {et~al.}(2009{\natexlab{b}})\citenamefont {Caprini}, \citenamefont {Durrer},\
  and\ \citenamefont {Servant}}]{Caprini:2009yp}%
  \BibitemOpen
  \bibfield  {author} {\bibinfo {author} {\bibfnamefont {C.}~\bibnamefont
  {Caprini}}, \bibinfo {author} {\bibfnamefont {R.}~\bibnamefont {Durrer}},\
  and\ \bibinfo {author} {\bibfnamefont {G.}~\bibnamefont {Servant}},\
  }\bibfield  {title} {\bibinfo {title} {{The stochastic gravitational wave
  background from turbulence and magnetic fields generated by a first-order
  phase transition}},\ }\href {https://doi.org/10.1088/1475-7516/2009/12/024}
  {\bibfield  {journal} {\bibinfo  {journal} {JCAP}\ }\textbf {\bibinfo
  {volume} {12}},\ \bibinfo {pages} {024}},\ \Eprint
  {https://arxiv.org/abs/0909.0622} {arXiv:0909.0622 [astro-ph.CO]}
  \BibitemShut {NoStop}%
\bibitem [{\citenamefont {Sakellariadou}(1990)}]{PhysRevD.42.354}%
  \BibitemOpen
  \bibfield  {author} {\bibinfo {author} {\bibfnamefont {M.}~\bibnamefont
  {Sakellariadou}},\ }\bibfield  {title} {\bibinfo {title} {Gravitational waves
  emitted from infinite strings},\ }\href
  {https://doi.org/10.1103/PhysRevD.42.354} {\bibfield  {journal} {\bibinfo
  {journal} {Phys. Rev. D}\ }\textbf {\bibinfo {volume} {42}},\ \bibinfo
  {pages} {354} (\bibinfo {year} {1990})}\BibitemShut {NoStop}%
\bibitem [{\citenamefont {Vachaspati}\ and\ \citenamefont
  {Vilenkin}(1985)}]{PhysRevD.31.3052}%
  \BibitemOpen
  \bibfield  {author} {\bibinfo {author} {\bibfnamefont {T.}~\bibnamefont
  {Vachaspati}}\ and\ \bibinfo {author} {\bibfnamefont {A.}~\bibnamefont
  {Vilenkin}},\ }\bibfield  {title} {\bibinfo {title} {Gravitational radiation
  from cosmic strings},\ }\href {https://doi.org/10.1103/PhysRevD.31.3052}
  {\bibfield  {journal} {\bibinfo  {journal} {Phys. Rev. D}\ }\textbf {\bibinfo
  {volume} {31}},\ \bibinfo {pages} {3052} (\bibinfo {year}
  {1985})}\BibitemShut {NoStop}%
\bibitem [{\citenamefont {Zhou}\ and\ \citenamefont
  {Bian}(2022)}]{ZhouRuiYu:2020bbs}%
  \BibitemOpen
  \bibfield  {author} {\bibinfo {author} {\bibfnamefont {R.}~\bibnamefont
  {Zhou}}\ and\ \bibinfo {author} {\bibfnamefont {L.}~\bibnamefont {Bian}},\
  }\bibfield  {title} {\bibinfo {title} {Gravitational waves from cosmic
  strings after a first-order phase transition *},\ }\href
  {https://doi.org/10.1088/1674-1137/ac424c} {\bibfield  {journal} {\bibinfo
  {journal} {Chinese Physics C}\ }\textbf {\bibinfo {volume} {46}},\ \bibinfo
  {pages} {043104} (\bibinfo {year} {2022})}\BibitemShut {NoStop}%
\bibitem [{\citenamefont {Chang}\ and\ \citenamefont
  {Cui}(2022)}]{Chang:2021afa}%
  \BibitemOpen
  \bibfield  {author} {\bibinfo {author} {\bibfnamefont {C.-F.}\ \bibnamefont
  {Chang}}\ and\ \bibinfo {author} {\bibfnamefont {Y.}~\bibnamefont {Cui}},\
  }\bibfield  {title} {\bibinfo {title} {{Gravitational waves from global
  cosmic strings and cosmic archaeology}},\ }\href
  {https://doi.org/10.1007/JHEP03(2022)114} {\bibfield  {journal} {\bibinfo
  {journal} {JHEP}\ }\textbf {\bibinfo {volume} {03}},\ \bibinfo {pages}
  {114}},\ \Eprint {https://arxiv.org/abs/2106.09746} {arXiv:2106.09746
  [hep-ph]} \BibitemShut {NoStop}%
\bibitem [{\citenamefont {Blachier}\ \emph {et~al.}(2023)\citenamefont
  {Blachier}, \citenamefont {Barrau}, \citenamefont {Martineau},\ and\
  \citenamefont {Renevey}}]{Blachier:2023ygh}%
  \BibitemOpen
  \bibfield  {author} {\bibinfo {author} {\bibfnamefont {B.}~\bibnamefont
  {Blachier}}, \bibinfo {author} {\bibfnamefont {A.}~\bibnamefont {Barrau}},
  \bibinfo {author} {\bibfnamefont {K.}~\bibnamefont {Martineau}},\ and\
  \bibinfo {author} {\bibfnamefont {C.}~\bibnamefont {Renevey}},\ }\bibfield
  {title} {\bibinfo {title} {{Competitive effects between gravitational
  radiation and mass variation for two-body systems in circular orbits}},\
  }\href@noop {} {\  (\bibinfo {year} {2023})},\ \Eprint
  {https://arxiv.org/abs/2306.09069} {arXiv:2306.09069 [gr-qc]} \BibitemShut
  {NoStop}%
\bibitem [{\citenamefont {Gertsenshtein}(1962)}]{Gertsenshtein:1962}%
  \BibitemOpen
  \bibfield  {author} {\bibinfo {author} {\bibfnamefont {M.~E.}\ \bibnamefont
  {Gertsenshtein}},\ }\bibfield  {title} {\bibinfo {title} {{Wave Resonance of
  Light and Gravitational Waves}},\ }\href@noop {} {\bibfield  {journal}
  {\bibinfo  {journal} {Sov. Phys. JETP 14}\ }\textbf {\bibinfo {volume} {84}}
  (\bibinfo {year} {1962})}\BibitemShut {NoStop}%
\bibitem [{\citenamefont {Boccaletti}\ \emph {et~al.}(1970)\citenamefont
  {Boccaletti}, \citenamefont {De~Sabbata}, \citenamefont {Fortini},\ and\
  \citenamefont {Gualdi}}]{osti_4070463}%
  \BibitemOpen
  \bibfield  {author} {\bibinfo {author} {\bibfnamefont {D.}~\bibnamefont
  {Boccaletti}}, \bibinfo {author} {\bibfnamefont {V.}~\bibnamefont
  {De~Sabbata}}, \bibinfo {author} {\bibfnamefont {P.}~\bibnamefont
  {Fortini}},\ and\ \bibinfo {author} {\bibfnamefont {G.}~\bibnamefont
  {Gualdi}},\ }\bibfield  {title} {\bibinfo {title} {Conversion of photons into
  gravitons and vice versa in a static electromagnetic field.},\ }\bibfield
  {journal} {\bibinfo  {journal} {Nuovo Cim., 60B: 129-46(11 Dec 1970).}\
  }\href {https://doi.org/10.1007/BF02710177} {10.1007/BF02710177} (\bibinfo
  {year} {1970})\BibitemShut {NoStop}%
\bibitem [{\citenamefont {Zel'dovich}(1973)}]{osti_4377804}%
  \BibitemOpen
  \bibfield  {author} {\bibinfo {author} {\bibfnamefont {Y.~B.}\ \bibnamefont
  {Zel'dovich}},\ }\bibfield  {title} {\bibinfo {title} {Electromagnetic and
  gravitational waves in a stationary magnetic field},\ }\bibfield  {journal}
  {\bibinfo  {journal} {Zh. Eksp. Teor. Fiz., v. 65, no. 4, pp. 1311-1315}\
  }\textbf {\bibinfo {volume} {65}},\ \href
  {https://www.osti.gov/biblio/4377804} {} (\bibinfo {year} {1973})\BibitemShut
  {NoStop}%
\bibitem [{\citenamefont {De~Logi}\ and\ \citenamefont
  {Mickelson}(1977)}]{PhysRevD.16.2915}%
  \BibitemOpen
  \bibfield  {author} {\bibinfo {author} {\bibfnamefont {W.~K.}\ \bibnamefont
  {De~Logi}}\ and\ \bibinfo {author} {\bibfnamefont {A.~R.}\ \bibnamefont
  {Mickelson}},\ }\bibfield  {title} {\bibinfo {title} {Electrogravitational
  conversion cross sections in static electromagnetic fields},\ }\href
  {https://doi.org/10.1103/PhysRevD.16.2915} {\bibfield  {journal} {\bibinfo
  {journal} {Phys. Rev. D}\ }\textbf {\bibinfo {volume} {16}},\ \bibinfo
  {pages} {2915} (\bibinfo {year} {1977})}\BibitemShut {NoStop}%
\bibitem [{\citenamefont {Domcke}(2023)}]{Domcke:2023qle}%
  \BibitemOpen
  \bibfield  {author} {\bibinfo {author} {\bibfnamefont {V.}~\bibnamefont
  {Domcke}},\ }\bibfield  {title} {\bibinfo {title} {{Electromagnetic
  high-frequency gravitational wave detection}},\ }in\ \href@noop {} {\emph
  {\bibinfo {booktitle} {{57th Rencontres de Moriond on Electroweak
  Interactions and Unified Theories}}}}\ (\bibinfo {year} {2023})\ \Eprint
  {https://arxiv.org/abs/2306.04496} {arXiv:2306.04496 [gr-qc]} \BibitemShut
  {NoStop}%
\bibitem [{\citenamefont {Ejlli}\ \emph {et~al.}(2019)\citenamefont {Ejlli},
  \citenamefont {Ejlli}, \citenamefont {Cruise}, \citenamefont {Pisano},\ and\
  \citenamefont {Grote}}]{Ejlli:2019bqj}%
  \BibitemOpen
  \bibfield  {author} {\bibinfo {author} {\bibfnamefont {A.}~\bibnamefont
  {Ejlli}}, \bibinfo {author} {\bibfnamefont {D.}~\bibnamefont {Ejlli}},
  \bibinfo {author} {\bibfnamefont {A.~M.}\ \bibnamefont {Cruise}}, \bibinfo
  {author} {\bibfnamefont {G.}~\bibnamefont {Pisano}},\ and\ \bibinfo {author}
  {\bibfnamefont {H.}~\bibnamefont {Grote}},\ }\bibfield  {title} {\bibinfo
  {title} {{Upper limits on the amplitude of ultra-high-frequency gravitational
  waves from graviton to photon conversion}},\ }\href
  {https://doi.org/10.1140/epjc/s10052-019-7542-5} {\bibfield  {journal}
  {\bibinfo  {journal} {Eur. Phys. J. C}\ }\textbf {\bibinfo {volume} {79}},\
  \bibinfo {pages} {1032} (\bibinfo {year} {2019})},\ \Eprint
  {https://arxiv.org/abs/1908.00232} {arXiv:1908.00232 [gr-qc]} \BibitemShut
  {NoStop}%
\bibitem [{\citenamefont {Chen}(1995)}]{PhysRevLett.74.634}%
  \BibitemOpen
  \bibfield  {author} {\bibinfo {author} {\bibfnamefont {P.}~\bibnamefont
  {Chen}},\ }\bibfield  {title} {\bibinfo {title} {Resonant photon-graviton
  conversion and cosmic microwave background fluctuations},\ }\href
  {https://doi.org/10.1103/PhysRevLett.74.634} {\bibfield  {journal} {\bibinfo
  {journal} {Phys. Rev. Lett.}\ }\textbf {\bibinfo {volume} {74}},\ \bibinfo
  {pages} {634} (\bibinfo {year} {1995})}\BibitemShut {NoStop}%
\bibitem [{\citenamefont {Pshirkov}\ and\ \citenamefont
  {Baskaran}(2009)}]{Pshirkov:2009sf}%
  \BibitemOpen
  \bibfield  {author} {\bibinfo {author} {\bibfnamefont {M.~S.}\ \bibnamefont
  {Pshirkov}}\ and\ \bibinfo {author} {\bibfnamefont {D.}~\bibnamefont
  {Baskaran}},\ }\bibfield  {title} {\bibinfo {title} {{Limits on
  High-Frequency Gravitational Wave Background from its interplay with Large
  Scale Magnetic Fields}},\ }\href {https://doi.org/10.1103/PhysRevD.80.042002}
  {\bibfield  {journal} {\bibinfo  {journal} {Phys. Rev. D}\ }\textbf {\bibinfo
  {volume} {80}},\ \bibinfo {pages} {042002} (\bibinfo {year} {2009})},\
  \Eprint {https://arxiv.org/abs/0903.4160} {arXiv:0903.4160 [gr-qc]}
  \BibitemShut {NoStop}%
\bibitem [{\citenamefont {Dolgov}\ and\ \citenamefont
  {Ejlli}(2012)}]{Dolgov:2012be}%
  \BibitemOpen
  \bibfield  {author} {\bibinfo {author} {\bibfnamefont {A.~D.}\ \bibnamefont
  {Dolgov}}\ and\ \bibinfo {author} {\bibfnamefont {D.}~\bibnamefont {Ejlli}},\
  }\bibfield  {title} {\bibinfo {title} {{Conversion of relic gravitational
  waves into photons in cosmological magnetic fields}},\ }\href
  {https://doi.org/10.1088/1475-7516/2012/12/003} {\bibfield  {journal}
  {\bibinfo  {journal} {JCAP}\ }\textbf {\bibinfo {volume} {12}},\ \bibinfo
  {pages} {003}},\ \Eprint {https://arxiv.org/abs/1211.0500} {arXiv:1211.0500
  [gr-qc]} \BibitemShut {NoStop}%
\bibitem [{\citenamefont {Ringwald}\ \emph {et~al.}(2021)\citenamefont
  {Ringwald}, \citenamefont {Sch\"utte-Engel},\ and\ \citenamefont
  {Tamarit}}]{Ringwald:2020ist}%
  \BibitemOpen
  \bibfield  {author} {\bibinfo {author} {\bibfnamefont {A.}~\bibnamefont
  {Ringwald}}, \bibinfo {author} {\bibfnamefont {J.}~\bibnamefont
  {Sch\"utte-Engel}},\ and\ \bibinfo {author} {\bibfnamefont {C.}~\bibnamefont
  {Tamarit}},\ }\bibfield  {title} {\bibinfo {title} {{Gravitational Waves as a
  Big Bang Thermometer}},\ }\href
  {https://doi.org/10.1088/1475-7516/2021/03/054} {\bibfield  {journal}
  {\bibinfo  {journal} {JCAP}\ }\textbf {\bibinfo {volume} {03}},\ \bibinfo
  {pages} {054}},\ \Eprint {https://arxiv.org/abs/2011.04731} {arXiv:2011.04731
  [hep-ph]} \BibitemShut {NoStop}%
\bibitem [{\citenamefont {Fujita}\ \emph {et~al.}(2020)\citenamefont {Fujita},
  \citenamefont {Kamada},\ and\ \citenamefont {Nakai}}]{Fujita:2020rdx}%
  \BibitemOpen
  \bibfield  {author} {\bibinfo {author} {\bibfnamefont {T.}~\bibnamefont
  {Fujita}}, \bibinfo {author} {\bibfnamefont {K.}~\bibnamefont {Kamada}},\
  and\ \bibinfo {author} {\bibfnamefont {Y.}~\bibnamefont {Nakai}},\ }\bibfield
   {title} {\bibinfo {title} {{Gravitational Waves from Primordial Magnetic
  Fields via Photon-Graviton Conversion}},\ }\href
  {https://doi.org/10.1103/PhysRevD.102.103501} {\bibfield  {journal} {\bibinfo
   {journal} {Phys. Rev. D}\ }\textbf {\bibinfo {volume} {102}},\ \bibinfo
  {pages} {103501} (\bibinfo {year} {2020})},\ \Eprint
  {https://arxiv.org/abs/2002.07548} {arXiv:2002.07548 [astro-ph.CO]}
  \BibitemShut {NoStop}%
\bibitem [{\citenamefont {Berlin}\ \emph {et~al.}(2022)\citenamefont {Berlin},
  \citenamefont {Blas}, \citenamefont {Tito~D'Agnolo}, \citenamefont {Ellis},
  \citenamefont {Harnik}, \citenamefont {Kahn},\ and\ \citenamefont
  {Sch\"utte-Engel}}]{Berlin:2021txa}%
  \BibitemOpen
  \bibfield  {author} {\bibinfo {author} {\bibfnamefont {A.}~\bibnamefont
  {Berlin}}, \bibinfo {author} {\bibfnamefont {D.}~\bibnamefont {Blas}},
  \bibinfo {author} {\bibfnamefont {R.}~\bibnamefont {Tito~D'Agnolo}}, \bibinfo
  {author} {\bibfnamefont {S.~A.~R.}\ \bibnamefont {Ellis}}, \bibinfo {author}
  {\bibfnamefont {R.}~\bibnamefont {Harnik}}, \bibinfo {author} {\bibfnamefont
  {Y.}~\bibnamefont {Kahn}},\ and\ \bibinfo {author} {\bibfnamefont
  {J.}~\bibnamefont {Sch\"utte-Engel}},\ }\bibfield  {title} {\bibinfo {title}
  {{Detecting high-frequency gravitational waves with microwave cavities}},\
  }\href {https://doi.org/10.1103/PhysRevD.105.116011} {\bibfield  {journal}
  {\bibinfo  {journal} {Phys. Rev. D}\ }\textbf {\bibinfo {volume} {105}},\
  \bibinfo {pages} {116011} (\bibinfo {year} {2022})},\ \Eprint
  {https://arxiv.org/abs/2112.11465} {arXiv:2112.11465 [hep-ph]} \BibitemShut
  {NoStop}%
\bibitem [{\citenamefont {Domcke}\ \emph {et~al.}(2022)\citenamefont {Domcke},
  \citenamefont {Garcia-Cely},\ and\ \citenamefont {Rodd}}]{Domcke:2022rgu}%
  \BibitemOpen
  \bibfield  {author} {\bibinfo {author} {\bibfnamefont {V.}~\bibnamefont
  {Domcke}}, \bibinfo {author} {\bibfnamefont {C.}~\bibnamefont
  {Garcia-Cely}},\ and\ \bibinfo {author} {\bibfnamefont {N.~L.}\ \bibnamefont
  {Rodd}},\ }\bibfield  {title} {\bibinfo {title} {{Novel Search for
  High-Frequency Gravitational Waves with Low-Mass Axion Haloscopes}},\ }\href
  {https://doi.org/10.1103/PhysRevLett.129.041101} {\bibfield  {journal}
  {\bibinfo  {journal} {Phys. Rev. Lett.}\ }\textbf {\bibinfo {volume} {129}},\
  \bibinfo {pages} {041101} (\bibinfo {year} {2022})},\ \Eprint
  {https://arxiv.org/abs/2202.00695} {arXiv:2202.00695 [hep-ph]} \BibitemShut
  {NoStop}%
\bibitem [{\citenamefont {Barrau}\ \emph {et~al.}(2023)\citenamefont {Barrau},
  \citenamefont {García-Bellido}, \citenamefont {Grenet},\ and\ \citenamefont
  {Martineau}}]{Barrau:2023kuv}%
  \BibitemOpen
  \bibfield  {author} {\bibinfo {author} {\bibfnamefont {A.}~\bibnamefont
  {Barrau}}, \bibinfo {author} {\bibfnamefont {J.}~\bibnamefont
  {García-Bellido}}, \bibinfo {author} {\bibfnamefont {T.}~\bibnamefont
  {Grenet}},\ and\ \bibinfo {author} {\bibfnamefont {K.}~\bibnamefont
  {Martineau}},\ }\href@noop {} {\bibinfo {title} {Prospects for detection of
  ultra high frequency gravitational waves from compact binary coalescenses
  with resonant cavities}} (\bibinfo {year} {2023}),\ \Eprint
  {https://arxiv.org/abs/2303.06006} {arXiv:2303.06006 [gr-qc]} \BibitemShut
  {NoStop}%
\bibitem [{\citenamefont {Domcke}\ \emph {et~al.}(2023)\citenamefont {Domcke},
  \citenamefont {Garcia-Cely}, \citenamefont {Lee},\ and\ \citenamefont
  {Rodd}}]{Domcke:2023bat}%
  \BibitemOpen
  \bibfield  {author} {\bibinfo {author} {\bibfnamefont {V.}~\bibnamefont
  {Domcke}}, \bibinfo {author} {\bibfnamefont {C.}~\bibnamefont {Garcia-Cely}},
  \bibinfo {author} {\bibfnamefont {S.~M.}\ \bibnamefont {Lee}},\ and\ \bibinfo
  {author} {\bibfnamefont {N.~L.}\ \bibnamefont {Rodd}},\ }\href@noop {}
  {\bibinfo {title} {Symmetries and selection rules: Optimising axion
  haloscopes for gravitational wave searches}} (\bibinfo {year} {2023}),\
  \Eprint {https://arxiv.org/abs/2306.03125} {arXiv:2306.03125 [hep-ph]}
  \BibitemShut {NoStop}%
\bibitem [{\citenamefont {Mourou}\ \emph {et~al.}(2006)\citenamefont {Mourou},
  \citenamefont {Tajima},\ and\ \citenamefont {Bulanov}}]{RevModPhys.78.309}%
  \BibitemOpen
  \bibfield  {author} {\bibinfo {author} {\bibfnamefont {G.~A.}\ \bibnamefont
  {Mourou}}, \bibinfo {author} {\bibfnamefont {T.}~\bibnamefont {Tajima}},\
  and\ \bibinfo {author} {\bibfnamefont {S.~V.}\ \bibnamefont {Bulanov}},\
  }\bibfield  {title} {\bibinfo {title} {Optics in the relativistic regime},\
  }\href {https://doi.org/10.1103/RevModPhys.78.309} {\bibfield  {journal}
  {\bibinfo  {journal} {Rev. Mod. Phys.}\ }\textbf {\bibinfo {volume} {78}},\
  \bibinfo {pages} {309} (\bibinfo {year} {2006})}\BibitemShut {NoStop}%
\bibitem [{\citenamefont {Mourou}\ \emph {et~al.}(2014)\citenamefont {Mourou},
  \citenamefont {Mironov}, \citenamefont {Khazanov},\ and\ \citenamefont
  {Sergeev}}]{Mourou}%
  \BibitemOpen
  \bibfield  {author} {\bibinfo {author} {\bibfnamefont {G.}~\bibnamefont
  {Mourou}}, \bibinfo {author} {\bibfnamefont {S.}~\bibnamefont {Mironov}},
  \bibinfo {author} {\bibfnamefont {E.}~\bibnamefont {Khazanov}},\ and\
  \bibinfo {author} {\bibfnamefont {A.}~\bibnamefont {Sergeev}},\ }\bibfield
  {title} {\bibinfo {title} {Single cycle thin film compressor opening the door
  to zeptosecond-exawatt physics},\ }\href
  {https://doi.org/10.1140/epjst/e2014-02171-5} {\bibfield  {journal} {\bibinfo
   {journal} {The European Physical Journal Special Topics}\ }\textbf {\bibinfo
  {volume} {223}},\ \bibinfo {pages} {1181} (\bibinfo {year}
  {2014})}\BibitemShut {NoStop}%
\bibitem [{\citenamefont {Li}\ and\ \citenamefont {Kawanaka}(2019)}]{Li:19}%
  \BibitemOpen
  \bibfield  {author} {\bibinfo {author} {\bibfnamefont {Z.}~\bibnamefont
  {Li}}\ and\ \bibinfo {author} {\bibfnamefont {J.}~\bibnamefont {Kawanaka}},\
  }\bibfield  {title} {\bibinfo {title} {Possible method for a single-cycle 100
  petawatt laser with wide-angle non-collinear optical parametric chirped pulse
  amplification},\ }\href {https://doi.org/10.1364/OSAC.2.001125} {\bibfield
  {journal} {\bibinfo  {journal} {OSA Continuum}\ }\textbf {\bibinfo {volume}
  {2}},\ \bibinfo {pages} {1125} (\bibinfo {year} {2019})}\BibitemShut
  {NoStop}%
\bibitem [{\citenamefont {Liu}\ \emph {et~al.}(2021)\citenamefont {Liu},
  \citenamefont {Shen}, \citenamefont {Du},\ and\ \citenamefont {Li}}]{Liu:21}%
  \BibitemOpen
  \bibfield  {author} {\bibinfo {author} {\bibfnamefont {J.}~\bibnamefont
  {Liu}}, \bibinfo {author} {\bibfnamefont {X.}~\bibnamefont {Shen}}, \bibinfo
  {author} {\bibfnamefont {S.}~\bibnamefont {Du}},\ and\ \bibinfo {author}
  {\bibfnamefont {R.}~\bibnamefont {Li}},\ }\bibfield  {title} {\bibinfo
  {title} {Multistep pulse compressor for 10s to 100s pw lasers},\ }\href
  {https://doi.org/10.1364/OE.424356} {\bibfield  {journal} {\bibinfo
  {journal} {Opt. Express}\ }\textbf {\bibinfo {volume} {29}},\ \bibinfo
  {pages} {17140} (\bibinfo {year} {2021})}\BibitemShut {NoStop}%
\bibitem [{\citenamefont {Han}\ \emph {et~al.}(2023)\citenamefont {Han},
  \citenamefont {Li}, \citenamefont {Zhang}, \citenamefont {Kong},
  \citenamefont {Cao}, \citenamefont {Jin}, \citenamefont {Leng}, \citenamefont
  {Li},\ and\ \citenamefont {Shao}}]{Han}%
  \BibitemOpen
  \bibfield  {author} {\bibinfo {author} {\bibfnamefont {Y.}~\bibnamefont
  {Han}}, \bibinfo {author} {\bibfnamefont {Z.}~\bibnamefont {Li}}, \bibinfo
  {author} {\bibfnamefont {Y.}~\bibnamefont {Zhang}}, \bibinfo {author}
  {\bibfnamefont {F.}~\bibnamefont {Kong}}, \bibinfo {author} {\bibfnamefont
  {H.}~\bibnamefont {Cao}}, \bibinfo {author} {\bibfnamefont {Y.}~\bibnamefont
  {Jin}}, \bibinfo {author} {\bibfnamefont {Y.}~\bibnamefont {Leng}}, \bibinfo
  {author} {\bibfnamefont {R.}~\bibnamefont {Li}},\ and\ \bibinfo {author}
  {\bibfnamefont {J.}~\bibnamefont {Shao}},\ }\bibfield  {title} {\bibinfo
  {title} {400nm ultra-broadband gratings for near-single-cycle 100 petawatt
  lasers},\ }\href {https://doi.org/10.1038/s41467-023-39164-3} {\bibfield
  {journal} {\bibinfo  {journal} {Nature Communications}\ }\textbf {\bibinfo
  {volume} {14}},\ \bibinfo {pages} {3632} (\bibinfo {year}
  {2023})}\BibitemShut {NoStop}%
\bibitem [{\citenamefont {Kleinman}(1962)}]{PhysRev.128.1761}%
  \BibitemOpen
  \bibfield  {author} {\bibinfo {author} {\bibfnamefont {D.~A.}\ \bibnamefont
  {Kleinman}},\ }\bibfield  {title} {\bibinfo {title} {Theory of second
  harmonic generation of light},\ }\href
  {https://doi.org/10.1103/PhysRev.128.1761} {\bibfield  {journal} {\bibinfo
  {journal} {Phys. Rev.}\ }\textbf {\bibinfo {volume} {128}},\ \bibinfo {pages}
  {1761} (\bibinfo {year} {1962})}\BibitemShut {NoStop}%
\bibitem [{\citenamefont {Brandenberger}\ \emph {et~al.}(2023)\citenamefont
  {Brandenberger}, \citenamefont {Delgado}, \citenamefont {Ganz},\ and\
  \citenamefont {Lin}}]{Brandenberger:2022xbu}%
  \BibitemOpen
  \bibfield  {author} {\bibinfo {author} {\bibfnamefont {R.}~\bibnamefont
  {Brandenberger}}, \bibinfo {author} {\bibfnamefont {P.~C.~M.}\ \bibnamefont
  {Delgado}}, \bibinfo {author} {\bibfnamefont {A.}~\bibnamefont {Ganz}},\ and\
  \bibinfo {author} {\bibfnamefont {C.}~\bibnamefont {Lin}},\ }\bibfield
  {title} {\bibinfo {title} {{Graviton to photon conversion via parametric
  resonance}},\ }\href {https://doi.org/10.1016/j.dark.2023.101202} {\bibfield
  {journal} {\bibinfo  {journal} {Phys. Dark Univ.}\ }\textbf {\bibinfo
  {volume} {40}},\ \bibinfo {pages} {101202} (\bibinfo {year} {2023})},\
  \Eprint {https://arxiv.org/abs/2205.08767} {arXiv:2205.08767 [gr-qc]}
  \BibitemShut {NoStop}%
  \bibitem [{\citenamefont {Herman}\ \emph {et~al.}(2023)\citenamefont
  {Herman}, \citenamefont {Lehoucq},\ and\
  \citenamefont {Fűzfa}}]{Herman:PhysRevD.108.124009}%
  \BibitemOpen
  \bibfield  {author} {\bibinfo {author} {\bibfnamefont {N.}~\bibnamefont
  {Herman}}, \bibinfo {author} {\bibfnamefont {L.}\ \bibnamefont
  {Lehoucq}},\ and\
  \bibinfo {author} {\bibfnamefont {A.}~\bibnamefont {Fűzfa}},\ }\bibfield
  {title} {\bibinfo {title} {{Electromagnetic antennas for the resonant detection of the stochastic gravitational wave background}},\ }\href {10.1103/PhysRevD.108.124009} {\bibfield
  {journal} {\bibinfo  {journal} {Phys. Rev. D}\ }\textbf {\bibinfo
  {volume} {108}},\ \bibinfo {pages} {124009} (\bibinfo {year} {2023})},\
  \Eprint {https://link.aps.org/doi/10.1103/PhysRevD.108.124009} 
  \BibitemShut %
  \bibitem [{\citenamefont {Herman}\ \emph {et~al.}(2023)\citenamefont
  {Herman}, \citenamefont {Lehoucq},\ and\
  \citenamefont {Fűzfa},\ and\ \citenamefont {Clesse}}]{Herman:PhysRevD.104.023524}%
  \BibitemOpen
  \bibfield  {author} {\bibinfo {author} {\bibfnamefont {N.}~\bibnamefont
  {Herman}}, \bibinfo {author} {\bibfnamefont {L.}\ \bibnamefont
  {Lehoucq}},
  \bibinfo {author} {\bibfnamefont {A.}~\bibnamefont {Fűzfa}},\ and\ \bibinfo {author} {\bibfnamefont {S.}~\bibnamefont {Clesse}},\ }\bibfield
  {title} {\bibinfo {title} {{Detecting planetary-mass primordial black holes with resonant electromagnetic gravitational-wave detectors}},\ }\href {10.1103/PhysRevD.104.023524} {\bibfield
  {journal} {\bibinfo  {journal} {Phys. Rev. D}\ }\textbf {\bibinfo
  {volume} {104}},\ \bibinfo {pages} {023524} (\bibinfo {year} {2021})},\
  \Eprint {https://link.aps.org/doi/10.1103/PhysRevD.104.023524} 
  \BibitemShut %

\bibitem [{\citenamefont {Capozziello}\ \emph {et~al.}(2023)\citenamefont
  {Herman} and\ \citenamefont {Bajardi}}]{Capozziello:2019klx}%
  \BibitemOpen
  \bibfield  {author} {\bibinfo {author} {\bibfnamefont {S.}~\bibnamefont
  {Capozziello}},\ and\ \bibinfo {author} {\bibfnamefont {F.}~\bibnamefont {Bajardi}},\ }\bibfield
  {title} {\bibinfo {title} {{Gravitational waves in modified gravity}},\ }\href {10.1142/S0218271819420021} {\bibfield
  {journal} {\bibinfo  {journal} {Int. J. Mod. Phys. D}\ }\textbf {\bibinfo
  {volume} {28}},\ \bibinfo {pages} {1942002} (\bibinfo {year} {2019})},\
  \Eprint {https://doi.org/10.1142/S0218271819420021} 
  \BibitemShut %
  \bibitem [{\citenamefont {Zel'dovich}(1974)}]{zeldovich1973}%
  \BibitemOpen
  \bibfield  {author} {\bibinfo {author} {\bibnamefont {Ya.~B.}~\bibnamefont {Zel'dovich}},\ }\bibfield
  {title} {\bibinfo {title} {{Electromagnetic and gravitational waves in a stationary magnetic field}},\ }\href {https://ui.adsabs.harvard.edu/abs/1974JETP...38..652Z} {\bibfield
  {journal} {\bibinfo  {journal} {Soviet Journal of Experimental and Theoretical Physics}\ }\textbf {\bibinfo
  {volume} {38}},\ \bibinfo {pages} {652} (\bibinfo {year} {1974})},\ \Eprint{https://ui.adsabs.harvard.edu/abs/1974JETP...38..652Z}
  \BibitemShut

  \bibitem [{\citenamefont {Anastassopoulos}\ \emph {et~al.}(2017)\citenamefont
  {Anastassopoulos} and\ \citenamefont {others}}]{CAST:2017uph}%
  \BibitemOpen
  \bibfield  {author} {\bibinfo {author} {\bibfnamefont {V.}~\bibnamefont
  {Anastassopoulos}} \emph {et~al.} (\bibinfo {collaboration} {CAST})},\ \bibfield
  {title} {\bibinfo {title} {{New CAST Limit on the Axion-Photon Interaction}},\ }\href {https://arxiv.org/abs/1705.02290} {\bibfield
  {journal} {\bibinfo  {journal} {Nature Phys.}\ }\textbf {\bibinfo
  {volume} {13}},\ \bibinfo {pages} {584--590} (\bibinfo {year} {2017})},\
  \Eprint {https://arxiv.org/abs/1705.02290} \BibitemShut

  \bibitem [{\citenamefont {Ballou}\ \emph {et~al.}(2015)\citenamefont
  {Ballou} and\ \citenamefont {others}}]{OSQAR:2015qdv}%
  \BibitemOpen
  \bibfield  {author} {\bibinfo {author} {\bibfnamefont {R.}~\bibnamefont
  {Ballou}} \emph {et~al.} (\bibinfo {collaboration} {OSQAR})},\ \bibfield
  {title} {\bibinfo {title} {{New exclusion limits on scalar and pseudoscalar axionlike particles from light shining through a wall}},\ }\href {https://arxiv.org/abs/1506.08082} {\bibfield
  {journal} {\bibinfo  {journal} {Phys. Rev. D}\ }\textbf {\bibinfo
  {volume} {92}},\ \bibinfo {number} {9},\ \bibinfo {pages} {092002} (\bibinfo {year} {2015})},\
  \Eprint {https://arxiv.org/abs/1506.08082} \BibitemShut

  \bibitem [{\citenamefont {Ejlli}\ \emph {et~al.}(2019)\citenamefont
  {Ejlli}, \citenamefont {Ejlli}, \citenamefont {Cruise}, \citenamefont {Pisano},\ and\ \citenamefont {Grote}}]{Ejlli:2019bqj}%
  \BibitemOpen
  \bibfield  {author} {\bibinfo {author} {\bibfnamefont {Aldo}~\bibnamefont
  {Ejlli}}, \bibinfo {author} {\bibfnamefont {Damian}~\bibnamefont {Ejlli}}, \bibinfo {author} {\bibfnamefont {Adrian~Mike}~\bibnamefont {Cruise}}, \bibinfo {author} {\bibfnamefont {Giampaolo}~\bibnamefont {Pisano}},\ and\ \bibinfo {author} {\bibfnamefont {Hartmut}~\bibnamefont {Grote}}},\ \bibfield
  {title} {\bibinfo {title} {{Upper limits on the amplitude of ultra-high-frequency gravitational waves from graviton to photon conversion}},\ }\href {https://arxiv.org/abs/1908.00232} {\bibfield
  {journal} {\bibinfo  {journal} {Eur. Phys. J. C}\ }\textbf {\bibinfo
  {volume} {79}},\ \bibinfo {number} {12},\ \bibinfo {pages} {1032} (\bibinfo {year} {2019})},\
  \Eprint {https://arxiv.org/abs/1908.00232} \BibitemShut%

  
  \bibitem [{\citenamefont {Beacham}\ \emph {et~al.}(2020)\citenamefont
  {Beacham} and\ \citenamefont {others}}]{Beacham:2019nyx}%
  \BibitemOpen
  \bibfield  {author} {\bibinfo {author} {\bibfnamefont {J.}~\bibnamefont
  {Beacham}} \emph {et~al.}},\ \bibfield
  {title} {\bibinfo {title} {{Physics Beyond Colliders at CERN: Beyond the Standard Model Working Group Report}},\ }\href {https://arxiv.org/abs/1901.09966} {\bibfield
  {journal} {\bibinfo  {journal} {J. Phys. G}\ }\textbf {\bibinfo
  {volume} {47}},\ \bibinfo {number} {1},\ \bibinfo {pages} {010501} (\bibinfo {year} {2020})},\
  \Eprint {https://arxiv.org/abs/1901.09966} \BibitemShut%
  
\bibitem{Alho:2022bki}
A.~Alho, J.~Nat\'ario, P.~Pani and G.~Raposo,
{title} {\bibinfo {title} {{Compactness bounds in general relativity}},
\ }\href {https://arxiv.org/abs/2202.00043}
{\bibfield
  {journal} {\bibinfo  {journal} {Phys. Rev. D}\ }\textbf {\bibinfo
  {volume} {106}},\ \bibinfo {number} {4},\ \bibinfo {pages} {L041502} (\bibinfo {year} {2022})},\
\Eprint {https://arxiv.org/abs/2202.00043}
\BibitemShut




\end{thebibliography}
\end{document}